\documentclass[useAMS,usenatbib]{mn2e}
\pdfoutput=1
\usepackage{journals}
\usepackage[pdftex]{graphicx,color}
\usepackage[T1]{fontenc}
\usepackage{graphics}
\usepackage{amsfonts}
\usepackage{amsmath}
\usepackage{multicol}
\usepackage{layout}
\usepackage{amssymb}
\usepackage{hhline}
\usepackage[a4paper,colorlinks=true,pdfstartview=FitV,
linkcolor=red,citecolor=blue,urlcolor=magenta,draft]{hyperref}
   \title[Halo   Mass   Function]   {The
universality of the virial halo mass function and models for non-universality of other halo definitions}  \author[Despali et al.   2015] {\parbox{\textwidth}{Giulia Despali$^{1,2}$\thanks{E-mail:         \href{mailto:gdespali@gmail.com}
{gdespali@gmail.com}},   Carlo   Giocoli$^{1}$,   Raul E.  Angulo$^{3}$,
Giuseppe  Tormen$^2$, Ravi  K. Sheth$^4$,  Giacomo Baso$^2$  and Lauro
Moscardini$^{5,6,7}$ \\}\\
$^1$ Aix
Marseille  Universit\'e,  CNRS,  LAM (Laboratoire  d'Astrophysique  de
Marseille)  UMR 7326,  13388, Marseille,  France \\ 
$^{2}$ Dipartimento  di Fisica  e Astronomia,  Universit\`{a} degli  Studi di
Padova, vicolo dell'Osservatorio 3,  35122, Padova, Italy \\  
$^{3}$  Centro de
Estudios de F\'isica del Cosmos de Arag\'on (CEFCA), Plaza San Juan 1,
Planta-2,  44001, Teruel,  Spain\\  
$^{4}$ Center for Particle Cosmology, 
University   of   Pennsylvania,   209  South   33rd   St,
Philadelphia, PA 19104 \\
$^{5}$ Dipartimento di Fisica e Astronomia, Alma Mater Studiorum Universit\'a di Bologna, Viale Berti Pichat 6/2, 40127 Bologna, Italia \\
$^{6}$  INAF, Osservatorio Astronomico di Bologna, via Ranzani 1, 40127 Bologna, Italia\\
$^{7}$ INFN, Sezione di Bologna, viale Berti Pichat 6/2, 40127 Bologna, Italia\\}

\begin{document} \date{}
\maketitle
\label{firstpage}   \pagerange{\pageref{firstpage}--\pageref{lastpage}}
\pubyear{2015}
\begin{abstract}
  The abundance of galaxy clusters can constrain both the geometry and
  growth of  structure in our  Universe. However, this probe  could be
  significantly  complicated by  recent claims  of nonuniversality  --
  non-trivial dependences  with respect to the  cosmological model and
  redshift.   In this  work  we  analyse the  dependance  of the  mass
  function on the way haloes are  identified and establish if this can
  cause  departures  from  universality.   In order  to  explore  this
  dependance,  we   use  a   set  of  different   N-body  cosmological
  simulations (Le SBARBINE simulations),  with the latest cosmological
  parameters  from  the  Planck  collaboration; this  first  suite  of
  simulations is followed by a  lower resolution set, carried out with
  different cosmological  parameters.  We identify dark  matter haloes
  using  a Spherical  Overdensity algorithm  with varying  overdensity
  thresholds   (virial,   2000$\rho_c$,   1000$\rho_c$,   500$\rho_c$,
  200$\rho_c$ and 200$\rho_b$) at all redshifts.  We notice that, when
  expressed in term of the  rescaled variable $\nu$, the mass function
  for virial  haloes is a nearly  universal as a function  of redshift
  and cosmology,  while this  is clearly  not the  case for  the other
  overdensities we  considered.  We provide fitting  functions for the
  halo mass  function parameters as  a function of  overdensity, that
  allow to  predict, to within a  few percent accuracy, the  halo mass
  function  for  a  wide  range of  halo  definitions,  redshifts  and
  cosmological  models.    We  then  show  how   the  departures  from
  universality associated  with other halo definitions  can be derived
  by  combining the  universality of  the virial  definition with  the
  expected shape of the density profile of halos.

\end{abstract}

\section{introduction}

In the Cold Dark Matter (CDM) model, structures -- up to protogalactic
scales -- form through the amplification of small density fluctuations
via                      gravitational                     instability
\citep{longair98,springel05b,mo10,angulo12}.   Dark matter  haloes are
objects  which  have  been  able  to break  away  from  the  expanding
background,  and collapse  \citep{press74,springel01b}.  Small  haloes
form first, before merging with one  another to form ever more massive
ones in  a hierarchical process \citep{lacey93,lacey94}.   As a result
of  repeated mergers,  dark matter  haloes grow  more massive  in time
\citep{tormen04}.  Haloes  hosting galaxy clusters represent  the most
massive and recently formed structures  in our Universe. The formation
and  the merger  rates  of dark  matter haloes  are  sensitive to  the
expansion history  of the  Universe and  so can  be used  to constrain
cosmological parameters \citep{lacey93,lacey94,moreno08}.

In particular, different theoretical studies  have shown that the halo
mass function and its evolution are important probes of the very early
Universe,   its  expansion   history,  and   the  nature   of  gravity
\citep{press74,bond91,lacey93,sheth01b}.   These have  shown that,  in
appropriately scaled  units, the  mass function can  be written  in an
approximately universal  form which  is independent of  power spectrum
and expansion history.  Although this universality is only expected to
be approximate  \citep{musso12,paranjape13}, it vastly  simplifies the
process  of constraining  cosmological  parameters from  observational
datasets, so  it has served as  the basis for fitting  functions whose
parameters  are calibrated  using  numerical simulations  of the  dark
matter
\citep{sheth99b,jenkins01,warren06,sheth02,delpopolo98,delpopolo99}.
As simulated datasets  have grown, it has become  possible to quantify
(small)            departures             from            universality
\citep{tinker08,crocce10,manera10,wu10,courtin11,corasaniti11,murray13,watson13}.
Several  recent  works  have  also   concentrated  on  the  small  but
significant modifications  by neutrinos  \citep{castorina14}, coupling
between dark matter-dark  energy \citep{cui12,giocoli13} and different
baryon physics \citep{cui14,bocquet15}.

Many cosmological  constraints have been obtained  from cluster counts
\citep{vikhlinin09,rozo10,planckxx}    and   other    galaxy   cluster
properties  \citep{evrard08,ettori09,giocoli12c}.   For  an  extensive
review,  see  \citet{borgani11}.  However  in  the  near future,  many
wide-field surveys  are expected to  use the cluster mass  function to
constrain                    cosmological                   parameters
\citep{pillepich12,euclidredbook,sartoris15,boldrin15}.  In  the light
of these, a mass function calibrated  to an accuracy of a few percent,
and  flexible  enough to  account  for  different halo  identification
definitions, is of primary importance.

Departures of the  halo mass function from universality  may depend on
how haloes  are defined.  One  of the main goals  of this paper  is to
explore  this dependance.   Our major  result is  that, if  haloes are
defined using the  virial density, and the fitting  formula includes a
parameter which is related to this  -- as was done by \citet{sheth99b}
-- then  the  mass function can  be  considered universal  to
  within a few percent.  In  this respect, our findings confirm those
of  \citet{courtin11}:   departures  from  universality   result  from
ignoring the  redshift and  cosmology dependance of  these quantities.
Moreover, the departures from  universality associated with other halo
definitions  can be  derived from  combining the  universality of  the
virial  definition  with knowledge  of  the  enclosed density  profile
around haloes.

The  paper  is  organised  as follows:  In  Sec.~\ref{simulations}  we
describe the cosmological  simulations we use for our  study.  It also
describes our  halo finder.   We present our  reference model  for the
halo mass function  in Sec.~\ref{secModelHMF}.  In Secs.~\ref{secuniq}
and~\ref{secrescaling}  we discuss  the  universality associated  with
using  the  virial overdensity  to  define  haloes  and show  how  the
parameters of the halo mass function depend on how haloes are defined.
The bulk of  this analysis is for  spherical halos: Appendix~\ref{EOs}
shows  how  our results  are  modified  if  halos  are allowed  to  be
ellipsoidal,  and Appendix~\ref{ICs}  describes how  a number  of more
technical details affect our  measurements.  In Sec.~\ref{matching} we
show  how  the nonvirial  halo  mass  functions  can be  derived  from
combining knowledge of the dark matter density profile with the virial
mass  function.   In  Sec.~\ref{comparison}  we  present  some
  comparisons with previous works. We discuss our results and conclude
  in  Sec.~\ref{conclusions}.  Our  analysis  suggests  that the  most
  massive end of the halo mass function is particularly simple, in the
  sense  that it  can  be  described by  a  function  with fewer  free
  parameters.  Appendix~\ref{cmf}  describes how this  impacts cluster
  cosmology. All  logarithms where not  explicitly stated in  the text
  are in base ten.

\section{The Numerical Simulations}\label{simulations}
\subsection{Le SBARBINE simulations}

Le SBARBINE simulations are a set of six dark-matter-only cosmological
simulations  run by  the  Padova cosmology  group.  These  simulations
follow  the  evolution  of  $1024^{3}$ particles,  whose  motions  are
assumed to be driven by  gravitational instability, using the publicly
available code  GADGET-2 \citep{springel05a}.  The  assumed background
cosmology and  initial conditions for  these runs are  consistent with
recent  Planck  results  \citep{planckxvi} (hereafter  Planck13).   In
particular we have  set: $\Omega_{m}=0.307$, $\Omega_{\Lambda}=0.693$,
$\sigma_{8}=0.829$ and $H  = 100h$km~s$^{-1}$~$h^{-1}Mpc$ with $h=0.677$.   
The initial power spectrum was generated using the CAMB code 
\citep{camb}, and initial conditions  were  produced by  perturbing  
a  glass distribution  with N-GenIC 
(\url{http://www.mpa-garching.mpg.de/gadget}); the realisations have
been carefully chosen, in order to follow the initial power spectrum
even at large scales and thus to reduce the differences between the
linear spectrum and that measured from the simulations.

\begin{table*} \centering
\begin{tabular}{|c|c|c|c|c|c|c|c|}   \hline 
\multicolumn{8}{c}{Main set of simulations}  \\ \hline  
name &  box [ $h^{-1}$Mpc] &  $z_{i}$ &
$m_{p}$[$M_{\odot}h^{-1}$] & soft [kpc $h^{-1}$] & $N_{h-tot} (z=0)$ &
$N_{h>300}  (z=0)$ &  colour \\  \hline \textbf{Ada}  & 62.5  &  124 &
$1.94\times 10^{7}$ & 1.5 &  2264847 &103852 & green\\ \textbf{Bice} &
125  & 99  & $1.55  \times 10^{8}$  & 3  & 2750411  & 129674  & cyan\\
\textbf{Cloe} & 250 & 99 & $1.24 \times 10^{9}$ & 6 & 3300880 & 161580
&  blue\\ \textbf{Dora}  & 500  & 99  & $9.92  \times 10^{9}$  &  12 &
3997898 & 191793 & magenta\\ \textbf{Emma}  & 1000 & 99 & $7.94 \times
10^{10}$ & 24  &4739379 & 176633 & red\\ \textbf{Flora} &  2000 & 99 &
$6.35 \times 10^{11}$ & 48 & 5046663 & 75513 & orange\\ \hline
\end{tabular}
\caption{Features  of  Le  SBARBINE   simulations  run  with  Planck13
  parameters       $\Omega_{m}=0.307$,       $\Omega_{\Lambda}=0.693$,
  $\sigma_{8}=0.829$  and  $h=0.677$  and containing  $1024^{3}$  dark
  matter particles.  The  last two columns report the  total number of
  haloes identified with the Spherical Overdensity at redshift $z = 0$
  that   are  resolved   with  more   than  10   and  300   particles,
  respectively.\label{tab_sim1}}
\end{table*}
\begin{table*} \centering
\begin{tabular}{|c|c|c|c|c|c|c|} \hline 
\multicolumn{7}{c}{Secondary set of simulations}  \\ \hline  
 name &
$\Omega_{m}$  &  $\Omega_{\Lambda}$  &  $\sigma_{8}$  &  box  [$h^{-1}$Mpc]  &
$m_{p}$[$M_{\odot}h^{-1}$] & colour\\ 
\hline  
\textbf{Tea} & 0.2 & 0.8
& 0.7 & 150 & $1.396 \times 10^{9}$ & gray-square\\ 
\textbf{Tea-big} &
0.2  & 0.8  & 0.7  &  1000 &  $4.135 \times  10^{11}$ &  gray-square\\
\textbf{Tina}  & 0.2  &  0.8 &  0.9 &  150  & $1.396  \times 10^{9}$  &
gray-triangle \\  
\textbf{Tina-big} & 0.2 &  0.8 & 0.9 &  1000 & $4.135
\times 10^{11}$& gray-triangle\\ 
\textbf{Vera} &  0.4 & 0.6 & 0.7 & 150
& $2.791 \times 10^{9}$ &  brown-square\\ 
\textbf{Vera-big} & 0.4 & 0.6
& 0.7 & 1000 & $8.271 \times 10^{11}$& brown-square\\ 
\textbf{Viola}
& 0.4  & 0.6 &  0.9 & 150  & $2.791 \times 10^{9}$  & brown-triangle\\
\textbf{Viola-big} & 0.4 & 0.6  & 0.9 & 1000 & $8.271 \times10^{11}$
&  brown-triangle\\ 
\textbf{Wanda (wmap7)}  & 0.272  & 0.728  & 0.81  &  150 &
$1.898  \times 10^{9}$  & blue-circle\\  
\textbf{Wanda-big (wmap7)} &  0.272 &
0.728 & 0.81 & 1000 & $5.624 \times 10^{11}$&blue-circle\\ \hline
\end{tabular}
\caption{Details of  the small set  of 10 simulations  with different
  cosmological  parameters.   Each   contains  $512^{3}$  dark  matter
  particles with initial conditions  generated at redshift $z=99$. For
  all the  models the Hubble parameter is $h=0.6777$,  apart from
  the WMAP7 cosmology for which $h=0.704$.\label{tab_sim2}}
\end{table*}

The   parameters  of   our   main  simulation   set   are  listed   in
Table~\ref{tab_sim1}.  We used a different  seed for the random number
generator which sets the initial  conditions of each simulation, so as
to  have a  sample  of independent  realisations.   Although each  box
contains the  same number of  dark matter particles  ($1024^{3}$), the
comoving box lengths are different, so the mass resolution in each
box is different.  The  box sizes were  chosen so that the set provides 
good mass resolution down to $10^7\,h^{-1}M_{\odot}$.

Le SBARBINE simulations are complemented  by a set of lower resolution
runs  having   different  cosmological  parameters.  These   all  have
$512^{3}$  dark matter  particles.   In particular,  for  each set  of
cosmological  parameters we  ran  two simulations:  one with  box-size
$150h^{-1}$Mpc and another with $1000h^{-1}$Mpc.  These were chosen to
ensure good  resolution both  for intermediate  and high  mass haloes.
This  lower  resolution set  was  produced  specifically to  test  the
universality  of   the  halo  mass   function  with  respect   to  the
cosmological model  (see Section  \ref{sec_cosmo}). The  parameters of
these other simulations are listed in Table \ref{tab_sim2}.

In addition,  we also re-ran  three simulations for which  the initial
conditions were generated using a second order Lagrangian Perturbation
Theory                         (2LPT)                        algorithm
\footnote{\url{http://cosmo.nyu.edu/roman/2LPT}}     \citep{crocce06}.
These are copies  of Bice ($1024^3$ particles) and of  the two $512^3$
simulations with  the WMAP7 cosmology \citep{komatsu11}  namely (wmap7
and  wmap7-big).   As we  discuss  in  Appendix  B4, there  are  small
differences --  not exceeding $5\%$  -- between the mass  functions in
simulations with Zel'dovich versus 2LPT initial conditions.  But these
differences appear only for high  $\nu$ and high redshifts, which play
little role in our calibration of the mass function parameters.

All runs were  performed in Padova on ``Nemo'': a  SuperServer Twin 2U
Dual Xeon Sandy Bridge composed  by four independend node servers each
equipped with two Xeon Sandy Bridge 8  Core E5-2670 and 128 GB of RAM,
for a total of 64 cores or 128 CPU-threads and 512 GB of RAM.

\subsection{Halo  catalogues}  
For each stored particle snapshot we identify haloes using a Spherical
Overdensity                       (SO)                       algorithm
\citep[e.g.][]{tormen98a,tormen04,giocoli08a}.   We chose this rather 
than the Friends-of-Friends  (FoF) method of \citet{davis85}, because 
we believe it to be slightly closer to physical models of halo formation, 
and because it is quite similar to how mass is defined in observational 
data.  

For  each particle  distribution, we  estimate the  local dark  matter
density at the  position of each particle by  calculating the distance
$d_{i,10}$ to the tenth nearest neighbour.   In this way, we assign to
each particle a local density  $\rho_i \propto d_{i,10}^{-3}$. We then
sort particles in  density and choose as centre of  the first halo the
position  of the  densest  particle.   We grow  a  sphere around  this
centre, and stop when the mean density within the sphere falls below a
desired critical value.  At this  point we assign all particles within
the sphere  to the  newly identified  halo, and  remove them  from the
global list of particles.  We then  choose the densest particle of the
ones remaining,  and repeat (i.e., grow  a sphere around it  until the
mean enclosed  density falls below  threshold, etc.).  We  continue in
this manner until none of the  remaining particles has a local density
large enough to be  the center of a $10$ particle  halo (as we discuss
shortly,  we apply  a more  stringent  cut when  we fit  for the  mass
function); particles  not assigned to  any halo are called  `field' or
`dust' particles.

For  the  critical overdensity  we  adopt  six different  definitions:
$2000$, $1000$,  $500$ and $200\rho_c(z)$, $200\rho_b$  and the virial
value.  We chose  these values of overdensity since they  are (or they
are very close to) the  commonly used ones: $200\rho_{b}$ is motivated
by the spherical collapse model  in an Einstein-de-Sitter universe and
-- together   with    $200\rho_{c}$   --    is   a    popular   choice
\citep{tinker08};  moreover, $200\rho_{c}$  is  often  used to  define
galaxy cluster  masses; $500\rho_{c}$  (and the  higher overdensities)
are used  in X-ray analyses, and  in general in observations  that are
able  to  resolve  only  the   inner  parts  of  haloes.   The  virial
overdensity    depends   on    redshift    and   cosmological    model
\citep[e.g.][]{bryan98};   we   use   the   numerical   solutions   of
\citet{eke96}.

The comoving density of the background is
\begin{equation}
 \rho_{\rm com}\equiv\rho_{b} = \rho_{c}(0)\Omega_{m}(0) = \rho_{c}(z)\Omega_{m}(z),
\end{equation}
where 
$\rho_{c}(0)\equiv 3H_0^{2}/8\pi G 
           \simeq 2.775\times10^{11}h^{-1}M_{\odot}h^{3}{\rm  Mpc}^{-3}$ 
is the critical density.  
For the Planck cosmology adopted in the main set of simulations,
$\Delta_{vir}(z=0)\simeq  319\rho_b$ is greater than $200\rho_{b}$ 
(corresponding to $\simeq  98\rho_{c}$) and  lower than all  the  
other  thresholds  we  consider.   At  high  redshifts  both
$200\rho_b$ and  $200 \rho_c$ converge to the virial definition, 
the first from below and the second from above.

\begin{figure}
 \includegraphics[width=\hsize]{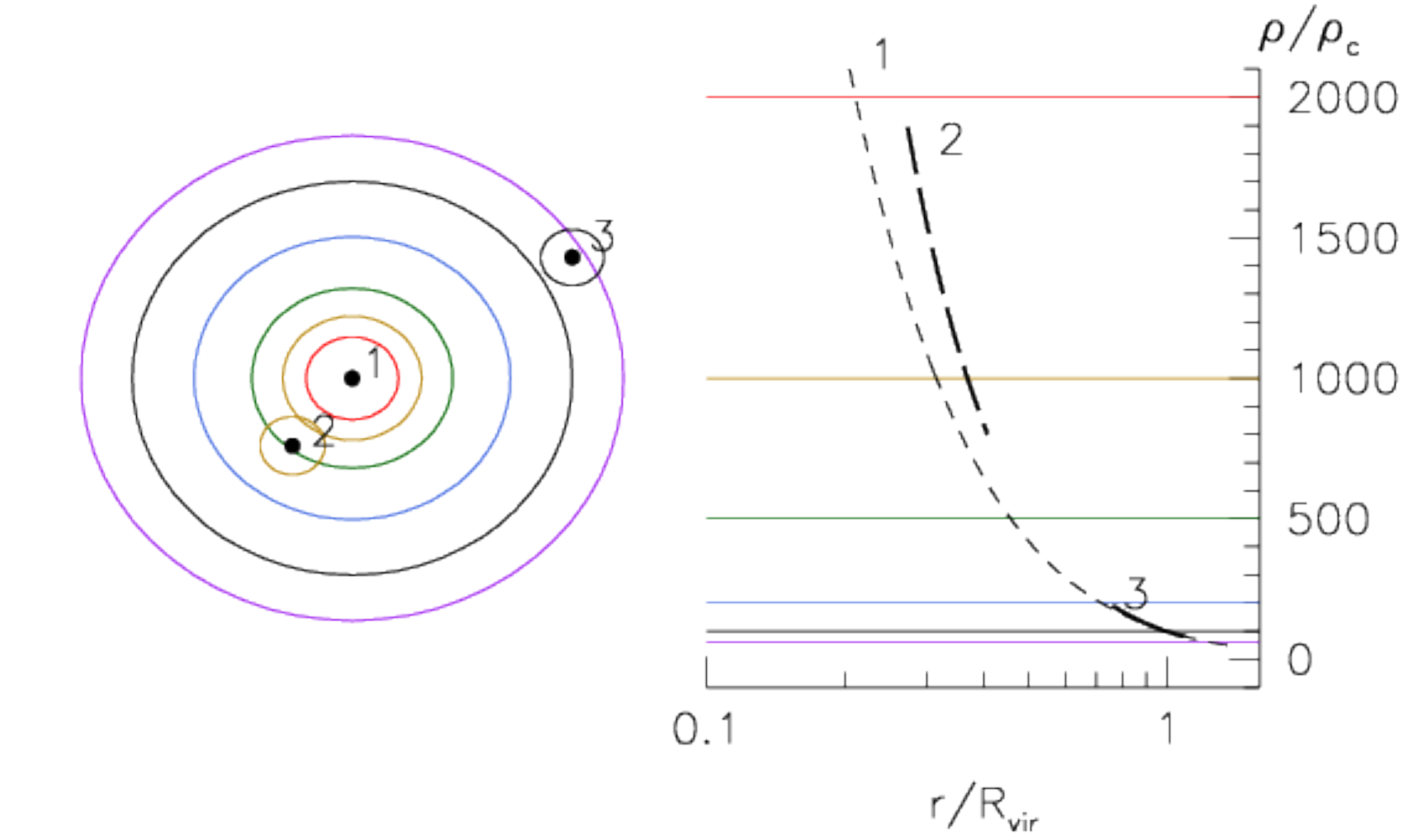}
 \caption{Schematic representation of the  halo identifications in the
   particle   density  distribution   at  $z=0$.    Different  colours
   represent  the various  overdensities,  from  $2000\rho_c$ down  to
   $200\rho_b$.  Since  our SO halo  finder starts to grow  the sphere
   starting  from the  densest particle,  the halo  centered at  $1$ is
   common  to all  the catalogues,  whereas the  ones at  $2$ and  $3$
   belong  only   to  $1000\rho_c$   and  to  the   virial  catalogue,
   respectively.   This is  clearer in  the  right hand  panel of  the
   figure which shows the density profiles of the three systems.
   \label{figSO}}
\end{figure}

Fig.~\ref{figSO}  shows  a  schematic  representation  of  the  haloes
identified  with our  spherical  overdensity finder.   In contrast  to
previous work,  in which  an FoF  catalogue is used  as the  basis for
subsequent  SO  identifications, we  run  our  halo finder  code  from
scratch for each  threshold.  Hence, although a halo  in one catalogue
may  be  present  in  another,  this  is  not  necessarily  true.   In
particular, while the halo centered at $1$ is common to all catalogues,
the ones at $2$ and $3$ belong  only to $1000\rho_c$ and to the virial
catalogue, respectively.  This  happens, as can be  noticed from their
density profiles  presented in the  right part of the  figure, because
while halo $1$ is dense enough  to go from overdensity 2000$\rho_c$ to
200$\rho_b$ -- so is in common to all catalogues -- haloes $2$ and $3$
reach only $1000\rho_c$ and  the virial overdensity, respectively, and
are present only in those catalogues.

Halo identification in each simulation  snapshot was done with the aim
of  studying  the   evolution  the  halo  mass   function.   For  each
snapshot-density   threshold  combination,   we   saved  a   catalogue
containing all the information about the identified haloes.  So as not
to  be biased  by  the mass  and force  resolution,  we only  consider
systems     resolved     with     at     least     $300$     particles
\citep{maccio07,maccio08,velliscig15}.   Therefore,  while  the  $z=0$
catalogues of  each simulation  contain many haloes  over a  wide mass
range,  only the  high  mass end  of the  higher  redshift outputs  is
reliably measured.

We    also    ran    an   Ellipsoidal    Overdensity    (EO)    finder
\citep{despali13,despali14,bonamigo15}, which  we used  for estimating
the triaxial properties of the  collapsed systems.  In what follows we
concentrate  on the  results for  SO haloes;  a brief  summary of  the
corresponding  results   for  ellipsoidal  haloes  can   be  found  in
Appendix~\ref{EOs}.  These  are broadly  similar to  spherical haloes,
although the  best fit mass  function parameters differ  slightly from
those for the SO case.

\section{Model  for the  halo mass  function} \label{secModelHMF}  
Let ${\rm d}n/{\rm d}\ln M$ denote the comoving number density of 
haloes in a logarithmic  bin ${\rm  d}\ln M$  around mass  $M$.  
Then,  the mass fraction in such haloes is
\begin{equation}
 f(M)\,{\rm d}M = \frac{M}{\rho_b}\,\frac{{\rm d}n}{{\rm d}\ln M}\,{\rm d}\ln M. 
\end{equation}

It is usual to define 
\begin{equation}
 \sigma^2(M,z) \equiv \int \frac{{\rm d}k}{k}\, 
             \frac{k^3P_{\rm lin}(k,z)}{2\pi^2}\, W^2[kR(M)] ,\label{eqsigma}
\end{equation}

\noindent  where  $P_{\rm lin}(k,z)$  is  the  initial power  spectrum
extrapolated  to  redshift  $z$  using  linear  theory,  $W(x)  \equiv
3\,j_1(x)/x$,  and $R(M)$  is  given by  requiring  $M/(4\pi R^3/3)  =
\rho_b$ (recall  that $\rho_b$  is comoving, so  it is  independent of
redshift).  
In principle, $\sigma(M,z)$ depends only on $P_{\rm lin}(k,z)$ and 
the smoothing filter $W$.  
In practice,  the number of Fourier modes that can be effectively sampled 
in the initial  conditions depends on some  computational limits (e.g. 
box size, number of particles, etc).  
As a result, there are differences  between the  actual power spectra 
in a box and the theoretical mean value.  Appendix B illustrates the 
impact  on $\sigma^2(M)$.

We compute $\sigma^2(M)$ for each  box using the actual realisation of
$P(k)$ in  it.  This reduces  bias and  scatter in the  mass function,
especially in  the high mass tail  of each simulation, allowing  us to
reach higher precision  (down from ten to a  few percent) particularly
in  the smaller  boxes within  which cosmic  variance would  otherwise
contribute substantially.   In particular, the difference  between the
theoretical  input  power spectrum  and  the  one measured  from  each
simulation can be  seen in Figure~B1, while  the associated difference
in  the mass  function  is shown  in the  bottom  panel of  Figure~B2.
Appendix~\ref{ICs} also discusses  the impact of box size  on the mass
functions,  and  compares a  number  of  different prescriptions  that
account for finite box-size effects on $\sigma^2(M)$.

Although  a  number of  workers  so  far  have parametrised  the  mass
function in terms of $\sigma$  alone (i.e., they write $f(M)\,{\rm d}M
= f(\sigma){\rm d}\sigma$ and work with $f(\sigma)$), \citet{sheth99b}
were careful to parametrize in terms of
\begin{equation}
 \nu \equiv \delta^2_{c}(z)/\sigma^2(M) ,
\end{equation}
where  $\delta_c(z)$   is  the  critical  linear   theory  overdensity
$\delta_{lin}(z)$ required for spherical collapse divided by the growth
factor  \citep{carroll92}.  This  quantity,  which  depends weakly  on
$\Omega$ for  the $\Lambda$CDM family of  models, is well-approximated
by
\begin{equation}
  \delta_{lin}(z) \approx \frac{3}{20}\,(12\pi)^{2/3}[1 + 0.0123\,\lg\Omega(z)]
\end{equation} 
\citep{kitayama96}.  The rationale for including it came from the fact
that they used  the virial density to find haloes,  and the same model
which  predicts   this  virial   density  also   predicts  $\delta_c$.
\citet{courtin11}  highlight   the  fact  that  this   factor  becomes
increasingly  important at  high  masses; failure  to  include it  can
masquerade as non-universality.

\begin{figure*}
\includegraphics[width=\hsize]{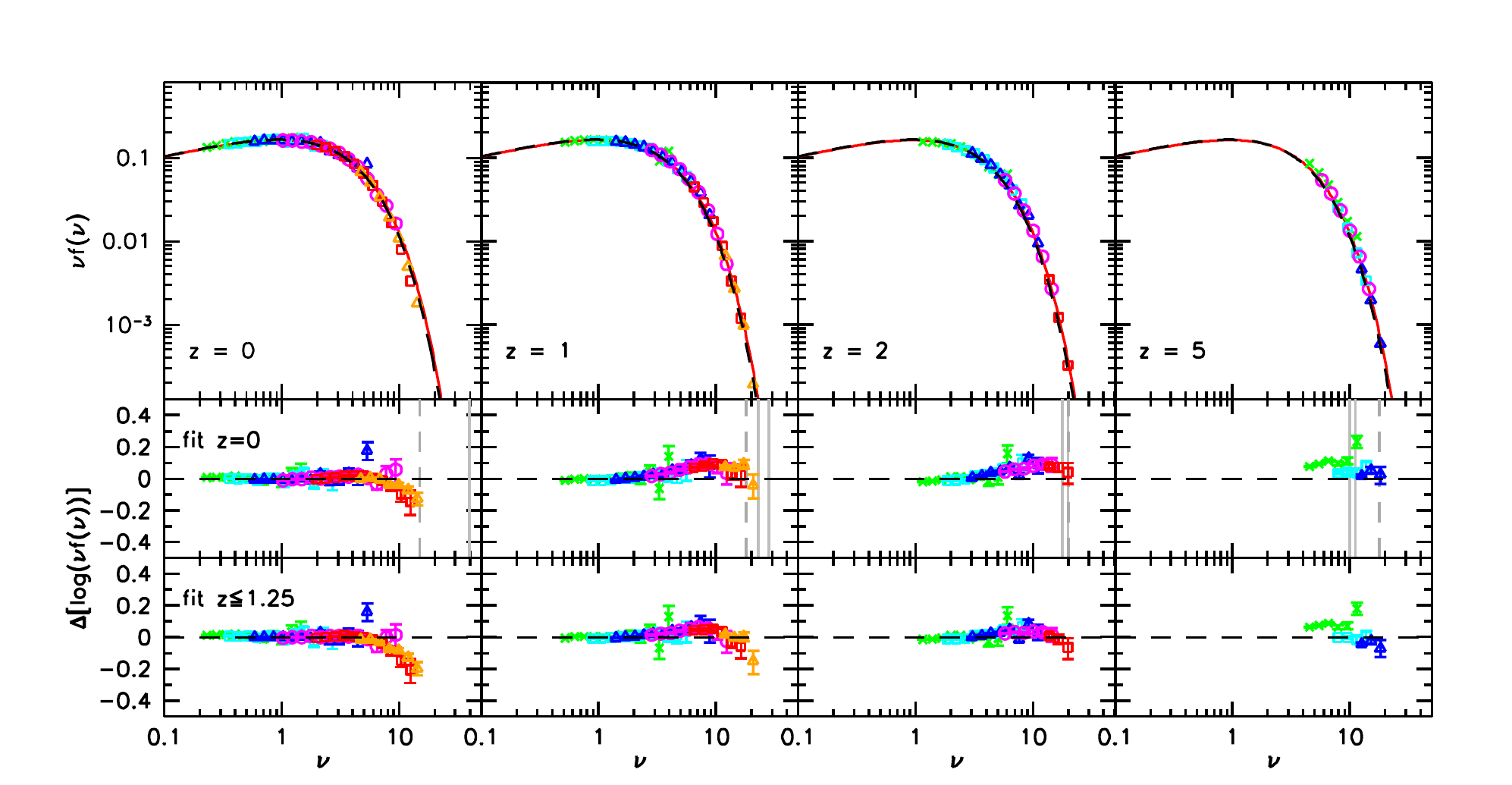}
\caption{Halo mass function at four  redshifts, for all SO haloes with
  more than 300 dark matter particles in the SO catalogues, identified
  using  the redshift  dependent virial  density.  Green,  cyan, blue,
  magenta red  and orange  symbols show results  for Ada,  Bice, Cloe,
  Dora, Emma, and Flora. Black  dashed line, same in all panels, shows
  the result  of fitting the $z=0$ points  to Eq.~(\ref{st99mf}).  The
  middle panels show the residuals  (in log space) from this best fit.
  The dashed vertical lines show  the maximum $\nu$ for which the bins
  contain at least 100 objects.  The two solid vertical lines show the
  minimum $\nu$ at  which we may expect differences  of at least $5\%$
  in  the halo mass  function accounting  for 2LPT  initial condition,
  adopting the  rescaling of equation~(11) by  \citet{reed13}, for the
  simulations starting  at $z=99$ and  $z=124$ (right and  left lines,
  respectively).   Red solid  line shows  the best  fit  mass function
  obtained  using   the  data  from  all  the   snapshots  with  $z\le
  1.25$.  The bottom  panels  show residuals from the fit to 
  the scaled counts from all snapshots up to $z=1.25$.  The departures 
  at high redshift are reduced, so universality is even more pronounced.
  \label{sovir}}
\end{figure*}

For this reason, our reference parametrisation for the halo mass 
function is that of \citet{sheth99b}:
\begin{equation}
 f(M)\,{\rm d}M = f(\nu)\,{\rm d}\nu 
\end{equation}
where
\begin{equation}                                                    
 \nu f(\nu) = A\, \left(1+\frac{1}{\nu^{'p}}\right)\,
                  \left(\frac{\nu^{'}}{2\pi}\right)^{1/2}\,
                  {\rm e}^{-\nu^{'}/2}
\label{st99mf}
\end{equation}     
with $\nu^{'}=a\nu$.  The parameters ($a,p,A_0$) define the high-mass 
cutoff, the shape at lower masses, and the normalisation  of the curve, 
respectively.  In addition, because $P_{\rm lin}(k)$ determines the value 
of the mass variance $S(M) = \sigma^2(M)$ which enters in the 
definition of $\nu$, the initial power spectrum plays an important 
role.  Since it enters in the denominator of $\nu$, one might say that 
the mass function is `non-perturbative' in $P_{\rm lin}(k)$.

\section{The uniqueness of the virial overdensity}\label{secuniq}

This section  studies the universality  of the halo mass  function for
haloes identified using  the virial overdensity.  We do  so by finding
the set of  parameters ($a$,$p$,$A_{0}$) that best fit  the $z=0$ data
from  the Planck13  simulations.  We  then use  measurements at  other
redshifts and cosmologies to test for universality.  We start with the
virial  overdensity  because we  believe  it  to be  the  most
  physically motivated  choice for identifying haloes.   Later in the
paper we study haloes identified using the other SO catalogues.

Our model  for the  mass function, equation~(\ref{st99mf}),  has three
free parameters ($a,p,A_0$), whose values  we adjust so as to minimise
the Chi square with respect to the measured binned mass function:
\footnote{Note  that \citet{sheth99b} only varied  $a$ and $p$: 
  they fixed the value of $A$ by requiring that the integral over 
  all masses give $\rho_b$.}
\begin{equation}  
 \label{chi2}
 \chi^2(a,p,A_0)  = 
   \sum_i \dfrac{\left(\log(\nu f(\nu))_{i} - \log(\nu f(\nu))_{fit} \right)^2}
                         {\epsilon^2_{\log (\nu f(\nu))_{i}}}\,
\end{equation}  
\noindent where the  sum is over binned counts.  The  bins are equally
spaced in $\log_{10}(\nu)$, with $\Delta \log_{10}(\nu) = 0.05$ and we
neglect covariances between the binned counts when fitting.
\footnote{We do not include a bin that counts the mass fraction 
which is not assigned to haloes.  See Manera et al. (2010) for an
algorithm which does  not   require  binned  counts, and does 
account for the unassigned mass fraction.} 
We  discard bins with  fewer than  30 objects (typically the high-$\nu$ 
bins), to  set a limit on the Poissonian error $\epsilon_{\log (\nu f(\nu))}$ in a bin. 
There were no significant differences in the best-fit parameters when 
we repeated the analysis using only bins with at least 100 objects.

\subsection{The halo mass function at different redshifts and
  overdensities}

\subsubsection{\textbf{Virial haloes}}
The best-fit values for the $z=0$ virial overdensity halo counts are
\begin{equation}
 (a,p,A_0) = (0.794\pm  0.005, 0.247\pm0.009, 0.333\pm 0.001).
\label{fit_z0}
\end{equation}
The dashed  black curves,  same in  all panels  of Figure~\ref{sovir},
show equation~(\ref{st99mf})  with these  parameters.  The  symbols in
the different  panels show the  measured virial halo mass  function at
four different redshifts as labelled.  Green, cyan, blue, magenta, red
and orange  points show results from  the six simulations of  the main
set:  Ada, Bice,  Cloe,  Dora,  Emma and  Flora.   Note the  excellent
agreement among the simulations despite the very different volume that
each  samples.   This  is   because  the  way   we  calculate
  $\nu(\delta_c,\sigma)$  largely  eliminates  the  impact  of  cosmic
  variance.

The middle panels of Fig.  \ref{sovir} show logarithmic residuals from
the best-fit to the $z=0$ counts.   These indicate the goodness of our
fit, which  was calibrated in log  space.  The residuals are  close to
zero at  small and intermediate values  of $\nu$.  They are  larger at
high $\nu$,  where the best-fit predicts  more haloes than we  find in
the  simulations  at $z=0$.   Some  of  this  is  due to  the  Poisson
uncertainty in the small number counts; at higher redshifts the effect
is reduced since  the high $\nu$ tail becomes populated  by lower mass
haloes.  The smallness of the  residuals in the other panels indicates
that the  $z=0$ model is  a good description  of the virial  halo mass
function at higher redshifts as well.

The vertical lines  in the bottom panels show  the regime of influence
of two numerical  effects: 
$(i)$ First, the dashed  gray vertical line shows the maximum $\nu$ 
for which  there are at least 100 (rather than 30) haloes in the bin.  
This matters only at high redshifts and in the high mass regime and, 
as mentioned before, the best fit parameters are not significantly 
different; 
$(ii)$ Second,  we addressed the effect of  not using 2LPT initial 
conditions  for our simulations: we rescaled  our points using
equation~(11)  of \citet{reed13} (calibrated  only for  $1\lesssim \nu
\lesssim5$),  which gives an estimate of  the bias  in the  halo mass
function between ZA and 2LPT methods; the  two solid  vertical lines
show the minimum $\nu$ at which  we may expect differences of at least
$5\%$, for the simulations starting at $z=99$ (right) and $z=124$ (left).  
Such discrepancies would  mainly affect the high-$z$ and high-$\nu$ 
data points;  they do not affect the counts in the $z\le 1.25$ range 
we will adopt  to calibrate the  parameters of equation~(\ref{st99mf}).  
The rescaling equation by \citet{reed13} and the ZA/2LPT difference 
are discussed in detail in Appendix~B4.

From the figure and the discussion  above we conclude that there is no
significant  systematic deviation from  universality within  $8\%$ for
$\nu \lesssim 10$ at redshifts $z\le 5$.  Therefore, we can combine 
measurements at many different redshifts to increase the precision in 
the estimate of $(a,p,A_{0})$. In addition to increasing the 
  statistics, adding the high-redshift data fills-in the high-$\nu$ 
  tail, allowing it to play a greater role in determining the best-fit 
  parameters.
We  determined best-fit  parameters using  the virial  halo data 
from all the snapshots  between $z=0$  and $1.25$ (a total of 15 
snapshots).  At $z\ge 1.25$, the data no longer samples the whole 
mass function, preventing  a reliable fit.  The resulting best fit 
parameters are:
\begin{eqnarray}
 a &=& 0.7663\pm0.0013 \nonumber\\
 p &=& 0.2579\pm0.0026 \label{fit_allz} \\
 A_0 &=& 0.3298\pm0.0003 \nonumber.
\end{eqnarray}
The differences between this all-$z$ fit and the previous $z=0$ fit 
are of order $\Delta a\simeq 3\%$, $\Delta p\simeq 4\%$ and 
$\Delta A_{0}\simeq 1\%$.  This leads to percent level differences 
in the mass function which reduce the residuals at high redshift.
The red solid line in each panel of Figure~\ref{sovir} shows this 
all-$z$ best fit: it can hardly be distinguished from that for $z=0$ 
only (black  dashed  curve) but, as shown by the residuals
in the lower panels, it traces the high-$\nu$ part of the mass 
function more closely.   

\begin{figure}
\includegraphics[width=\hsize]{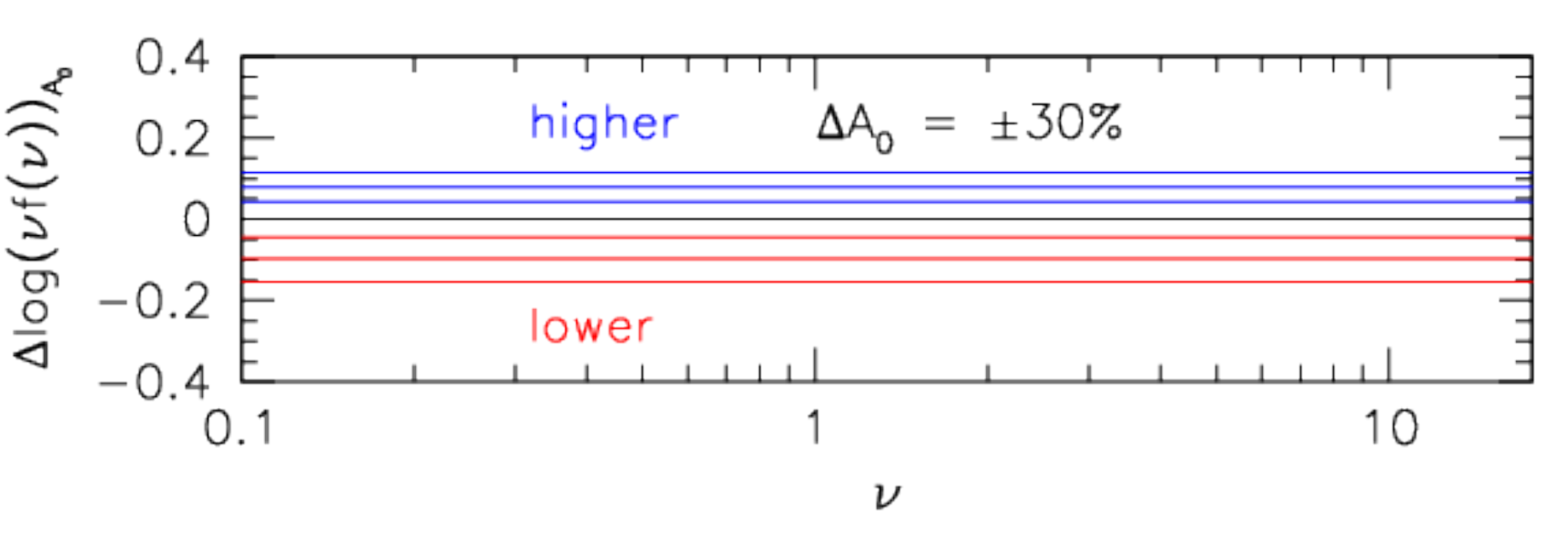}
\includegraphics[width=\hsize]{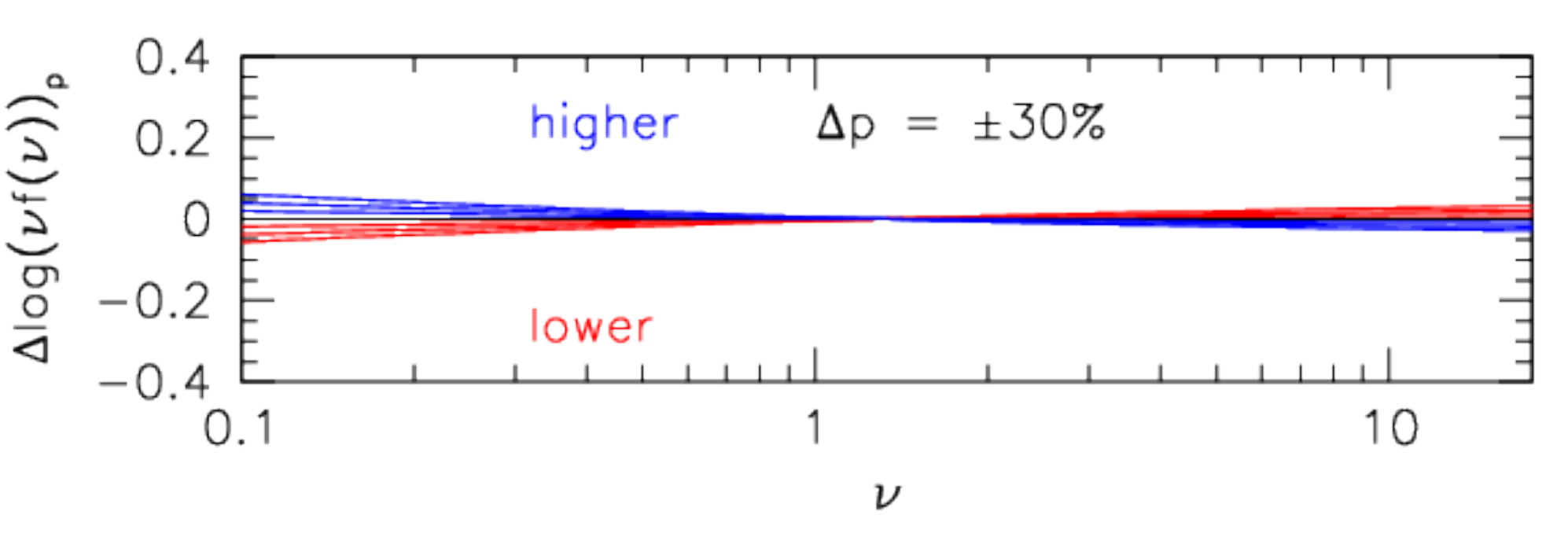}
\includegraphics[width=\hsize]{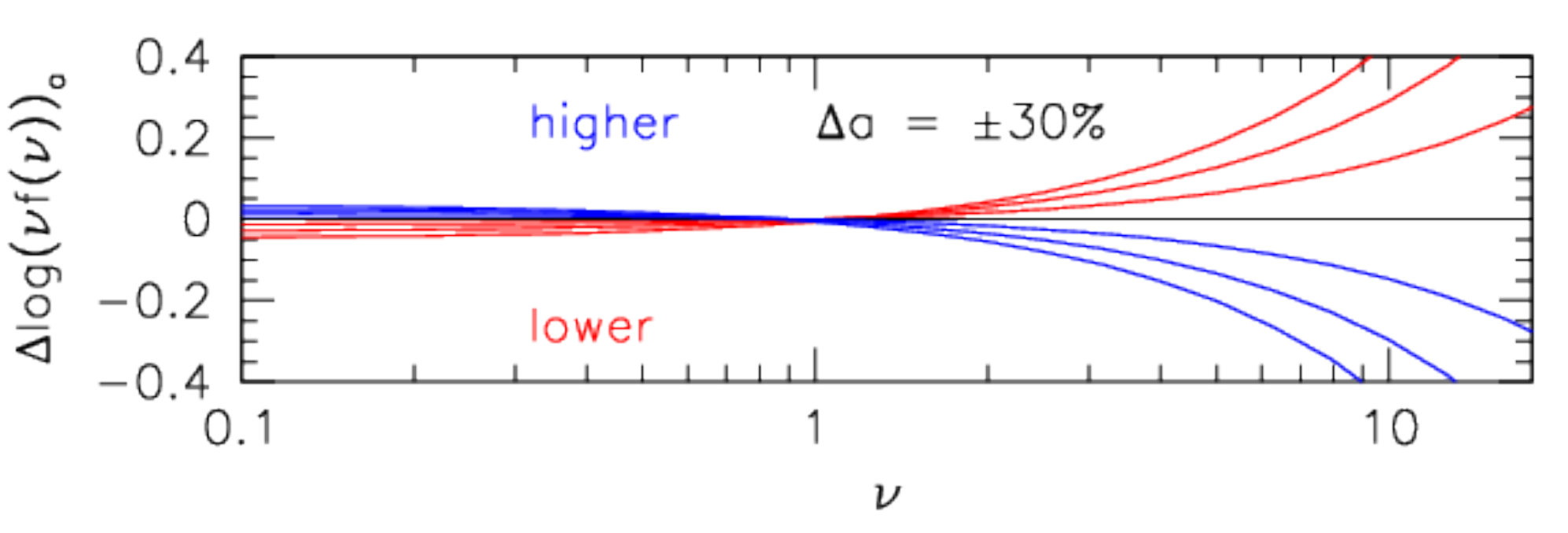}
\caption{Change in the best-fit mass function when one parameter at a time 
  is  changed by up to $\pm  30\%$ (in increments of $10\%$). Blue and
  red  curves show cases where the parameter values are larger or 
  smaller than the reference one. \label{MFpfig}}
\end{figure} 

Figure~\ref{MFpfig} shows how the virial  mass function changes as each 
of the parameters $(a,p,A_0)$ is varied by $\pm 30\%$ while the other 
two are kept fixed.  While the mass function is sensitive to both 
$A_0$  and $a$ --  which modify  the normalisation  and the  high mass
cut-off, respectively -- this is less true for $p$.  Even if $p$ changes 
by $\pm 30\%$, the mass  function changes by $5\%$ at most, for 
$\nu\lesssim 0.1$ and $\nu>10$.   In what follows, we use the values in 
equation~(\ref{fit_allz}) as our reference model.

\subsubsection{\textbf{Other overdensity thresholds}}
We now study the mass functions associated with the other five density
thresholds.   It is  tempting  to identify  these  thresholds with  an
`effective formation  redshift', and hence with an  effective value of
$\delta_c(z)$  when  defining  $\nu$.    Since  the  role  of  $a$  in
equation~(\ref{st99mf}) is  simply to rescale $\nu$,  one might wonder
if these other halo definitions lead to universal mass functions which
differ from those for the virial overdensity only in the value of $a$.

Table~\ref{tab_massf} shows the values  of the best-fitting parameters
at $z=0$  for the  different threshold  densities.  Clearly  the other
parameters, $p$ and $A_0$, also depend strongly on how the haloes were
identified.   Moreover, as  we will  show below,  universality clearly
does not  hold for any of  these other definitions: in  all cases, the
best   fit  parameters   for   the  higher   redshift  counts   depart
significantly from those at $z=0$.

\begin{figure*}
\includegraphics[width=0.99\hsize]{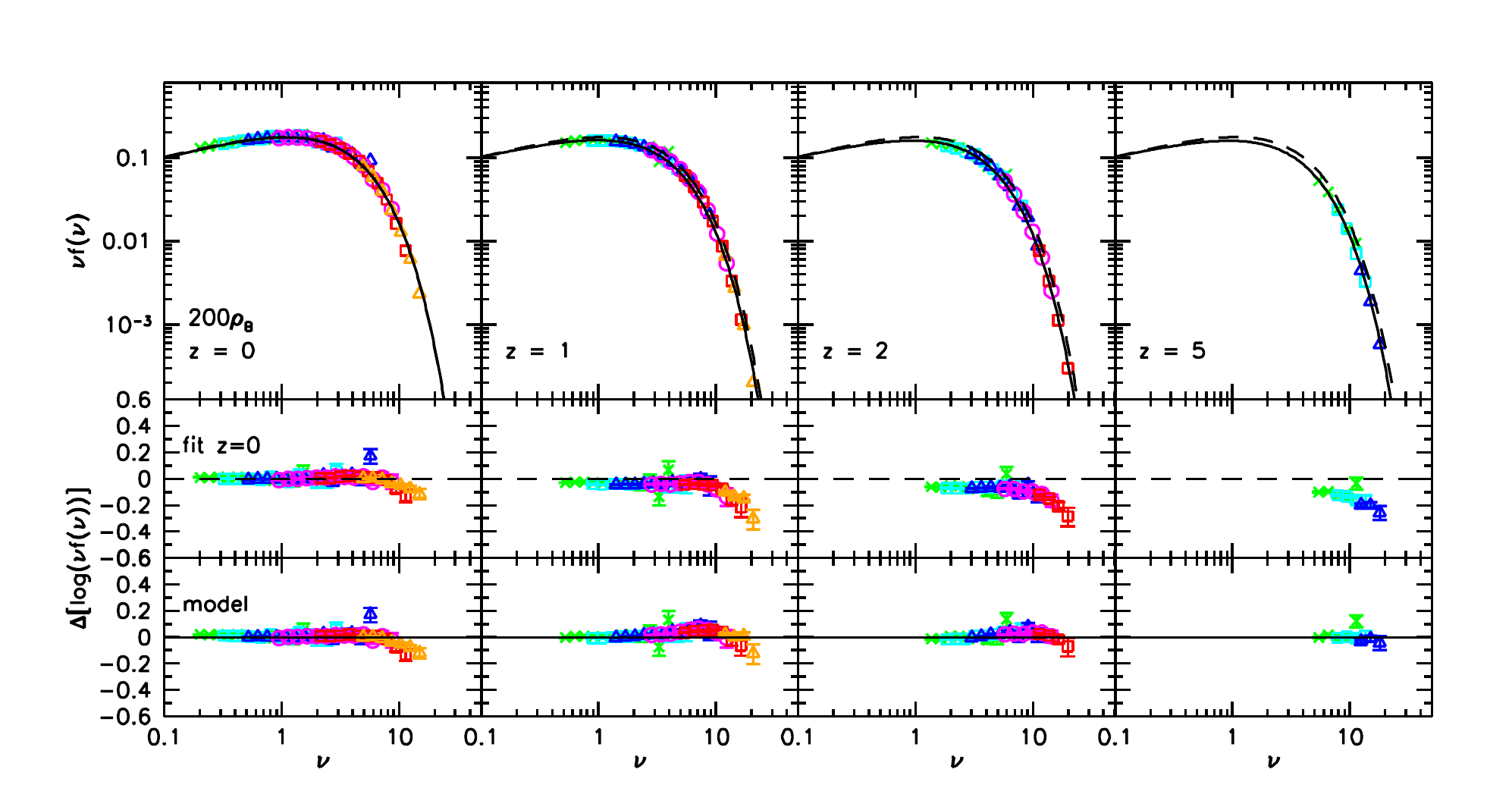}
\caption{Same  as Fig.~\ref{sovir},  but for  haloes identified  using
  $200\rho_{b}$  instead   of  $\Delta_{vir}$.   Middle   panels  show
  residuals with respect to the  best-fit at $z=0$ (dashed curve, same
  in  all top  panels);  these  show that  the  mass  function is  not
  universal across  all redshifts.   Lower panels show  residuals with
  respect to our rescaled model (equation~(\ref{model}); solid curves,
  different in each top panel).
\label{so200b}}
\end{figure*}

To illustrate this, Figure~\ref{so200b}  shows the halo mass functions
at $z=0$, $1$, $2$, and $5$ for haloes identified using $200\rho_b$ as
density threshold.   The dashed curve, which  is the same in  all four
top panels, shows  the best-fit to the $z=0$ data  points.  The middle
panels, which show  the residuals with respect to this  fit, show that
it overestimates the counts at all higher redshifts.  These departures
from universality are  in agreement with previous  work on $200\rho_b$
haloes \citep[e.g.][]{tinker08}.  The solid curves in the upper panels
show the result of rescaling  our universal virial counts as described
in the next section; the bottom panels show the residuals with respect
to it, which provides a much better fit.

\begin{figure*}
\includegraphics[width=0.99\hsize]{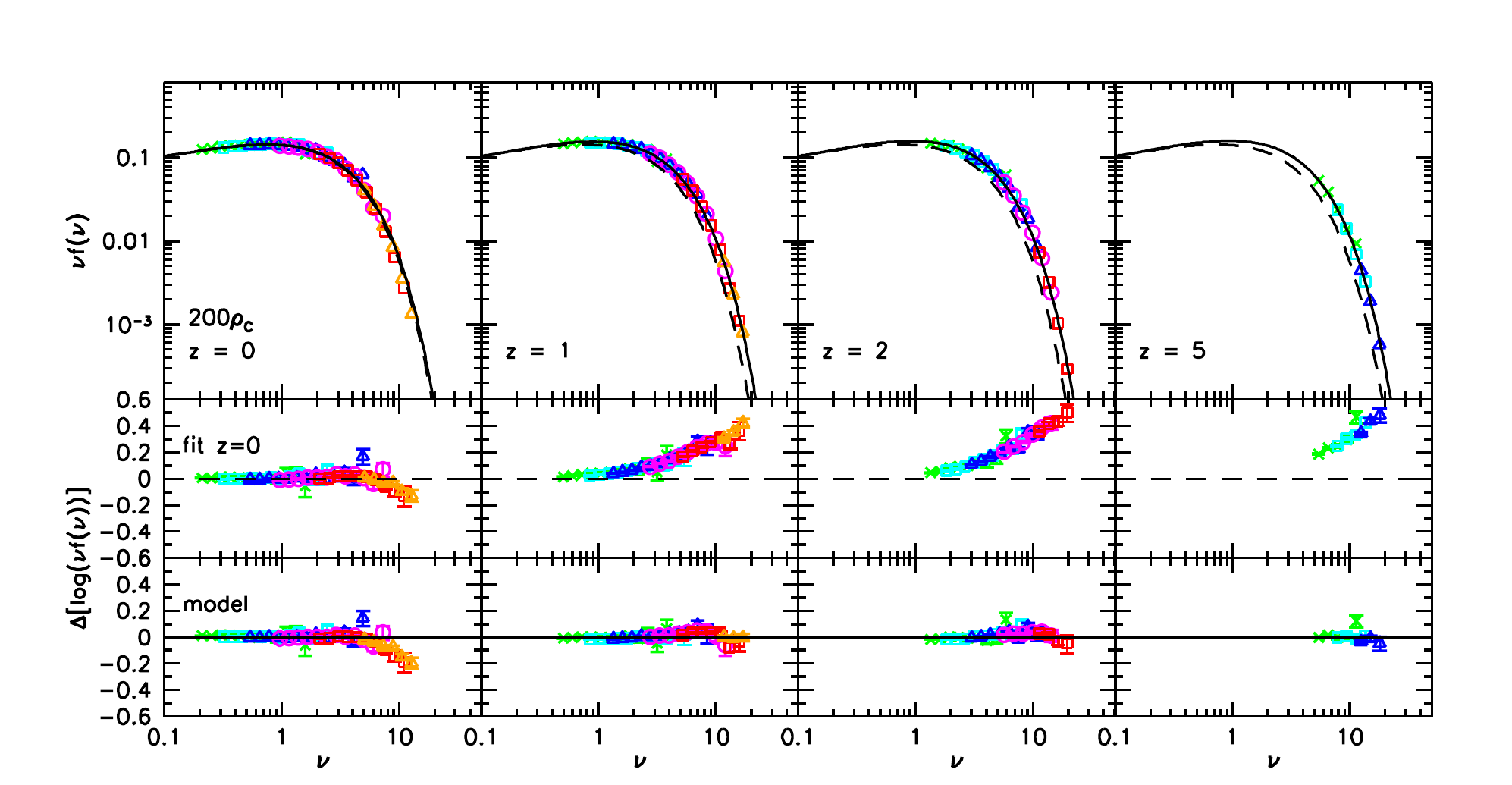}
\caption{Same  as Fig.~\ref{sovir},  but for  haloes identified  using
  $200\rho_{c}$; middle panels show that the universality is broken in
  the  opposite sense  to when  the threshold  was $200\rho_{b}$,  and
  bottom panels show  that our rescaled model accounts  for this quite
  well.\label{so200c}}
\end{figure*}

\begin{figure}
\includegraphics[width=\hsize]{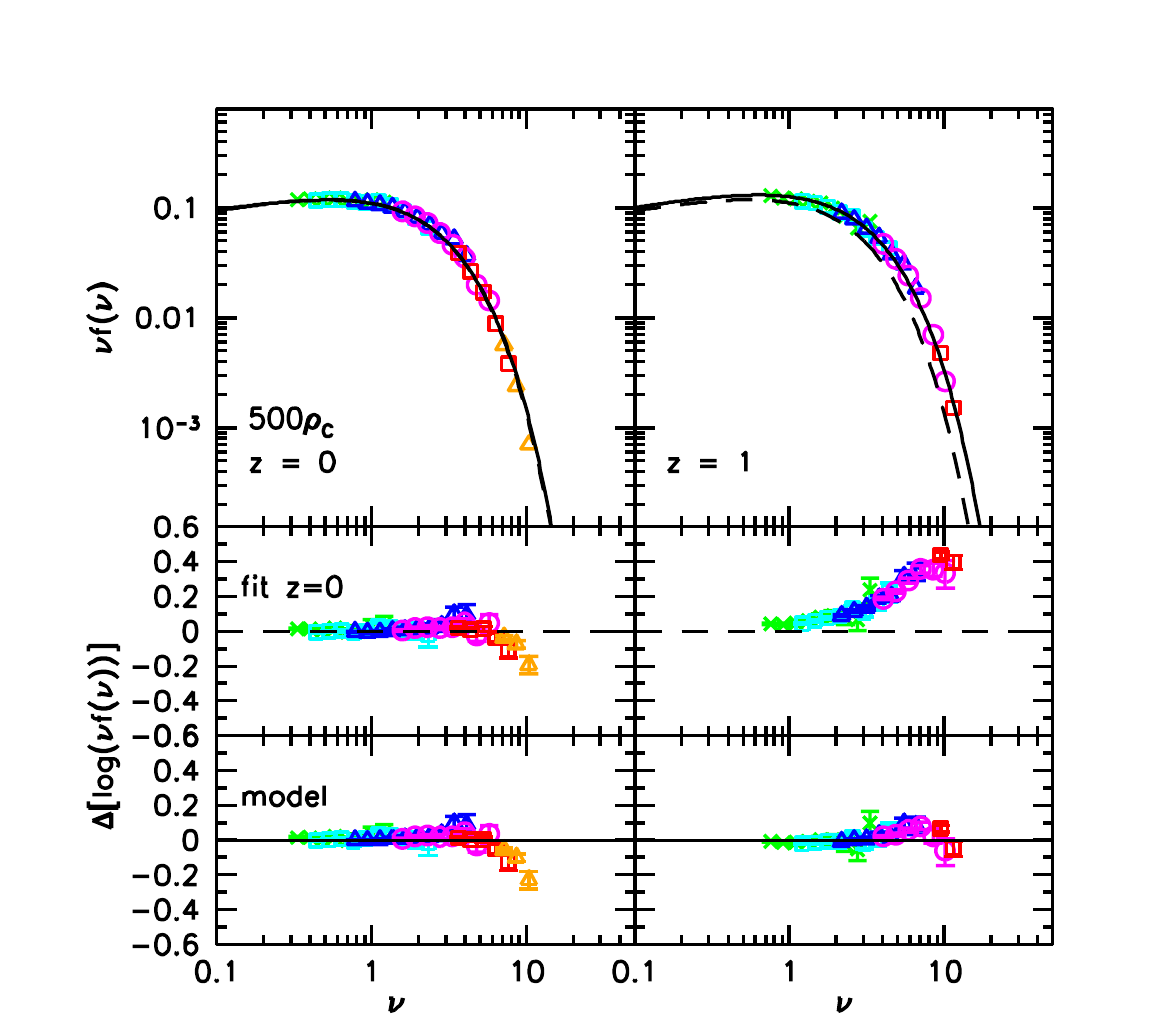}
\includegraphics[width=\hsize]{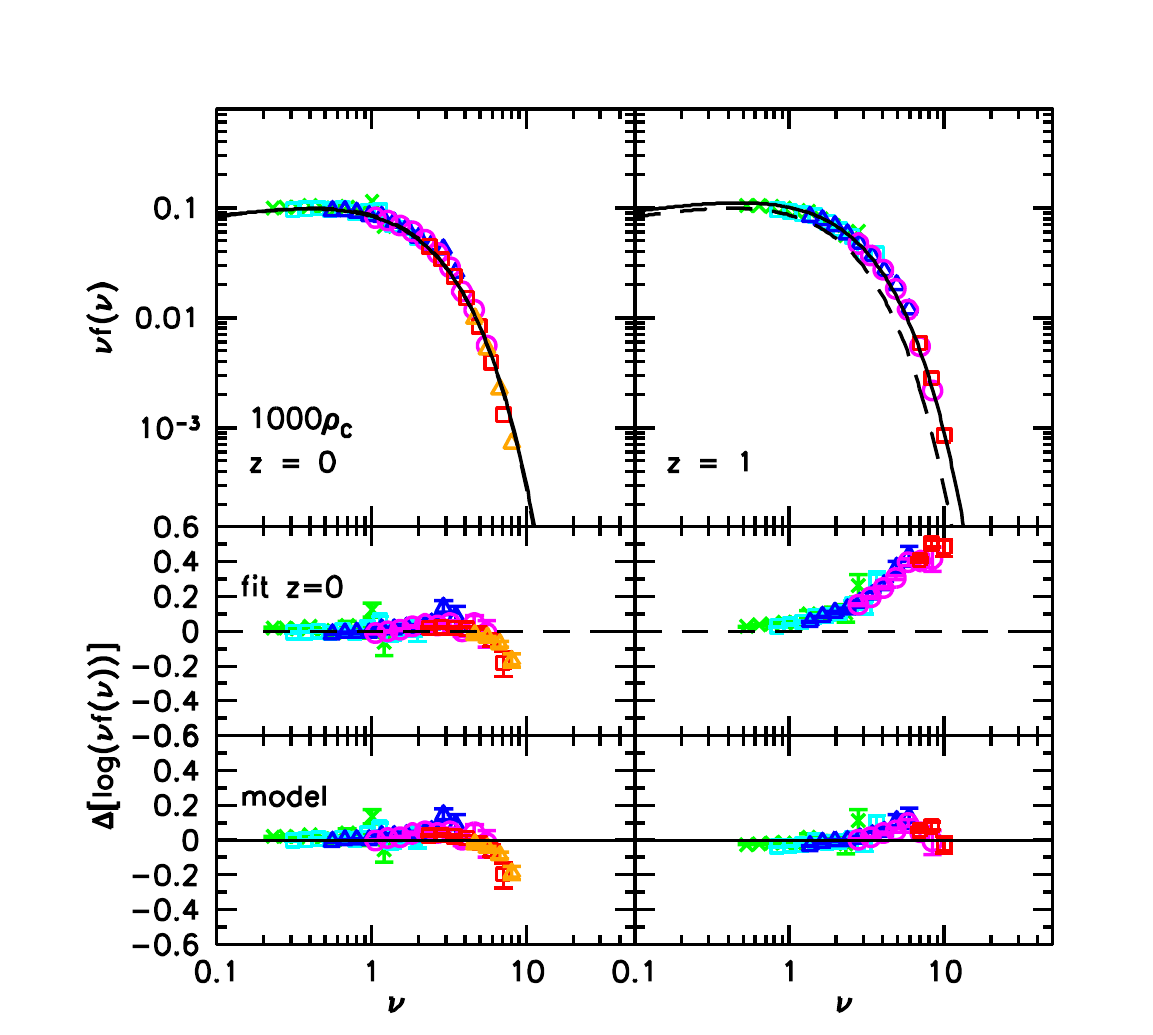}
\includegraphics[width=\hsize]{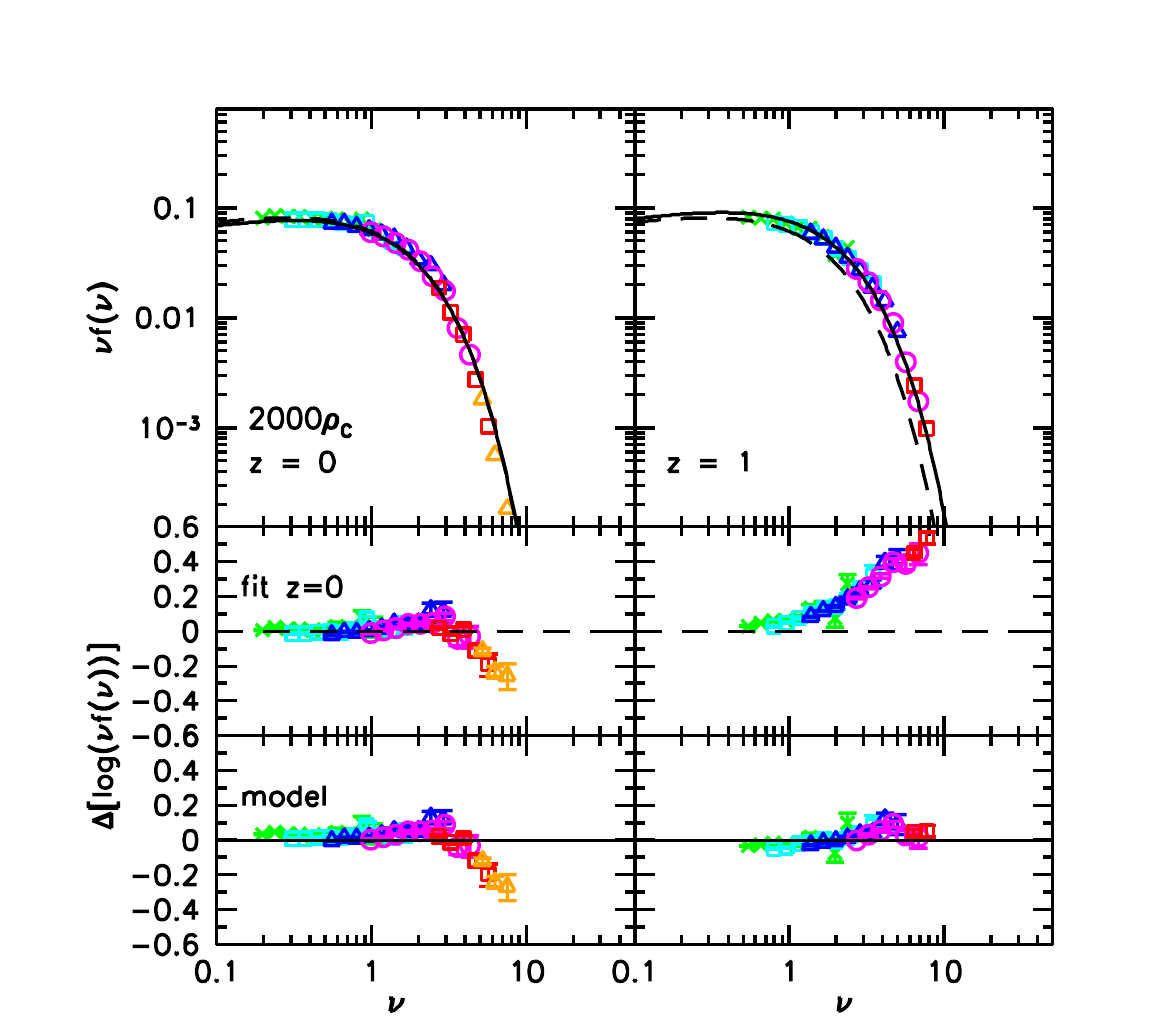}
\caption{Same as Fig. \ref{sovir}, but for haloes identified 
using 500,  1000  and  2000 $\rho_{c}$ instead of $\Delta_{vir}$, 
at $z=0$ and $1$.\label{so_highc}}
\end{figure}

Figure~\ref{so200c} shows a similar analysis of haloes identified using
$200\rho_c(z)$.   In  this  case,  the best-fit  relation  at  $z=0$
underestimates the counts at higher $z$.  Finally, Fig.~\ref{so_highc}
shows haloes  identified using threshold  values of $500$,  $1000$, and
$2000\rho_c(z)$,  at  redshifts  $z=0$  and  $z=1$.   The  trends  are
qualitatively similar to those for $200 \rho_c(z)$, with the $z=0$ fit
underestimating the counts at higher $z$.

\begin{table*} \centering
\begin{tabular}{|c|c|c|c|}  \hline 
 $\rho$  (SO)& a  & p  &  A\\ \hline \hline
 & \multicolumn{3}{c}{z = 0}  \\ \hline

$200\rho_{b}$  &0.739  $\pm 0.005$  & 0.206  $\pm 0.008$&  0.360 $\pm 0.001$\\ 
$\Delta_{vir}$ & 0.794 $\pm 0.005$ & 0.247 $\pm 0.009$ & 0.333 $\pm 0.001$\\ 
$200\rho_{c}$ & 0.903  $\pm 0.006$ & 0.322 $\pm 0.009$ & 0.287 $\pm  0.001$\\ 
$500\rho_{c}$  & 1.166 $\pm  0.009$ &  0.344 $\pm 0.012$ & 0.236 $\pm 0.001$\\ 
$1000\rho_{c}$ & 1.462 $\pm 0.012$& 0.349 $\pm 0.015$& 0.197 $\pm 0.001$\\  
$2000\rho_{c}$ & 1.821 $\pm 0.017$ & 0.413 $\pm 0.017$& 0.158 $\pm 0.001$\\ 
  \hline
 & \multicolumn{3}{c}{All z - Planck cosmology}  \\ 
\hline
$\Delta_{vir}$ & 0.7663 $\pm 0.0013$ & 0.2579 $\pm
0.0026$ & 0.3298 $\pm 0.0003$\\ 
  \hline
  & \multicolumn{3}{c}{All z \& cosmologies}  \\ \hline
$\Delta_{vir}$  &  0.7689 $\pm 0.0011$ & 0.2536 $\pm
0.0026$  & 0.3295 $\pm 0.0003$ \\ \hline
  & \multicolumn{3}{c}{All z \& cosmologies -- Cluster Counts: $M_{vir} > 3 \times 10^{13} M_{\odot}/h$}  \\ \hline
$\Delta_{vir}$  &  0.8199 $\pm 0.0010$ & 0 & 0.3141 $\pm 0.0006$\\
%  & & 
 \hline 
\end{tabular}
\caption{Dependence of best-fit parameters  on the overdensity used to
  identify SO  haloes. In the  top part  we show the  three parameters
  calculated at  $z=0$; then we  report those obtained by  fitting the
  virial halo counts from all snapshots up to $z=1.25$ $(i)$ of  the Planck
  simulations  and  $(ii)$  of  all  the  simulations  with  different
  cosmologies  together.  The bottom row  shows the parameters
  which best-fit the $M_{\rm vir}\ge 3 \times 10^{13} h^{-1}M_{\odot}$ 
  counts from  all cosmologies and all redshifts.
\label{tab_massf}}
\end{table*}

\subsection{The halo mass function for different cosmologies}
\label{sec_cosmo}

Having  established that the  virial mass  function is universal with
respect to redshift, we now test if it is universal across other
background cosmological  models.  To  do  so, we  use  the halo  
catalogs in  our secondary set of $512^3$  simulations, whose 
properties are summarised in Table~\ref{tab_sim2}.

\begin{figure*}
\includegraphics[width=\hsize]{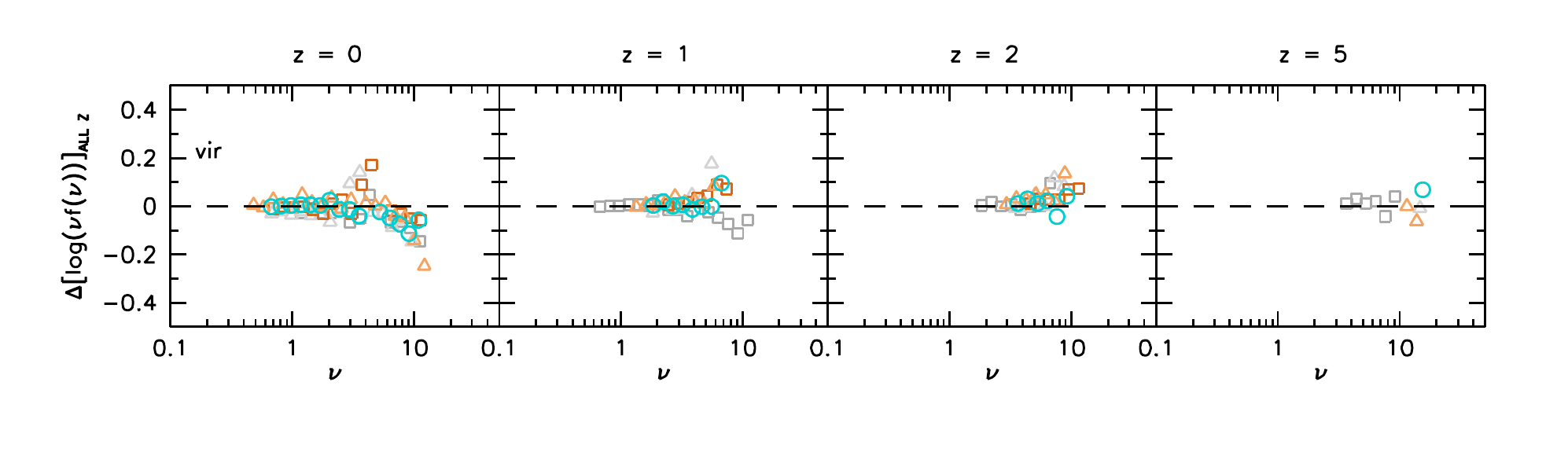}
\caption{ Cosmology independence of the  virial halo mass function. We
  show the logarithmic  difference from the best  fit calculated using
  the Planck  cosmology for redshifts  $z=0, 1, 2$, and  5.  Following
  the colour  code of  Table~\ref{tab_sim2}, gray points  indicate the
  results  of   the  simulations  with  $\Omega_{m}=0.2$,   brown  for
  $\Omega_{m}=0.4$ and light blue  for $\Omega_{m}=0.272$; squares and
  triangles    represent   $\sigma_{8}=0.7$    and   $\sigma_{8}=0.9$,
  respectively.\label{resid_cosmo_vir}}
\end{figure*}

First,  for  virial  haloes,  we  used  the  best  fit  parameters  of
equation~(\ref{fit_allz})  which were  calibrated  using the  $1024^3$
particle  simulations of  a Planck13  cosmology.  The  four panels  in
Figure~\ref{resid_cosmo_vir} show the logarithmic difference from this
best fit  at redshifts $z=0, 1,  2, 5$.  Following the  colour code of
Table~\ref{tab_sim2}, gray points are  for $\Omega_{m}=0.2$, brown for
$\Omega_{m}=0.4$  and light  blue  for  $\Omega_{m}=0.272$ (the  WMAP7
cosmology);  squares and  triangles  represent  $\sigma_{8} =0.7$  and
$\sigma_{8}=0.9$, respectively.   For halo  mass definitions  that are
close  to  the  virial  value  the  residuals  are  relatively  small,
indicating  that  our  relation  -- calculated  exclusively  from  the
Planck13 cosmology  data -- remains  valid for the  other cosmological
models as well.  In addition, although we do not show this explicitly,
for  these  cosmological  models  also, universality  is  broken  when
considering  other overdensities.   This demonstrates  universality of
the  virial  relation  with  respect  to  other  cosmological  models.
However, departures  from this  universality may arise  if one
  considers   more   extreme    departures   from   Planck13.    E.g.,
  \citep{courtin11} explore models in  which the growth factor, growth
  history and  virial density differ  more radically from that  of the
  Planck cosmology,  and find  correspondingly larger  departures from
  universality.

\begin{figure*}
\includegraphics[width=0.6\hsize]{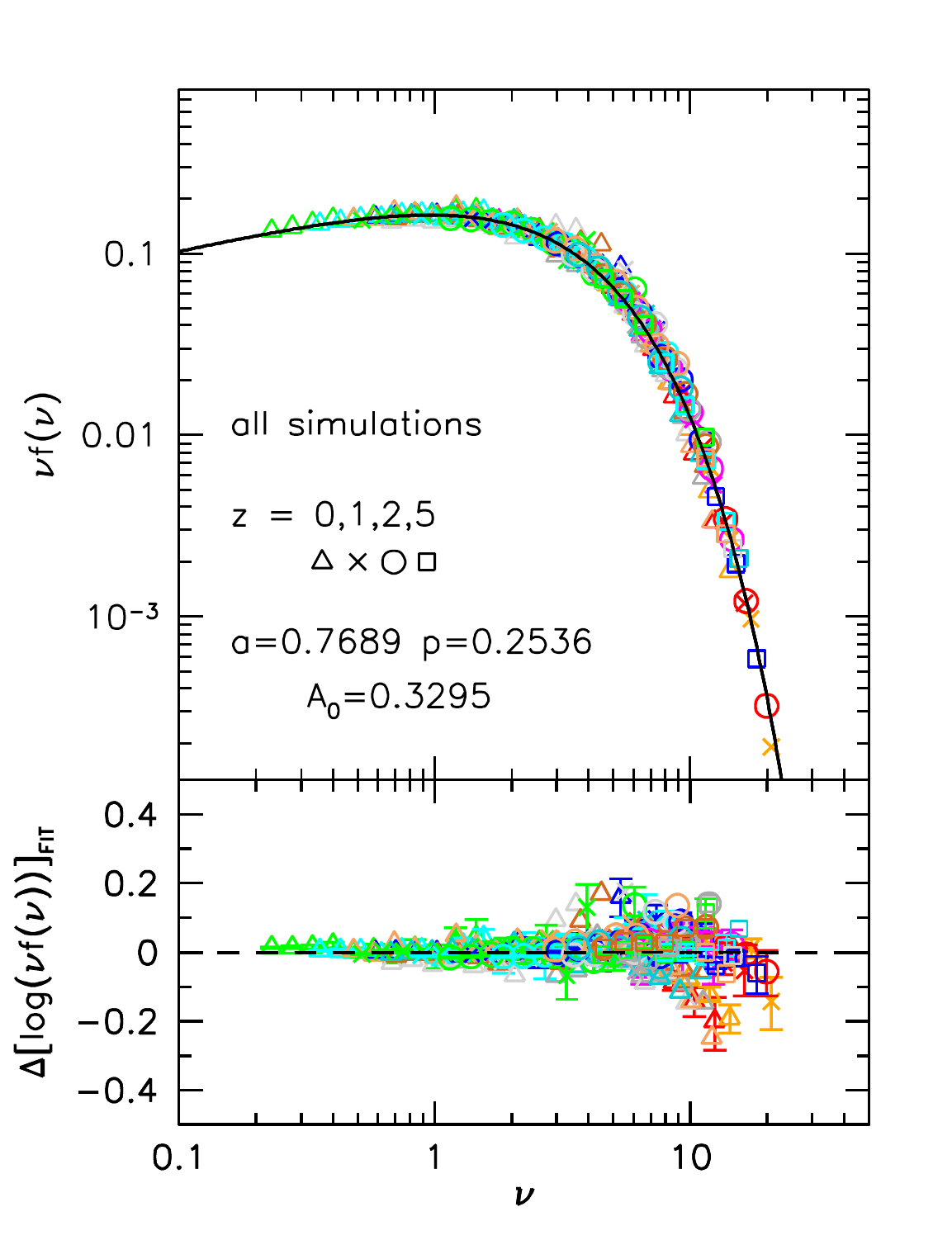}
\caption{Universality  of the virial halo  mass function.  Symbols show 
measurements from all our simulations  -- both from the main and secondary
set -- at  four  redshifts:  $z=0,1,2,5$.  Although we show measurements 
at these four redshifts, the fit, which was calibrated on all simulations, 
used only the $z\le 1.25$ snapshots.
 \label{univ_all}}
\end{figure*}

Figure~\ref{univ_all} shows  the results from all  our simulations and
redshifts   together.  Different   colours  represent   the  different
simulations (with  the same colour  code as in previous  figures), and
different  symbols  distinguish  the   four  redshifts  (and  not  the
different simulations). The best fit to all the virial halo catalogues
out to $z=1.25$ has
\begin{eqnarray}
 a &=& 0.7689 \pm 0.0011\nonumber\\
 p &=& 0.2536 \pm 0.0026 \label{fit_cosmo} \\
 A_{0} &=& 0.3295 \pm 0.0003, \nonumber
\end{eqnarray}
and  is  shown  by  the  solid black  line.   Lower  panel  shows  the
logarithmic residuals.   These best-fit  parameter values equal  -- at
the per mil level -- the values  obtained in the last Section from our
first  set   of  simulations,  which  assumed   a  Planck13  cosmology
(equation~\ref{fit_allz}), thus further confirming the universality as
a  function  of  the background  cosmological  model.   

Table~\ref{tab_massf} compares  the three sets of  best-fitting values
for the  virial haloes.  The bottom  row shows  the parameters
  which    best   fit    the    counts    of   cluster-mass    haloes:
  $M_{vir}\ge 3 \times 10^{13}h^{-1}M_{\odot}$.   Notice that at these
  high  masses $p=0$,  so  that  the mass  function  is  like that  of
  \citep{press74}.  We  discuss this further in  Appendix~\ref{cmf} in
  the context of cosmological constraints in the $\Omega_m$-$\sigma_8$
  plane from cluster counts.

\begin{figure*}
\includegraphics[width=0.3\hsize]{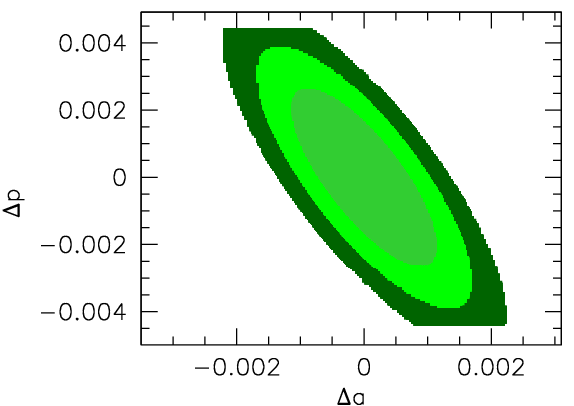}
\includegraphics[width=0.3\hsize]{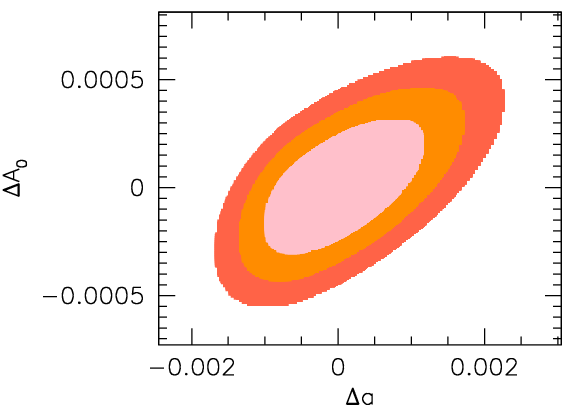}
\includegraphics[width=0.3\hsize]{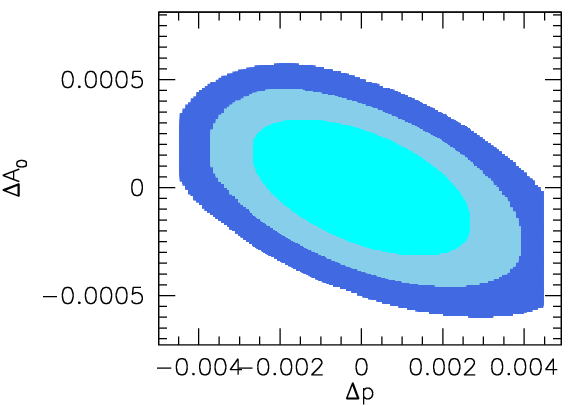}
\caption{Covariance between fitted parameters  of the virial halo mass
  function  (the  $1024^3$ and  $512^3$  runs,  and all  redshifts  to
  $z\le 1.25$): Contours  show 1-, 2- and  3-$\sigma$ reference levels
  for each pair of parameters.
\label{contours}}
\end{figure*}

Covariances   between   the   best-fit   parameters   are   shown   in
Fig.~\ref{contours}.  Contours  show 1-, 2- and  3-$\sigma$ levels for
each  pair  of parameters.   While  $A_0$  and  $a$ are  not  strongly
degenerate, $p$ correlates with  both $a$ and $A_0$.  \citet{manera10}
show  that  such correlations  can  be  understood as  resulting  from
requiring the model  to reproduce the total measured  mass fraction in
haloes (the mass fraction is measured with much greater precision than
is   the   detailed   shape   of   the   mass   function;   also   see
Figure~\ref{MFpfig}).

\section{Rescaling the mass function: a universal parametrisation} \label{secrescaling}

We now describe a  simple method that allows one to derive the halo 
abundances associated with other overdensities at any redshift by 
a straightforward rescaling of the virial halo mass function.

\begin{figure}
\includegraphics[width=\hsize]{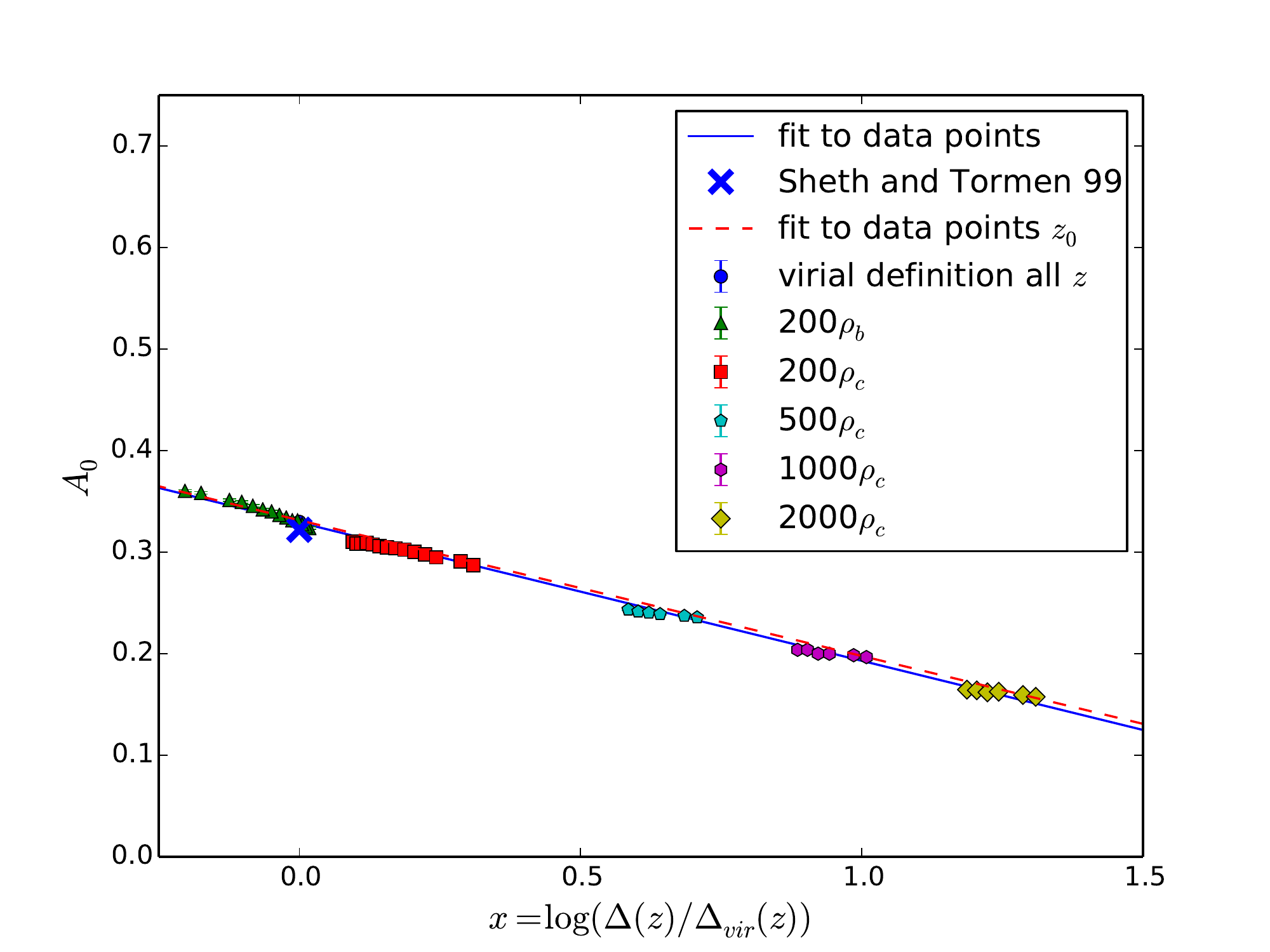}
\includegraphics[width=\hsize]{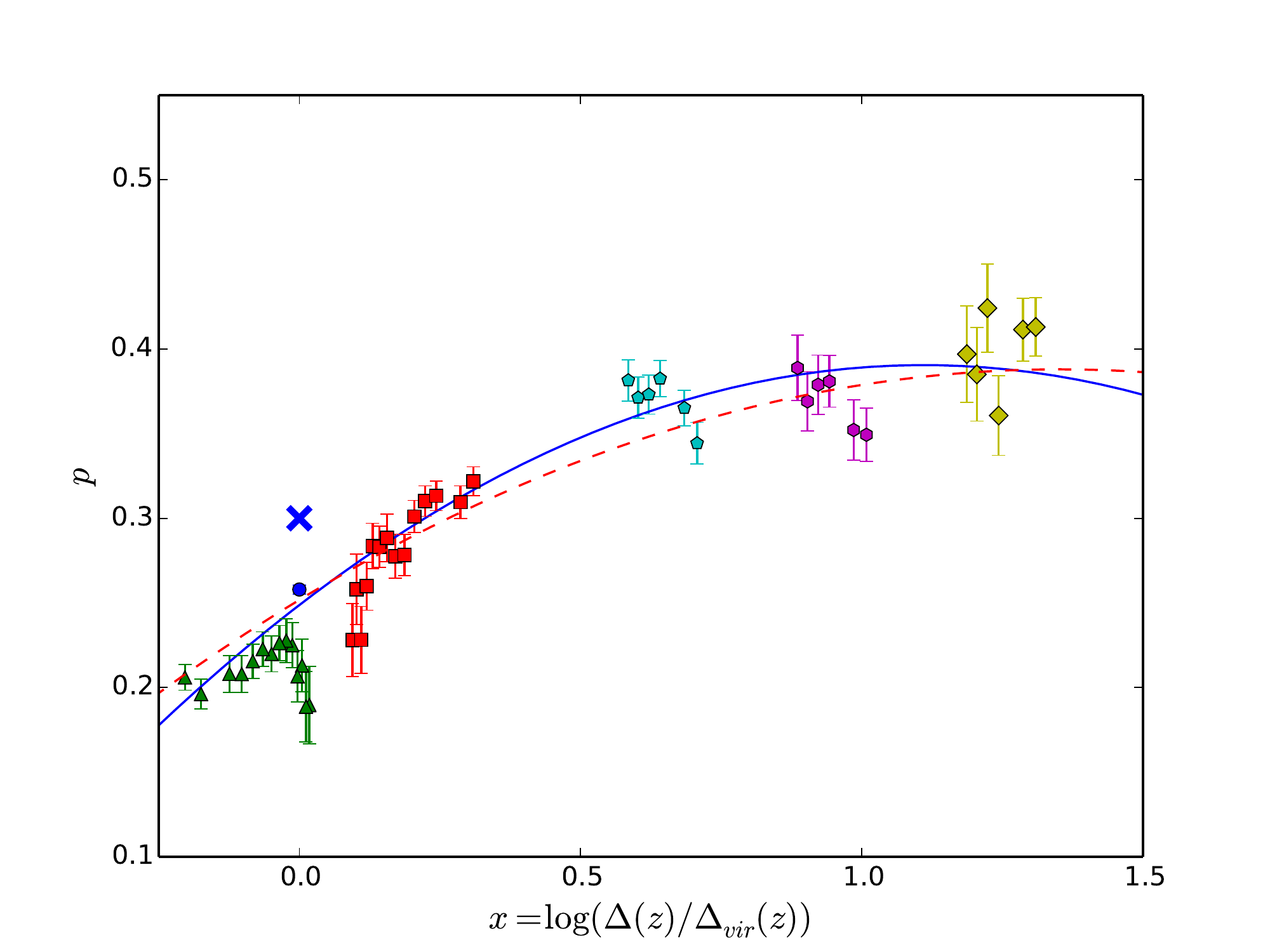}
\includegraphics[width=\hsize]{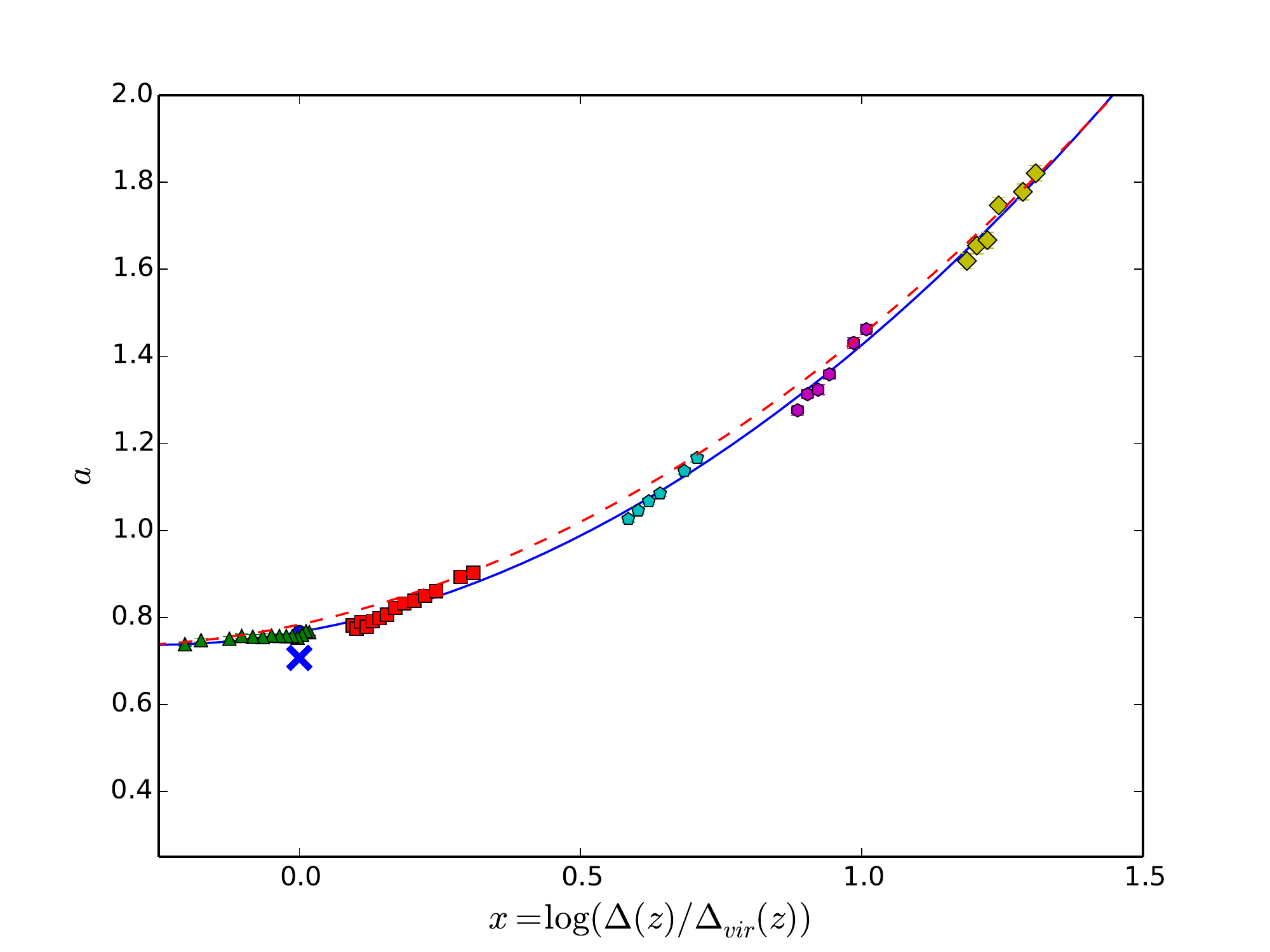}
\caption{Dependance of best-fit parameters $a,p,A_0$ on    
$x = \log(\Delta(z)/\Delta_{vir}(z))$ for $z=0$ to $1.25$.     
Different coloured symbols represent different overdensities;  
smooth curve shows equation~(\ref{model}).
\label{fit_univ}}
\end{figure}

We  calculated  the  best-fitting  parameters for  all  the  six  halo
catalogues  at  all  outputs  between   $z=0$  and  $1.25$.   For  the
overdensities $\ge 500\rho_c(z)$ we only fitted the data points out to
$z\sim 0.4$ so  as to be sure  of having good statistics  for both the
shape  and the  cut-off  of  the mass  function,  i.e.   $p$ and  $a$,
respectively.  At  higher redshifts the  small number of  haloes which
probe  a smaller  range  of  $\nu$ cannot  break  degeneracies in  the
determination  of  the  three parameters.   Nevertheless,  even  after
restricting  to $z\sim  0.4$, we  obtain  very good  results.  As  the
reference   virial  halo   mass  function   we  used   the  fit   from
eq.~(\ref{fit_cosmo}).   Fig.~\ref{fit_univ} shows  how  the best  fit
parameters vary as a function of $x=\log(\Delta(z)/\Delta_{vir}(z))$.

As  noted  in~\citet{tinker08}  (albeit for  a  different functional 
form),  the  best  fit  parameters  ($a,p,A_0$)  are  smooth functions 
of the critical density threshold.   As can be  seen in the
top  panel of Fig.~\ref{fit_univ},  the normalisation $A_0$ decreases
linearly with the overdensity -- a natural consequence of the decrease 
in halo mass -- while  $a$ and $p$ both increase  with threshold 
overdensity.  The trends can be described using linear or quadratic 
polynomials:
\begin{eqnarray}
  a &=&  0.4332\, x^{2} + 0.2263\, x +  0.7665, \nonumber  \\
  p &=& -0.1151\, x^{2} + 0.2554\, x + 0.2488, \label{model} \\
  A &=& -0.1362\, x + 0.3292, \nonumber
\end{eqnarray} 
shown by the blue solid  lines in Figure~\ref{fit_univ}.  
While $a$  and  $A$ behave  very regularly, the determination of $p$ 
is less certain.  This does not  have a  big influence  on the 
rescaling method,  since  the mass function is less sensitive to 
$p$ than to the other parameters (see Figure~\ref{MFpfig}).   
The red  dashed curves in Figure~\ref{fit_univ}  show the trends 
obtained by fitting to the $z=0$ counts only:
\begin{eqnarray}
 a &=& 0.3881\, x_0^{2}  + 0.2776\, x_0 + 0.7837, \nonumber\\ 
 p &=& -0.07459\, x_0^{2}  + 0.2016\, x_0 +  0.2518, \label{fit_so1}\\ 
 A &=& -0.1337\, x_0 + 0.3315, \nonumber
\end{eqnarray}     
where $x_0\equiv\log(\Delta(z_0)/\Delta_{vir}(z_0=0))$.
These trends are very similar to the previous ones.  

The  above  relations  show  that  both  $a$  and  $p$  increase  with
increasing  threshold, qualitatively consistent  with \citet{manera10}
who studied  FOF rather than  SO haloes.  (They  found that $a$  and $p$
increase as  the FOF  linking length is decreased,  and it is  well known
that  shorter linking lengths  return  denser haloes.)   In addition,  upon
noting  that $a$ multiplies  $\delta_c(z)$ in  Eq.~(\ref{st99mf}), the
increase  of $a$  with density  threshold is  qualitatively consistent
with the  notion that the denser  inner parts of a halo virialized at
higher  redshift, so mass functions for higher density thresholds are 
qualitatively like those for higher redshifts.  Unfortunately, the 
agreement is  not quantitative.
To  see this,  consider  the thresholds  $2000\rho_c$ and  $200\rho_c$
which differ  by a factor  of $10$.  If  the mass associated  with the
denser  threshold virialized  at  $(1+z_{2000})^3 \sim  10$, we  would
expect  the  associated  $\delta_c$  to  be  larger  by  a  factor  of
$(1+z_{2000}) \sim 10^{1/3}$.   If this increase is to  be provided by
increasing  $a$, then  $a$  must  be larger  by  $10^{2/3}$.  This  is
considerably larger than the ratio  of $a$ values for $2000\rho_c$ and
$200\rho_c$ in Table~\ref{tab_massf}.

Even though  we do not  have a quantitative physical  understanding of
the  scaling  in   equations~(\ref{fit_so1})  and~(\ref{model})  --  a
question we return  to in Section~\ref{matching} -- they  can still be
used  to  predict  the  mass functions  at  overdensities  of  current
interest that are  not considered in our work.   These precise fitting
formulae  are able  to follow  the  evolution of  the non-virial  mass
functions at all redshifts,  at the level of a few  percent - which is
comparable to the intrinsic uncertainty in the mass function.

To show this, we return to  the solid lines and the bottom panels in 
Figure~\ref{so200b} - \ref{so_highc}.
The  black solid  curves in  Fig.~\ref{so200b}--\ref{so_highc} show
equation~(\ref{model})  (recall that  the  dashed curves  show the  best
$z=0$ fits) and the bottom panels show the logarithmic difference from
our model.  The agreement is indeed very good: the residuals are 
generally smaller than a few  percent; this is  quite  acceptable 
given the intrinsic scatter of the mass  function is of this order.  
The middle and bottom panels show that our model curves are able 
to account both for the change in the normalisation and the tilt of 
the mass function as $z$ changes.

To conclude:  our results provide  an efficient way for predicting 
the halo mass  function   for  any  redshift   and  overdensity  
in   a  Planck $\Lambda$CDM-like  model.  Once  
$\log(\Delta(z)/\Delta_{vir}(z))$ has been calculated, it is 
straightforward to obtain the other parameters.

\begin{figure}
\includegraphics[width=\hsize]{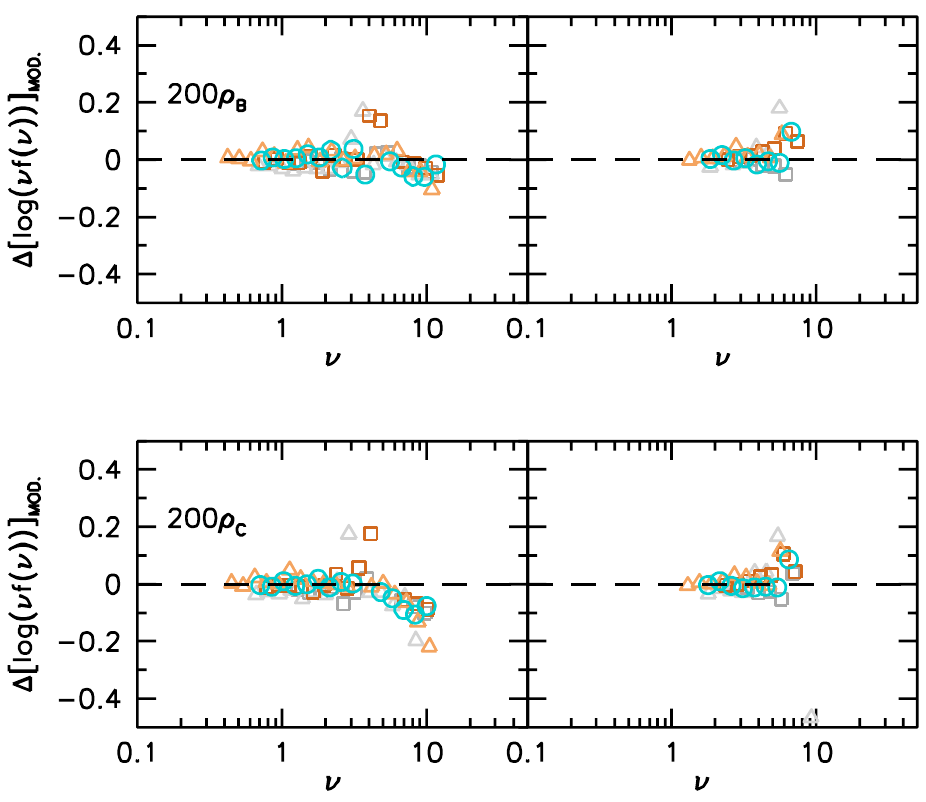}
\includegraphics[width=\hsize]{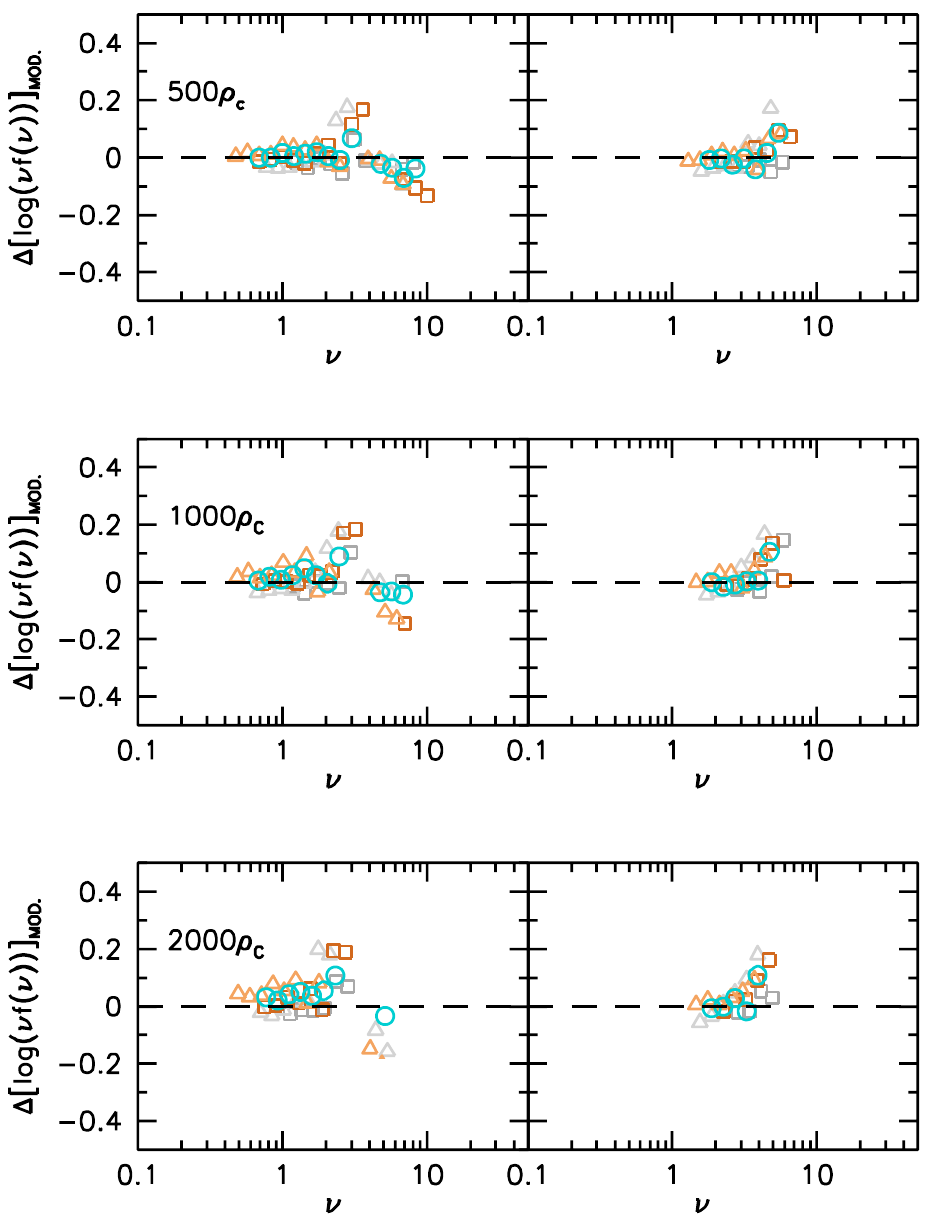}
\caption{Mass  functions for  other  cosmological models:  Logarithmic
  difference from our best fit models  in the secondary set of $512^3$
  simulations, at  two different redshifts  ($z=0,1$) and for  all the
  overdensities.    Gray,  brown,   light   blue   show  results   for
  $\Omega_{m}=0.2, 0.4$ and 0.272.  Squares and triangles show results
  for $\sigma_{8}=0.7$  and $0.9$.  For  the virial haloes we  use the
  fit obtained from stacking all  $z$ outputs of the Planck cosmology.
  For   the  other   cases,  the   residuals  are   with  respect   to
  equation~(\ref{model}).\label{resid_cosmo}}
\end{figure}

Returning to  the  other cosmological  models, Figure~\ref{resid_cosmo}
shows the logarithmic differences  from our reference rescaling models
for  all  the  mass  functions  of the  secondary  set,  at  redshifts
$z=0,1$.  At  both redshifts,  the residuals are  very small
showing that our fitting relations also work for any cosmological
model once mass and redshift are appropriately rescaled to $\nu$.

\section{Matched haloes}\label{matching}

So  far, we  used the  halo catalogues  derived independently  at each
overdensity threshold from the  particle density distribution saved in
the  snapshot  files.  As  already noted,  the number  of  haloes
identified using different overdensities at  a certain redshift is not
the  same.  This  means that  we are  not looking at a ``rescaled 
version'' of  the virial population, but at different halo samples 
(see Figure~\ref{figSO}).  For example, virial haloes may contain
many smaller higher density peaks, each of which could correspond 
to a massive substructure within the virial  radius.  On the other 
hand, the $200\rho_{b}$ threshold is lower than the virial one and 
so --  at low redshifts  -- the identified  haloes will be larger and
more massive then the virial ones. In general in this latter case we may
have fewer haloes than in the virial  catalogues, since the particles
corresponding to more than one smaller virial halo may be included in 
a single $200\rho_{b}$ halo (c.f. the virial halo $3$ in Figure~\ref{figSO}).
 
In  this  section we  analyse  the  halo  mass function  of  ``matched
haloes'' which we create as follows.  For each virial halo, we select 
from the different halo  overdensity  catalogues the  object  whose  
center of  mass  is closest to that  of the virial halo.  For the  
halos identified with a threshold that  is denser than virial,  we 
only keep the  object which corresponds most closely to the virial 
one.  To ensure good resolution (and hence  a match) in the  inner 
regions, we only  considered virial haloes with more than $10^3$ 
particles.  In this way, we define a series of approximately concentric 
spheres of ever higher density within the virial radius:  a density 
profile.  

If we average together all the profiles for a narrow bin in virial mass, 
then we can use this mean profile, and the scatter around it, to model 
the `matched halo' mass function associated with different overdensity 
thresholds $\Delta\ge\Delta_{vir}$, because 
\begin{equation}
 \frac{dn_\Delta}{dM_\Delta} =
    \int dM_{vir} \frac{dn_{vir}}{dM_{vir}}\,p(M_\Delta|M_{vir}).
 \label{convolve}
\end{equation}
If $p(M_\Delta|M_{vir})$ is sharply peaked around 
 $M_\Delta = g_{\Delta}(M_{vir})\,M_{vir}$ say, then 
\begin{equation}
 \frac{\mathrm{d}n_\Delta}{\mathrm{d}M_{\Delta}} = 
 \frac{\mathrm{d}n_{vir}}{\mathrm{d}M_{vir}}\,
 \frac{\mathrm{d}M_{vir}}{\mathrm{d}M_{\Delta}},
\end{equation} 
so that 
\begin{equation}    
 \left[\nu f(\nu)\right]_{\Delta}
 = g_{\Delta}(M_{vir})\,\left[\nu f(\nu)\right]_{vir}
 \label{rescale}
\end{equation}  
as in \citet[e.g.][]{bocquet15}.  But if the scatter around the mean 
profile is not negligible, then the full convolution of 
equation~(\ref{convolve}) must be performed.  

In principle, $p(M_\Delta|M_{vir})$ could be determined directly by 
fitting an NFW profile \citet{navarro96} to each virial halo in 
the stack.  This functional form has just one free parameter, the 
concentration $c$, so $p(M_\Delta|M_{vir})$ is simply related to the 
mean $c$ at fixed $M_{vir}$, and the scatter around this mean, which 
is known to be lognormal with width $\sigma_{\ln c|m} = 0.25$.

 % (m/Mvir)(rvir/r)^3 = [ln(1 + x cvir) - x cvir/(1 + x cvir)]/x^3/[ln(1+cvir) - cvir/(1+cvir)]

 % (D/Dvir) = [ln(1 + x cvir) - x cvir/(1 + x cvir)]/x^3/[ln(1+cvir) - cvir/(1+cvir)]

In practice, we have not measured the profile shapes.  Rather, we 
use the fact that the concentration of a halo is related to its 
mass accretion history \citep{zhao09}, which we also do not measure.  
Rather, we estimate it, and hence $c$, using the model of 
\citet{giocoli12c}.  We then use the mean $c$ along with the 
assumption that the scatter is negligible, to compute the value of 
$g_\Delta$ to insert in equation~(\ref{rescale}) for the desired $\Delta$.

\begin{figure*}
\includegraphics[width=0.48\hsize]{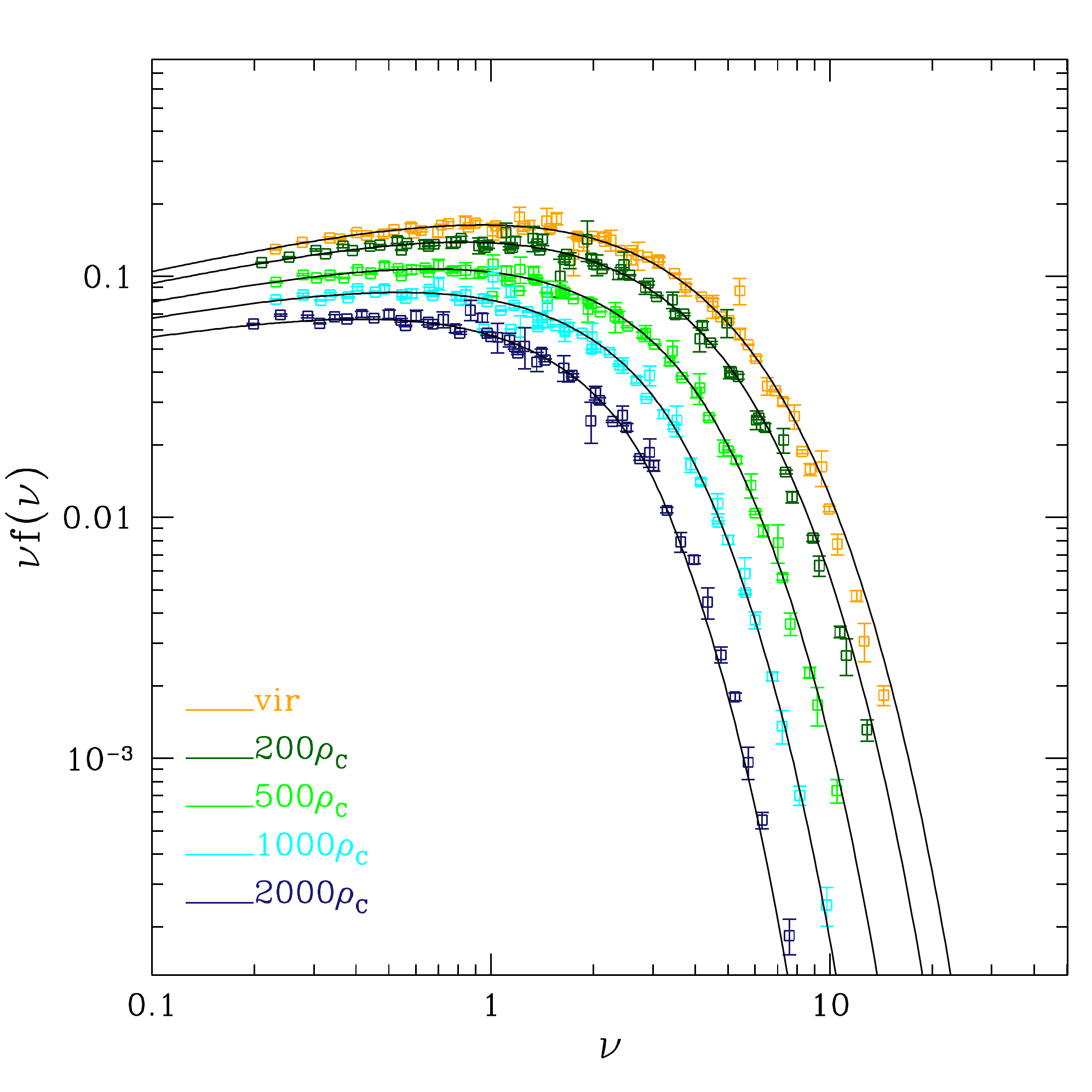}
\includegraphics[width=0.48\hsize]{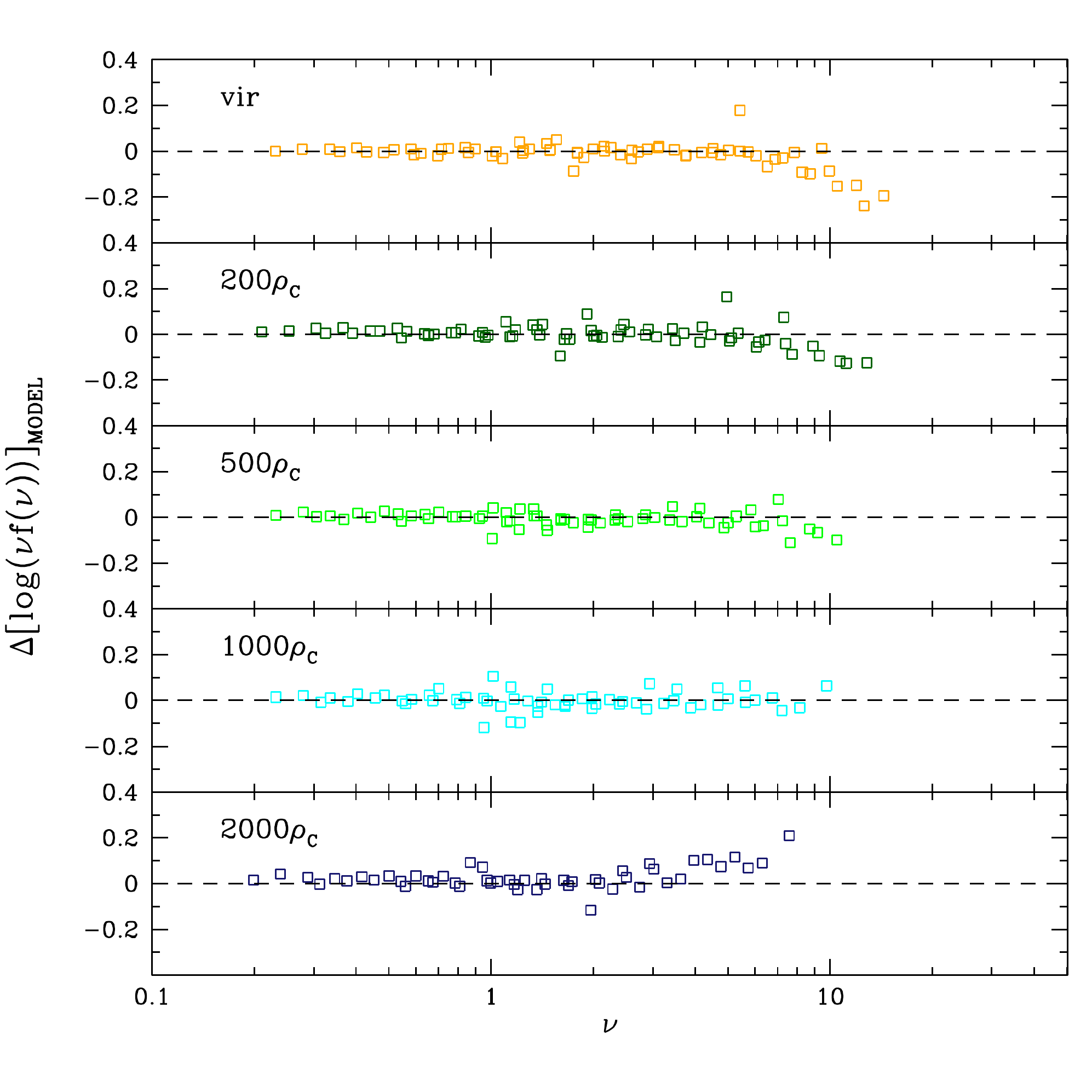}
\includegraphics[width=0.48\hsize]{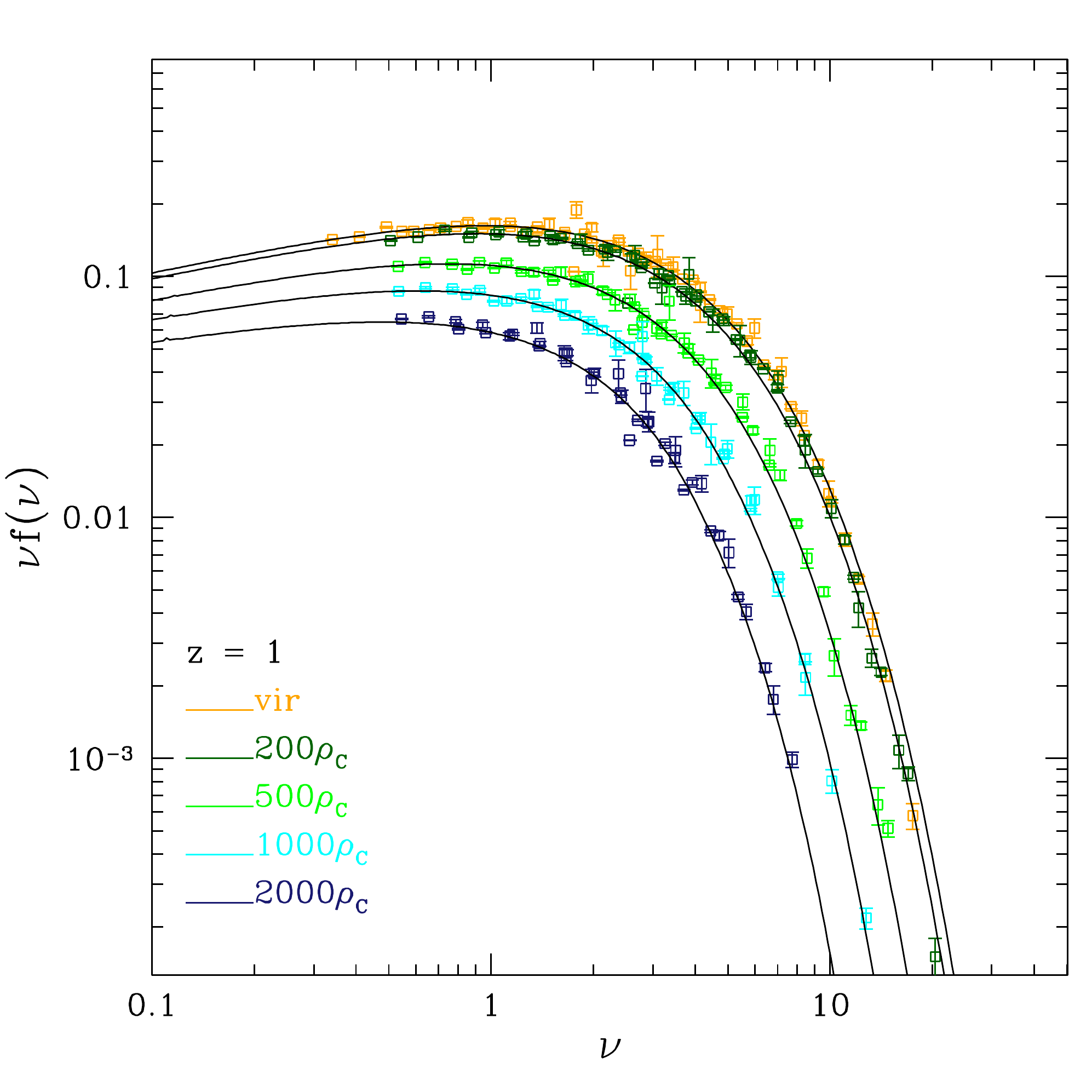}
\includegraphics[width=0.48\hsize]{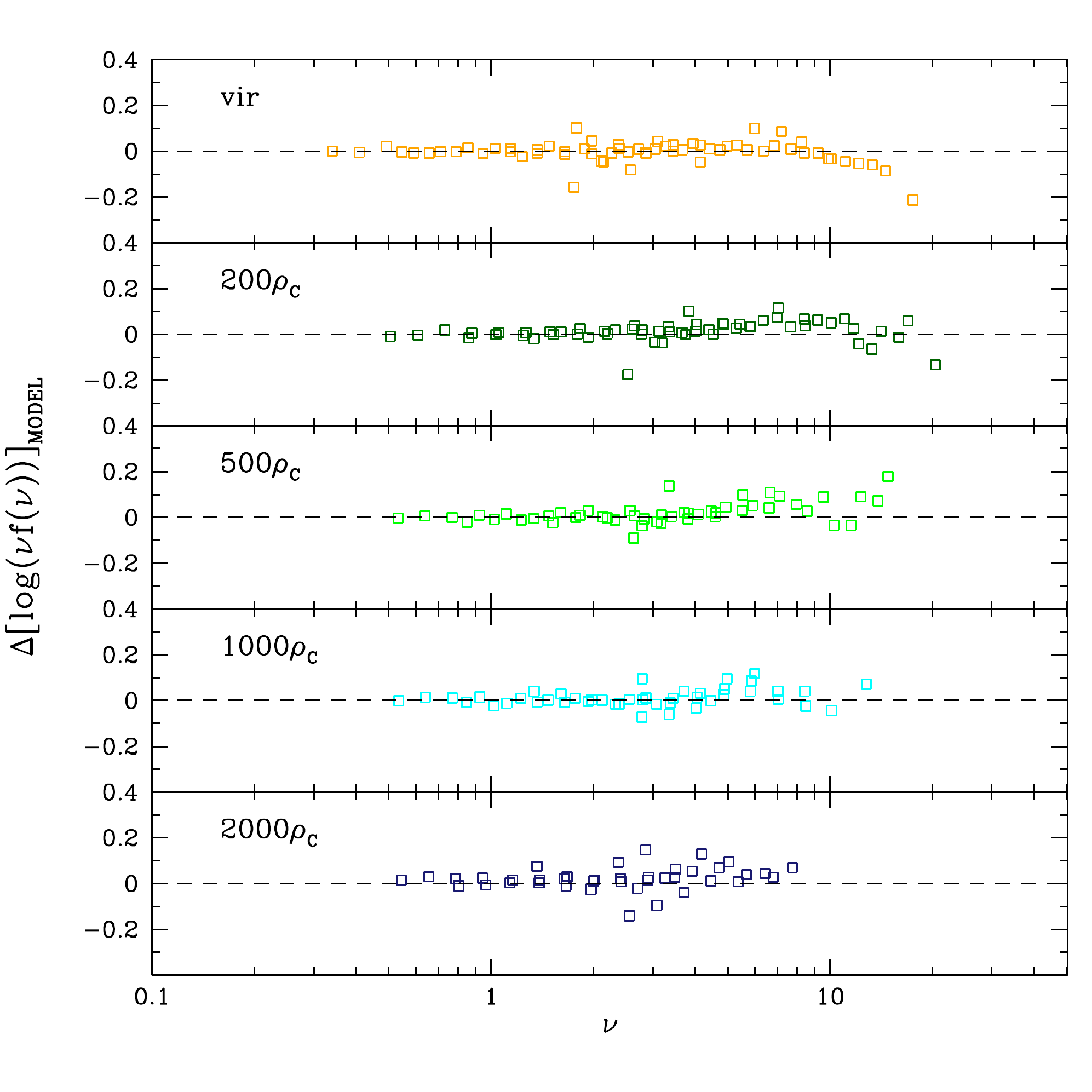}
\caption{$Top - left$: $z=0$ halo mass functions of the matched haloes.
  Different  colours  show different  overdensities,  while the  black
  solid  curves  show   the  corresponding  rescaled  mass  functions,
  obtained  from   Eq.   \ref{rescale}.   $Top   -right$:  logarithmic
  residuals from the  rescaled mass functions of the  upper panel.  
  The scatter with respect to the rescaled mass function is of the 
  same order  of the  intrinsic scatter  of the  mass function
  (shown   by   the   virial   case).  $Bottom$:  Same as top, but for
  $z=1$.\label{massf_match}}
\end{figure*}

The symbols in the left panels of Fig.~\ref{massf_match} show the mass
function of the haloes in the different catalogues matching the virial
systems, at  redshift $z=0$ and  $z=1$ for the  catalogues constructed
considering $200$,  $500$, $1000$ and $2000\rho_c$  as thresholds. The
solid curves, following the corresponding data points, show the virial
mass function rescaled to the different overdensities.

The  panels   on  the  right  of   Figure~\ref{massf_match}  show  the
logarithmic  residuals  of  the  data   points  with  respect  to  the
corresponding rescaled models.   They all remain well  below $10\%$ --
apart from the high-$\nu$ tail -- indicating that the precision of the
rescaled mass functions is of the same order as of the ``true'' virial
mass function.  Some of the discrepancy may be caused by the fact that
not all haloes follow NFW profiles all the way down to the very center
\citep{einasto65,retana-montenegro12,ludlow13,dutton14}.

The case  of  $200\rho_{b}$ must be treated separately, since the more 
massive haloes identified using $200\rho_{b}$ may include more than 
one smaller virial overdensity haloes in their outskirts which we 
exclude (e.g. object $3$ in Figure~\ref{figSO}).  
Nevertheless, in this case also, the rescaled  relation captures the
behaviour of  the mass function measured  in the simulations with a
precision comparable to the  other overdensities.  Recall 
that  the rescaled  mass function  at $200\rho_b$  at $z=1$  is almost
the same as the virial mass function, since $\Delta_{vir}$ at high 
redshift is nearly equal to $200\rho_b$.

{Figure  \ref{diff_match} shows  how  different these  rescaled
  mass   functions    are   from   the   original    fits   of   Table
  \ref{tab_massf}. The upper panels show  the two mass functions - the
  rescaled one in dashed lines and  the original ones in solid lines -
  for  all  the   density  thresholds;  the  lower   panel  shows  the
  differences in halo  counts between the two cases.   Notice that the
  matched mass  functions 'lose'  haloes: because  some of  the smaller
  denser systems  are halo  substructures.  This is  true for  all the
  overdensities except for $200\rho_{b}$: as  we said, in this case we
  are outside the virial haloes and so we may in fact have more haloes
  than those found at the virial overdensity.

\begin{figure}
\includegraphics[width=\hsize]{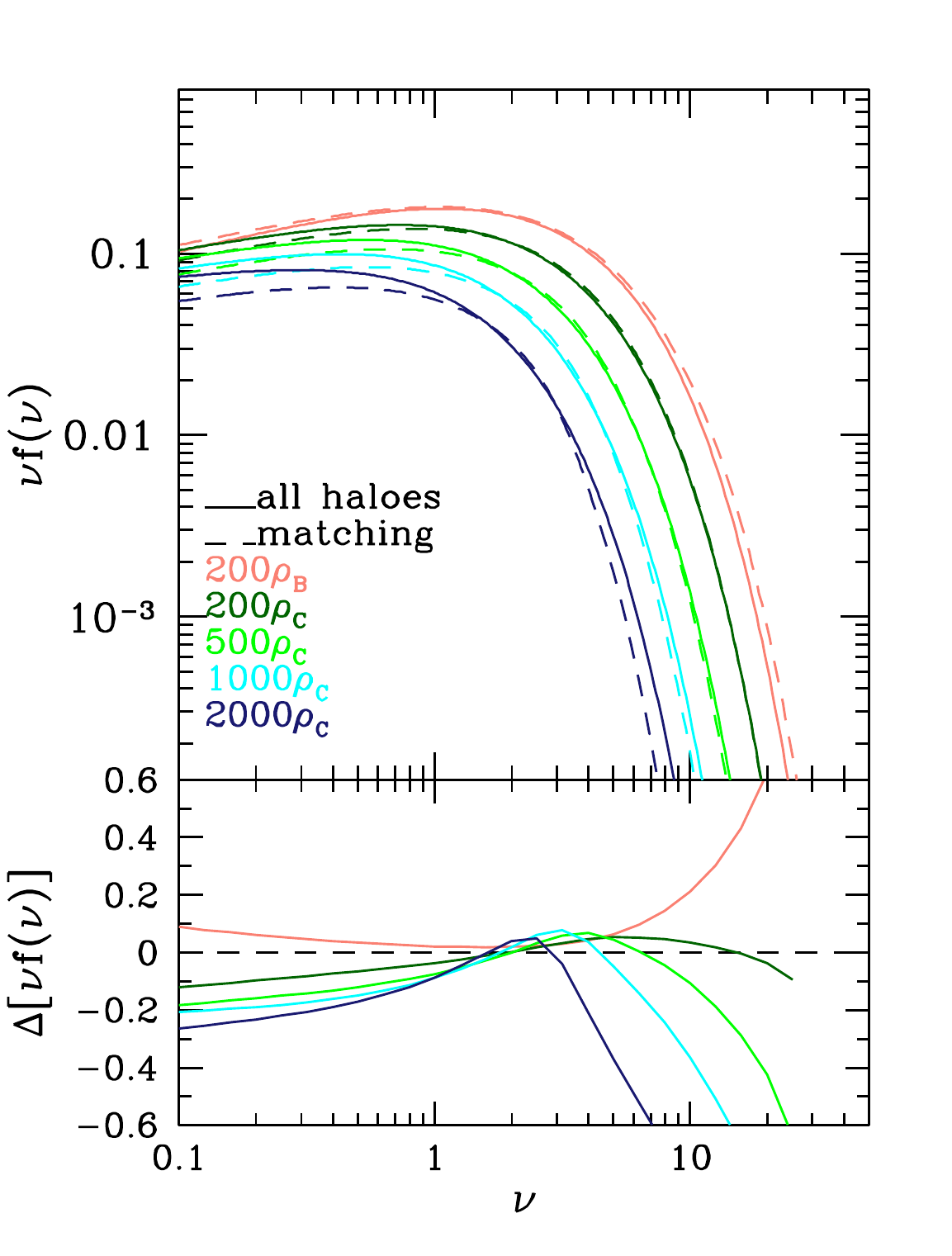}
\caption{Difference between the mass functions of Table
  \ref{tab_massf} (solid lines) and the rescaled ones from the 
  matched haloes as in Figure \ref{massf_match} (dashed lines); the
  lower panel shows the residuals between the two relations on the halo counts \label{diff_match}}
\end{figure}

\section{Comparison with previous work}\label{comparison}

Figure~\ref{other_models} shows the residuals between our virial mass  
function $\nu f(\nu)$ --  the one calibrated from all $z\le 1.25$ 
outputs of the $1024^3$ and the  $512^3$ simulations -- and the 
mass functions derived in previous work (as indicated).  The shaded 
region in each panel shows  the range in $\nu$, at $z=0$, over which 
the various authors have  calibrated their mass functions. 
The two vertical dot-dashed lines  show the range in $\nu$ over 
which we have performed our calibration.

To facilitate comparison, the main  features of these works are listed
in Table~\ref{tab_others}.  Most of these authors used a FOF algorithm
to identify the  haloes in their simulations: while b=0.2  is the most
common  choice  for the  linking  length,  \citet{manera10} use  three
different  values  (b=0.15,  0.168, 0.2);  moreover,  \citet{warren06}
discuss   how  to   correct  the   FOF  masses,   claiming  that   the
identification is influenced by how well  an halo is resolved and thus
the  same  linking  length  may  identify  structures  with  different
enclosed  overdensities.   The  same  argument is  discussed  also  in
\citet{courtin11} and \citet{more11} and they predict a linking length
equivalent  to our  virial overdensity  (at z=0)  of 0.193  and 0.206,
respectively.    \citet{sheth99b}   and    \citet{tinker08}   use   SO
algorithms,  while \citet{watson13}  compares the  results of  the two
methods (in Figure \ref{other_models} we show only their universal FOF
fit).  At intermediate masses all  the analytical mass functions agree
quantitatively.  Larger differences arise for more massive systems: in
this  range, the  precision of  the fit  is strongly  affected by  the
resolution of  the simulation and  consequently by the number  of high
mass haloes.

The main purpose of  our work is to analyse the  impact of the density
threshold chosen  to identify  the haloes on  the universality  of the
mass function:  the majority  of the  identification algorithms  use a
threshold    $\Delta$   of    178   or    200   for    all   redshifts
(e.g. \citet{tinker08}  and \citet{watson13}).   Since we  have argued
that ignoring  the redshift evolution  of $\Delta$ causes most  of the
observed non-universality one must  carefully match density thresholds
when comparing  different works.   This is shown  more clearly  in the
lower  panels  of  Figure~\ref{other_models},  where  we  compare  our
results with  those of  \citet{tinker08}: we tried  to match  their SO
thresholds   as  closely   as  possible   to  ours,   to  test   their
compatibility.   The  Figure  shows  --  from top  to  bottom  --  the
residuals between: ($i$) their $\Delta=200$ fits with our virial fits;
($ii$) their  $\Delta=300$ fits  with our  virial fits;  ($iii$) their
$\Delta=200$  fits with  our  $200\rho_{b}$ fits.   The agreement  for
cases $ii$  and $iii$ is much  better than for $i$.   Since our virial
overdensity   \textbf{at  z=0}   is   closer   to  $\Delta=300$   than
$\Delta=200$, this shows that different analyses converge provided one
compares  compatible  halo   identification  schemes  \citep{knebe13}.
Moreover, not only are our $200\rho_{b}$ fits close to the main result
of  \citet{tinker08},   but  the   non-universality  they   report  is
compatible  with  the redshift  evolution  of  our $200\rho_{b}$  mass
function   (Figure  \ref{so200b}):   the  systematic   deviation  from
universality  is  primarily  due  to the  choice  of  the  overdensity
threshold used in the halo  identification method. The agreement, once
overdensities  have been  correctly matched,  is even  more reassuring
when one notes that \citet{tinker08}  fit their counts to a functional
form  that  differs   from  our  equation~(\ref{st99mf}).   Additional
reassurance  comes from  the fact  that, while  one may  expect trends
arising  from  differences  in  halo identification  methods,  our  SO
results  lie in  between the  $l=0.168$  and $l=0.15$  FOF results  of
\citet{manera10}, just as they should.

As a  final remark we note  that, even when the  computational schemes
are theoretically similar, the mass function for structures identified
with different  codes (in the same  simulation!)  can differ by  up to
$10\%$ \citep{knebe11}.  This might  explain some of the discrepancies
between  our best  fit model  and those  of other  works. Finally,  we
underline that there  could be other effects depending  on the assumed
cosmological parameters as explored by \citet{murray13}.

\begin{table*} \centering
\begin{tabular}{|c|c|c|c|c|c|}  \hline & $N_{part}$ & box [$h^{-1}Mpc$]&
  cosmological models & algorithm & threshold \\ 
\hline
Sheth \& Tormen 1999  & (3 sim x) $256^{3}$& 85, 141&
S/O/$\Lambda$CDM & SO & $\Delta=\Delta_{vir}$\\ 
Jenkins et al. 2001 & $256^{3}$, $512^{3}$, $10^{9}$ & 84 -3000 &
$\tau$/$\Lambda$CDM & FOF &b=0.2\\ 
Warren et al. 2006 & $1024^{3}$ & 96 - 3072 & $\Lambda$CDM & FOF & b corrected\\  
Watson et al. 2013 & $3072^{3}$-$6000^{3}$ & 11 - 6000 & $\Lambda$CDM (WMAP5) & FOF
(SO)&b=0.2 ($\Delta$=178)\\ 
Manera et al. 2010  & (49 sim x) $640^{3}$  & 1280 & $\Lambda$CDM &
FOF & l=0.15, 0.168, 0.2\\
Tinker et al. 2008 & $512^{3}$-$1024^{3}$ & 80 - 1280 & $\Lambda$CDM
(WMAP1-3) & SO & $\Delta$=n x $\Delta_{bg}$\\ \hline
\end{tabular}
\caption{Schematic (i.e. incomplete) summary of the main features
    of the other works to which we compare our results in Figure
    \ref{other_models} \citep{sheth99b,jenkins01,warren06,watson13,manera10,tinker08}. 
    We list the resolution and scale of their simulations and the main algorithm with the associated threshold 
    used to identify the haloes.
 \label{tab_others}}
\end{table*}

\begin{figure} \centering
\includegraphics[width=\hsize]{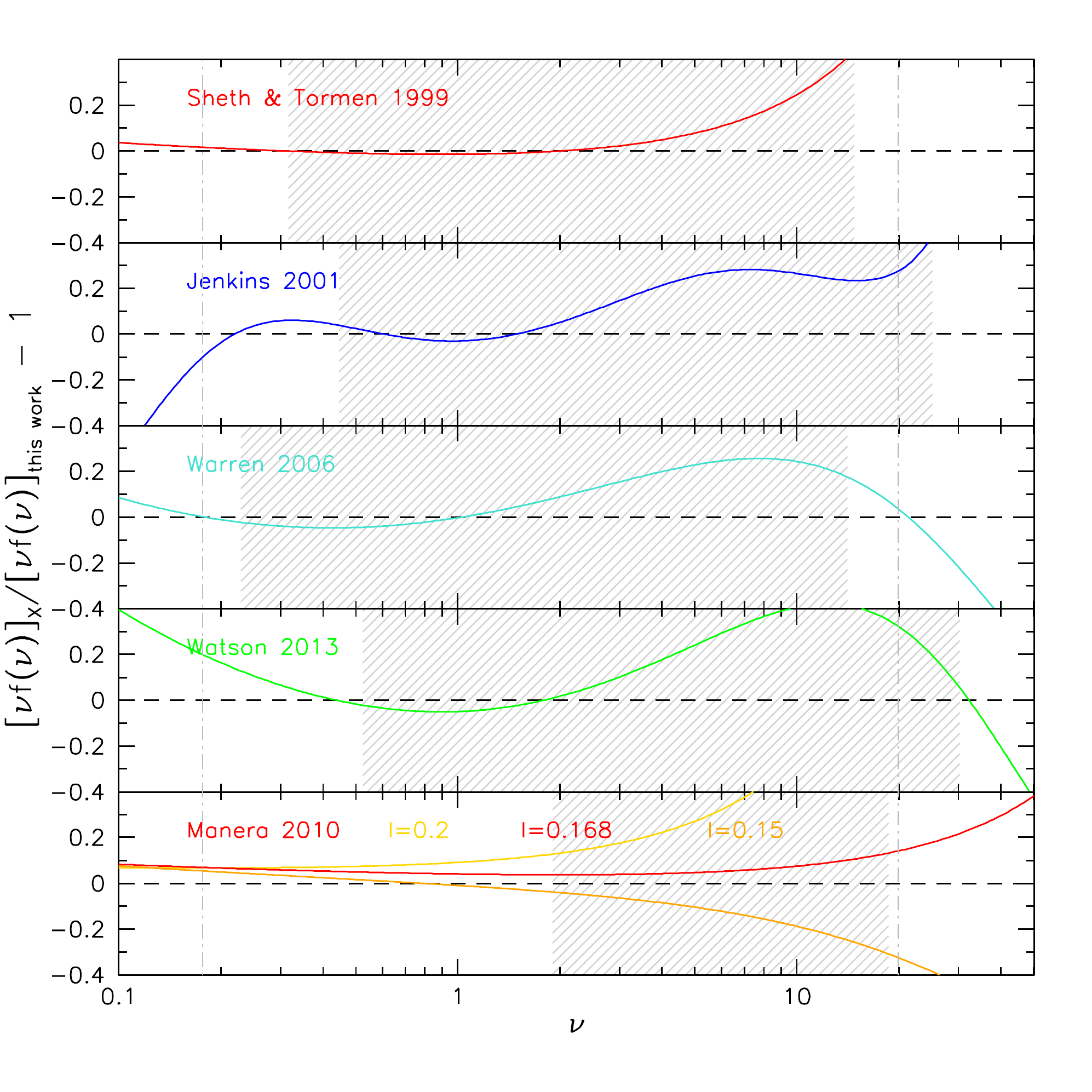}
\includegraphics[width=\hsize]{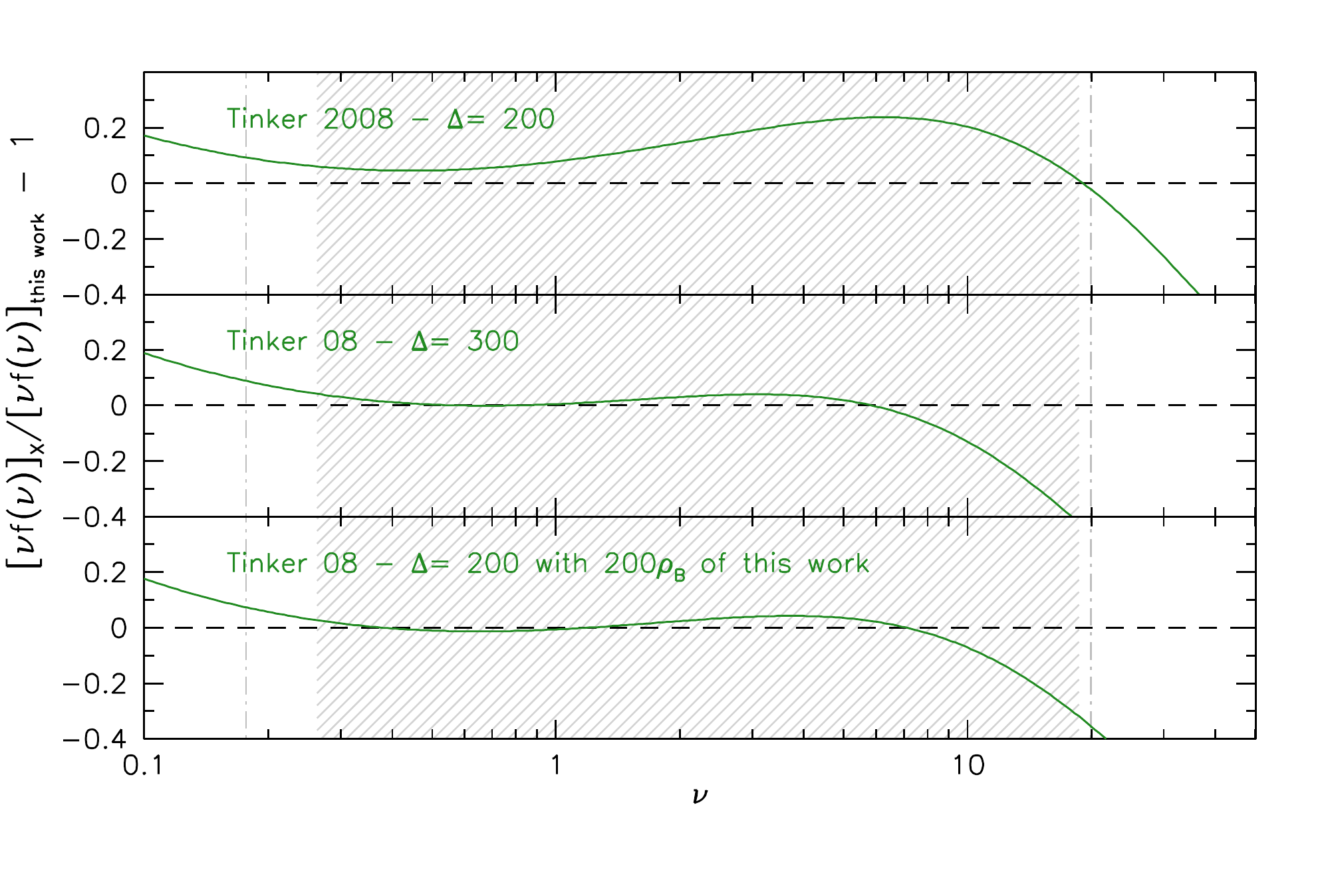}
\caption{Comparison with some previous work.  In  all panels,  the dotted
  vertical gray  lines show our  range in  mass (for virial  haloes at
  $z<1.25$) and the gray  region show  an estimate of the range in mass
  of the  other works; we  calculated this last converting  their mass
  ranges in $\nu$, so it may not be perfectly the same but it gives an
  estimate of  the overlapping  regions. $Top$:  we show  the relative
  residuals of our  virial mass function $\nu f(\nu)$  with respect to
  the models of other works.  From top to bottom, we compare our model
  with    \citet{sheth99b},    \citet{jenkins01},    \citet{warren06},
  \citet{watson13} and  \citet{manera10}.  For this last,  we show the
  results obtained  from FOF catalogues with linking lengths of 0.2, 
  0.168  and   0.15, which they fit to the same functional form we do.  
  Our results lie between their fits to the smaller linking catalogs, 
  as they should, over the whole range of masses.  $Bottom$: we compare
  our   results   with   those  of   \citet{tinker08}   at   different
  overdensities.   First  we  show  the relative  residuals  of  their
  $\Delta=200$ reference  model with  our best  fit calculated  at the
  virial density; secondly we compare their model with our best fit at
  $200\rho_{b}$. Finally,  we test  the opposite  situation, comparing
  our  virial  model with  their  mass  function at  $\Delta=300$,  an
  overdensity closer  to the value  of $\Delta_{vir}$ adopted  in this
  work.\label{other_models}}
\end{figure}

\section{Discussion and Conclusions}\label{conclusions}
In this paper we analysed a set of cosmological simulations in order
to study the halo mass function and its dependence on redshift,
cosmology and the halo identification method. In what follows we
summarise our main results.
\begin{enumerate}
\item We  demonstrated and  confirmed the  universality of  the virial
  halo mass function, by comparing  the measured mass function at many
  different redshifts  and in simulations with  different cosmological
  models.   As  stated  in  \citet{sheth99b, courtin11},  the  halo  mass
  function at  any time, and for  any cosmological model, can  be well
  described by  a single  functional form once  mass and  redshift are
  appropriately         parametrised        in         terms        of
  $\nu=\delta^2_c(z)/\sigma^2(M)$.
\item We showed that only the  virial overdensity leads to a universal
  halo  mass function:  most of  the non-universality  seen in
    other  works  arises   from  not  using  the   virial  value  when
    identifying halos.  Commonly used values of 178 or 200$\times$ the
    critical or background density induce non-universal trends.  
\item We derived  three different sets of best  fit parameters for the
 virial halo mass function:
\begin{itemize}
\item  $a=0.794$, $p=0.247$  and $A=0.333$  - from  the z=0  virial mass
function in the Planck cosmology;
\item $a=0.7663$, $p=0.2579$ and $A=0.3298$ - from the virial mass function
in the Planck cosmology, using all  the data points up to $z=1.25$ (15
snapshots $\times$ 6 runs);
\item $a=0.7689$,  $p=0.2536$ and $A=0.3295$  - from the mass  functions in
all  cosmologies,  using  all  the  data points  up  to  $z=1.25$  (15
snapshots  $\times$ 16 runs).
\end{itemize} 
The last two are  in remarkable agreement -- at the  per mil level for
the  best-fit parameters,  and  sub percent  for  the mass  function),
illustrating the level of the universality as a function of cosmology.
In general, the normalisation is  the most stable parameter, while $a$
and $p$ change  between the first and the other  two cases: when using
only the $z=0$ points, the high-$\nu$  tail is less well resolved, and
so the fit is  less precise in the determination of  the full shape of
the mass function.
\item We  presented a simple  rescaling method which allows one to estimate
  the three parameters of the fitting function through first or second
  order scaling  relations.  The three parameters  ($a,p,A_0$) are 
  smooth functions of the overdensity, for any redshifts.  
  Equations~(\ref{model}) are able to describe the change in slope 
  and normalisation of  the mass function (as a function of redshift 
  and  overdensity) with good precision (Figs. \ref{so200b}, 
  \ref{so200c}, \ref{so_highc} and \ref{resid_cosmo}).

\item Finally,  we studied the  mass function of  ``matched haloes'':
  the  counterparts of  virial haloes  at different overdensities. 
  We  provide an  efficient rescaling method, equation~(\ref{rescale}), 
  which uses knowledge of the mass density profile and the 
  concentration-virial mass relation, to estimate their mass function 
  with a high precision (Fig. \ref{massf_match}).

\end{enumerate}
These rescaling  methods are useful for comparing analyses which use 
different definitions of halo  mass.  In particular, they can be used 
to translate our universal virial halo mass functions to the nonuniversal 
form associated with halo definitions which are closer to those commonly 
used in observational studies -- ranging from the X-ray and SZ to 
the visible and near infrared.

We  conclude  that   --  over  the  range   of  redshifts  and
  cosmological models  in our simulation  set -- the virial  halo mass
  function  is, to  within  $5-8\%$  for a  wide  range  of masses,  a
  universal  function   of  redshift  and   cosmology.   Non-universal
  behaviour  can be  an artifact  induced by  the halo  identification
  method and by the choice  of the overdensity threshold.  Other, true
  departures from universality  may be sought in  the other components
  of  the universe  or in  more extreme  cosmological models.   Future
  extremely well  resolved simulations  should allow an  percent level
  estimate of the universality of the mass function.

\section{Acknowledgements} 
GD,  CG   and  RA   thank  Marceau   Limousin  and   LAM  (Laboratoire
d'Astrophysique de Marseille), for organising and allowing the meeting
in   Marseille  where   this   work  started,   and   Y.  Rasera   and
P.S. Corasaniti for comments on an  early version.  CG thanks CNES for
financial support.  GD  has been partially financed by  the ``Ing.  A.
Gini'' Fellowship.  This work has been  carried out thanks also to the
support of the OCEVU Labex  (ANR-11-LABX-0060) and the A*MIDEX project
(ANR-11-IDEX-0001-02) funded by  the "Investissements d'Avenir" French
government program managed  by the ANR. The Cosmology  Group in Padova
thanks  Vincenzo Mezzalira  for installing  and managing  the ``Nemo''
cluster. We thanks  the anonymous referee for  his/her useful comments
that  considerably  improve the  presentation  of  our results.  Great
gratitude goes to the cosmology group at UPENN -- Philadelphia -- from
GD and CG for the hospitality.

\appendix \newpage
\section{Results for ellipsoidal haloes}\label{EOs}

We   identified  haloes   using   the  Ellipsoidal   Halo  Finder   of
\citep{despali13}  and repeated  the  analysis described  in the  main
text.  Although the best-fit  parameters are systematically different,
the results are  otherwise consistent with those found  for SO haloes:
the virial overdensity yields universality, and others do not.

Figure~\ref{eovir} shows the measured mass  function for EO virial 
haloes and the residuals  with respect to the best  fit relation  
(calculated at $z=0$), similarly to what was done in Figure~\ref{sovir}.  
Table~\ref{tab_massfeo} summarises how the best fit $z=0$ parameters 
depend on overdensity threshold.  They behave regularly, just as 
for the SO haloes shown in Figure \ref{fit_univ},  and  are
well modelled by equations (\ref{fit_eo})  (equivalent to the
blue  curves in Figure \ref{fit_univ}).

\begin{table} \centering
\begin{tabular}{|c|c|c|c|}  \hline $\rho$  (EO)& a  & p  &  A\\ 
\hline
$200\rho_{b}$ & 0.6730 $\pm 0.004$& 0.1783 $\pm 0.007$ & 0.4237 $\pm 0.001$ \\ 
$\Delta_{vir}$ & 0.7369 $\pm 0.004$ & 0.2089 $\pm 0.007$ & 0.3894 $\pm
0.001$\\ 
$200\rho_{c}$ &  0.8286 $\pm 0.005$ &  0.2776 $\pm 0.009$ &  0.3335
$\pm 0.001$ \\  
$500\rho_{c}$ & 1.0223 $\pm 0.007$ &  0.3417 $\pm 0.009$ &  0.2672
$\pm 0.001$ \\ 
$1000\rho_{c}$  & 1.2576$\pm 0.009$  & 0.3803 $\pm 0.012$ &  0.2181
$\pm 0.001$ \\
$2000\rho_{c}$ & 1.6088$\pm 0.015$ & 0.3824$\pm 0.018$ & 0.1755 $\pm 0.001$ \\ \hline
\end{tabular}
 \caption{Parameters of the best-fitting  mass function for EO haloes at $z=0$.
 \label{tab_massfeo}}
\end{table}

For EO  haloes we have obtained analogous relations to those of 
equation~(\ref{fit_so1}):
\begin{eqnarray}  
 a &=&  0.7057+0.2125x+0.3268x^{2}  ,  \nonumber\\  
 p &=& 0.2206+0.0.1937x-0.04570x^{2}, \label{fit_eo}\\ 
 A_{0} &=& 0.3953-0.1768x. \nonumber
\end{eqnarray}

\begin{figure*}
\includegraphics[width=\hsize]{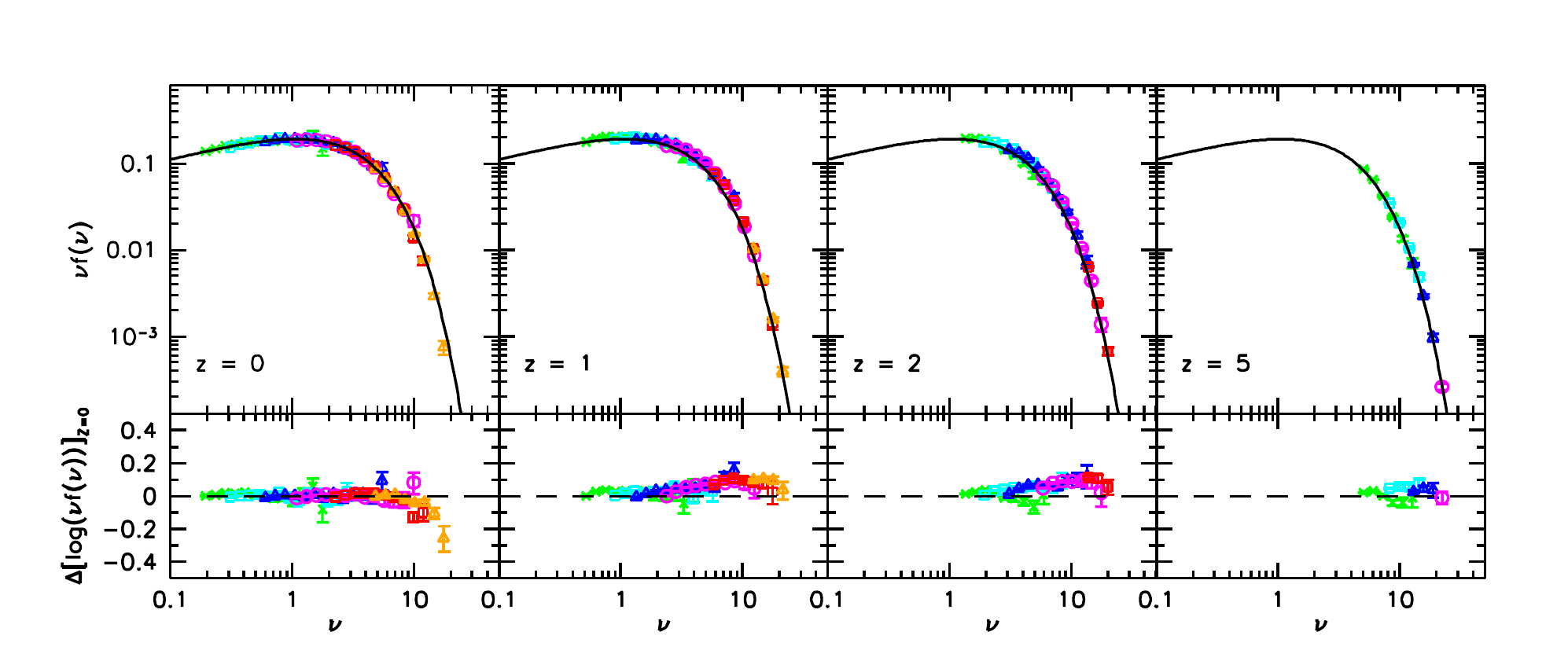}
\caption{Same  as  Figure  \ref{sovir},  but  for the  haloes  in  the
Ellipsoidal Overdensity catalogue.\label{eovir}}
\end{figure*}

\section{Issues from the Initial Conditions}\label{ICs}

The  initial power  spectrum $P(k)$ plays an important role when 
translating from halo mass $M$ to the scaled variable
 $\nu = \delta_c^2/\sigma^2(M)$, 
since it determines the value  of the mass variance  $S(M) = \sigma^2(M)$.

In  principle, $\sigma(M)$ should be the same for any simulation run 
with the same cosmological parameters.  In practice, the number of 
Fourier modes that can be effectively sampled in the  initial 
conditions depends on some computational limits: 
 ($i$) the box size determines the minimum mode in the power spectrum 
       that can be  sampled by a specific simulation; 
 ($ii$) each of  our simulations started from a  different random 
       realisation of  the displacement  field.  
If not properly accounted for, these effects may increase the scatter --
and  possibly  bias  --  the  halo mass  function  measured  in  the
simulation with respect to theoretical model predictions.  

E.g.,  using  the theoretical  linear  power  spectrum when  computing
$\sigma$  yields  a precision  which  cannot  be reduced  below  about
$10\%$.       However,      future       wide      field      missions
\citep[e.g. Euclid][]{euclidredbook} require  percent level precision.
As we describe  below, to achieve this, we  calculate $\sigma(M)$ from
the  actual  realization  of  the   initial  power  spectrum  in  each
simulation box -- scaled using linear  theory to $z=0$ -- and not from
the theoretical linear power spectrum.

\subsection{Power   spectrum   of    each   realisation}  
Figure~\ref{variance} shows some  detailed results  on the  computational
effects introduced  in the initial  $P(k)$ regarding ($i$)  the random
number seed choice and ($ii$) the box size.  The top panel shows
$S=\sigma^{2}(M)$ as a function of mass $M$, for all the simulations 
of the  main set; the black  line shows the  theoretical power spectrum 
from  CAMB.  To compute $S(M)$ for each simulation we integrated the 
initial power spectrum from the minimum mode resolvable in each box 
(see equation~(\ref{eqsigma}).  The bottom  panel shows the relative 
difference between  the measured $S(M)$  and the one 
calculated using the  theoretical $P(k)$.
For comparison the lower panel shows the initial power spectrum
for each simulation of our main set and the relative differences with
respect to the theoretical one.  The  use of the actual power spectrum
for each simulation allows us to achieve a more precise estimate of
$\nu$ for the box, and hence greater precision on the measured mass 
function.

\begin{figure}
\hspace{-2cm} \includegraphics[width=\hsize, angle=90]{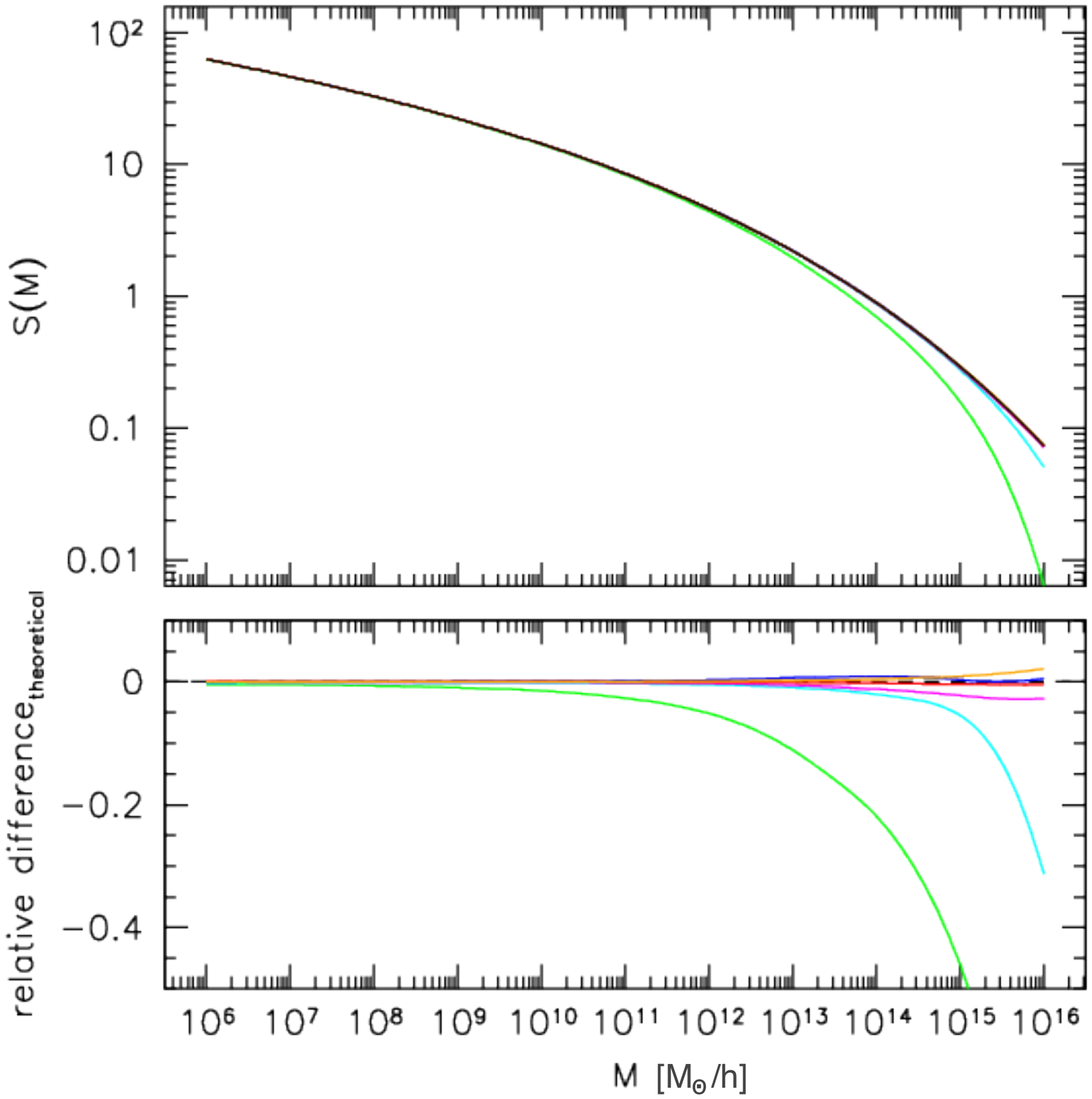}
\includegraphics[width=\hsize]{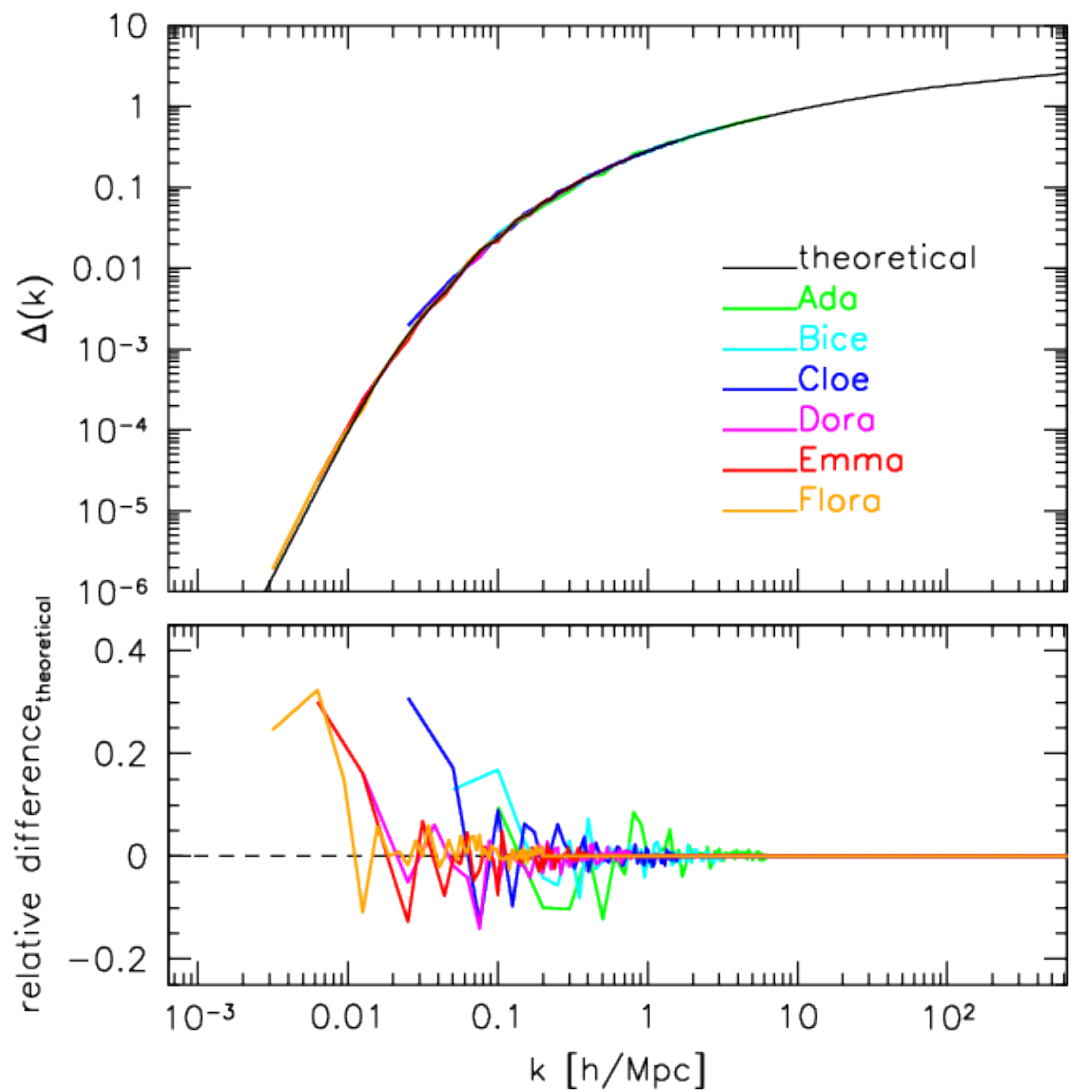}
\caption{ $Top$: Measured mass variance  and residuals with respect to
  the theoretical calculation.  $Bottom$: Initial power spectrum
  measured from  each simulation and relative  difference with respect
  to the theoretical case. \label{variance}}
\end{figure}

\subsection{The random  seed for the initial displacement field}  

The six  simulations from the  main set are independent,  meaning that
they  have  all been  generated  from  different realisations  of  the
displacement  field.  However,  these simulations  also differ  in box
size, etc.  To isolate the effect of  the random seed, we reran two of
the simulations of the secondary set (``uno'' and ``wmap7'') using two
different random seeds.  We found few percent-level differences in the
high  mass  tail,  with  one  of  the  two  seeds  generating  greater
departures from the universality.  See, e.g. the last blue triangle in
the $z=0$  panel of Figure \ref{sovir}  or the gray and  brown ones in
the  corresponding  panel  of Figure  \ref{resid_cosmo_vir}.  This  is
despite  the fact  that we  use the  actual realisation  of the  power
spectrum  when  computing  $\sigma(M)$.   Thus, even  when  all  other
parameters are  held fixed, some of  the scatter in the  measured halo
mass function is introduced by the initial seed.

\subsection{Mass variance definition}  
The top part of Figure~\ref{figS} compares different definitions for
the mass variance $S(M)$.  
The  relative differences  are presented  with respect  to  the 
theoretical  prediction  --  i.e.   integrating  the theoretical 
power spectrum ( black solid curve). The green line in the top panel 
(orange  line in the middle panel) shows  the prediction for $S(M)$ 
in  our smallest  (largest) simulation  Ada (Flora) where we 
sum over the modes in the initial conditions, starting from the 
$k$-mode corresponding  to the box-size. The blue short dashed 
line represents  one way of  accounting for  the limited
box-size of the simulation:  we subtract from the theoretical one
the  mass  variance  computed  using  a top-hat filter with scale
\begin{equation}
  R_{Box} = \left( \dfrac{3\,M_{Box}}{4 \pi \rho_b} \right)^{1/3} =L_{Box}\left( \dfrac{3}{4 \pi} \right)^{1/3}.
\end{equation}
The red long  dashed curve shows the case in  which we compute $S(M)$ 
accounting for the fact that the periodic boundary conditions of the 
simulation mean that the Fourier modes within the box are constrained 
to give the background density on the scale of the box.  This 
constraint modifies the expression for the variance to 
\[
 s  \rightarrow s  \left(1 - \dfrac{S_x^2}{s  S_{Box}} \right)
              = s  \left(1 - \dfrac{S_x^2}{S_{Box}^2}\frac{S_{Box}}{s} \right)
\]
where 
\begin{equation}
 S_x(M) = \int \frac{{\rm d}k}{k}\, 
                    \frac{k^3P_{\rm lin}(k,z)}{2\pi^2}\, W[kR(M)] W[kR_{box}]\,
\end{equation}
represents the cross correlation variance between the lagrangian 
scale  of the  halo and  that of  the simulation  box
\citep[e.g.][]{musso14a,musso14b}.  
For large boxes $S_{Box}\ll s$ so $S_x/S_{box}\approx 1$ and the 
correction to $s$ becomes vanishingly small.

\begin{figure}
\includegraphics[width=\hsize]{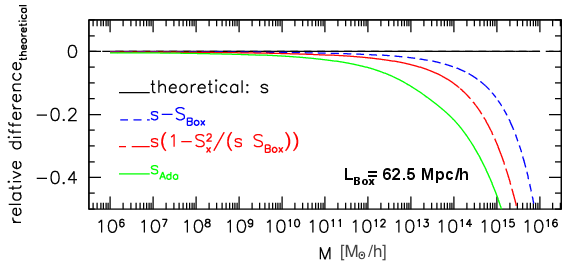}
\includegraphics[width=\hsize]{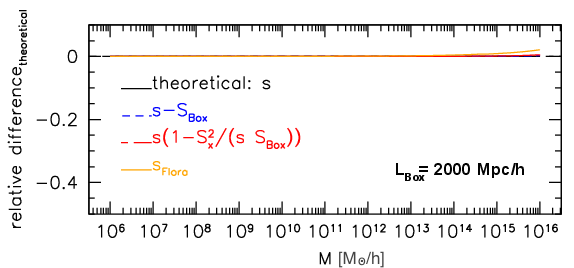}
\includegraphics[width=\hsize]{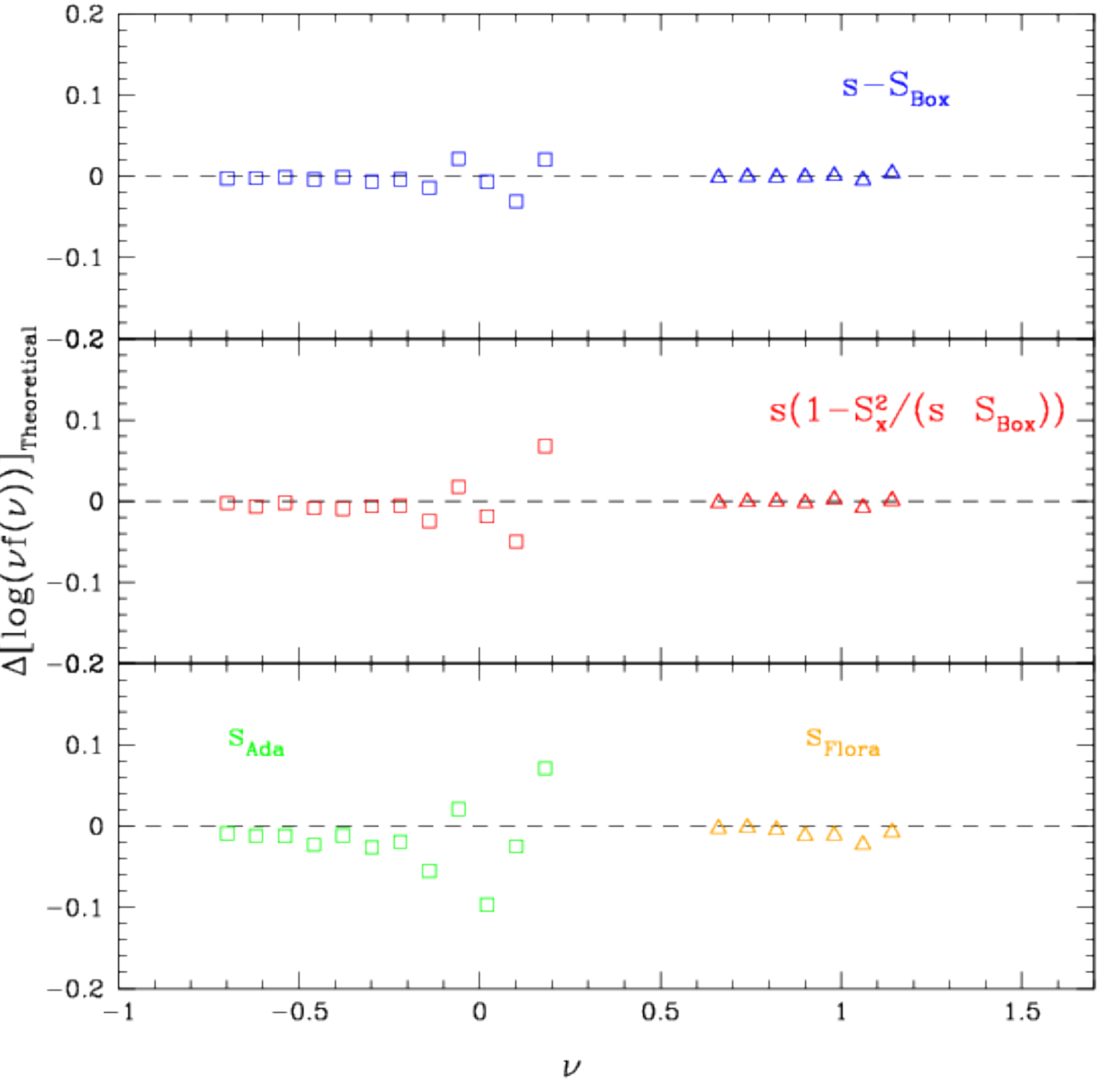}
\caption{Top: Comparison of  the variance as a function  of mass using
  the different prescriptions described in the text to account for the
  finite  box size,  for our  smallest and  largest simulation  boxes.
  Bottom:  Differences between  the measured  mass functions  when the
  prescriptions shown in  the top panel are implemented,  and when the
  box size is effectively infinite. \label{figS}}
\end{figure}

The bottom panels show the $z=0$ mass function residuals in Ada and 
Flora which result from using different prescriptions for accounting 
for the finite box size (with respect to the case in which the box 
size is infinite).  For the larger box, Flora, the effect is 
negligible, but for the smaller box, Ada, it is not.
If the mass variance does not properly account for  the  limited  
box-size, the rescaling to $\nu$ results in non-negligible bias 
at small masses and large scatter at large masses.

\subsection{ZA or 2LPT? } For all  our simulations, we used N-GenIC to
generate  the initial  conditions from  a glass  distribution; N-GenIC
uses  first  order  perturbation theory, calculated  through  the  
Zel'dovich Approximation.  Not including  second order perturbations 
in the  IC  has  a strong  impact  on the  results obtained at 
outputs ``close''  to  the initial conditions \citet{crocce06}:  the 
differences  are  larger  for simulations  with small  boxes and/or  
late starting  redshifts.  We  chose $z=99$  as a starting redshift 
for all our simulations  with a box size larger than 100 $h^{-1}Mpc$; 
for Ada, which has  a box of only 62.5 $h^{-1}Mpc$, we generated the 
ICs at $z=124$.  To test the impact of ICs on our results we ran  an 
identical copy of Bice  and  of  the  two  WMAP7 simulations,  using 
exactly  the  same parameters and seeds,  but generating  the IC  
with 2LPTic  instead of N-GenIC.  The symbols in Figure~\ref{2lpt} 
show the differences in the measured mass functions; different 
coloured solid curves show the relation of \citet{reed13}:
\begin{equation}
 \mathrm{d} n_{\rm ZA}/\mathrm{d} n_{\rm 2LPT} = e^{-0.12 \dfrac{a_i}{a_f} \nu_{1}^{2.5}},
\end{equation}
even      though       it      was      only       calibrated      for
$1 \lesssim  \nu_{1} \lesssim  5$.  Note that  their $\nu$  is defined
differently               than               ours:               their
$\nu_{1}\equiv\sqrt{\nu}=\delta_{c}(z)/\sigma(M,z)$.    At  the   four
redshifts shown,  the differences  are on average  very small  -- less
than 5\% at small  and intermediate masses and less than  10 \% in the
very  high-$\nu$  tail --  which  lies  within the  intrinsic  scatter
(e.g. that due to the random  initial seed) seen in the mass function.
A small  difference between the  two definitions  can be seen  at high
redshift (i.e. at $z=5$ in the  Figure), but this also lies within the
intrinsic  scatter of  the mass  function.   Since our  best fits  are
calibrated using data at $z\leq 1.25$, they are not affected by second
order effects  in the initial  condition density field.   For example,
the best fit parameters obtained  using Reed-rescaled points (all $z$,
all  cosmologies)  are  $a=0.7603$, $p=0.2549$  and  $A_{0}=0.330$  --
similar  at per-mil  level to  the ones  calculated with  the original
data.

\begin{figure}
\includegraphics[width=\hsize]{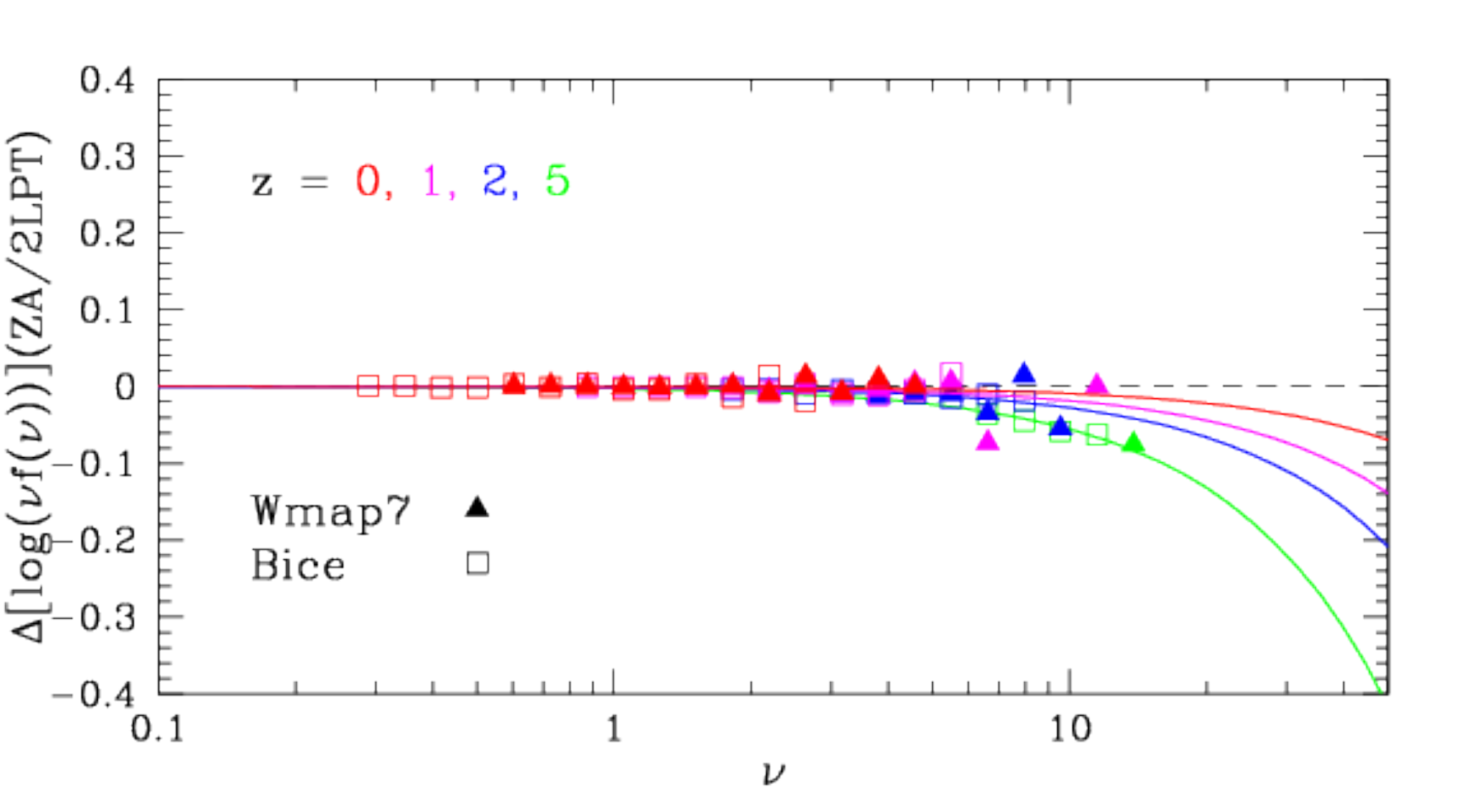}
\caption{Difference in  the measured mass function for different IC
  methods.   For four redshifts  (in different  colours), we  show the
  logarithmic residuals between  the results obtained using the Zel'dovich
  Approximation (ZA - obtained  with N-GenIC) and 2nd order lagrangian
  perturbation theory  (2LPT - obtained  with 2LPTic) to  generate the
  Initial   Conditions.   The   solid  curves   are   calculated  from
  \citet{reed13},  where the  authors present  a fit  to  the expected
  ratio between the  two methods, which is in  good agreement with our
  measurements. \label{2lpt}}
\end{figure}

\section{Cluster Mass Function}\label{cmf}

One of  the most important  application that necessite of  an accurate
and well calibrated mass function is the study of the observed cluster
counts   \citep{vikhlinin09,rozo10,planckxx,sartoris15}.  Degeneracies
between fitted  parameters mean that  the best fit parameters  to fits
restricted to  cluster mass  halos ($M  \ge 10^13  h^{-1}M_\odot$) may
differ from those  returned from fitting a larger range  of masses (we
refer  to  these  as the  CMF  and  HMF,  for  cluster and  halo  mass
functions, respectively).  This Appendix  discusses the expected level
of systematic bias this may induce on cosmological constraints derived
from the observed number of clusters per square degree.

The black  circles in the  top panel of Figure~\ref{figClMf}  show the
halo mass function  extracted from all cosmologies  and redshifts from
our simulation suite.  The main text shows that a good fit to the bins
that  contain  at  least  30  halos (red  triangles)  is  given  by  a
\citet{sheth99b}   mass  function   with  the   following  parameters:
$A_0=0.3295$,  $p=0.2536$   and  $a=0.7689$.   However  it   is  worth
mentioning that  according to the  points density distribution  in the
considered area,  this curve  may be less  accurate in  describing the
shape of the most massive haloes:  the cluster mass function.  To test
the difference, and  to better describe the objects  more massive than
$M_{vir}  \ge  3  \times 10^{13}h^{-1}M_{\odot}$  that  are  typically
associated with groups  and clusters of galaxies, we  have performed a
fit  to  the  orange  crosses  only.   The  best  fit  parameters  are
$A_0=0.8199$,    $a=0.3141$,    and     $p=0$,    as    reported    in
Table~\ref{tab_massf}. The value  $p=0$ is due to the fact  that it is
mainly the small masses which  determine $p$ \citep{sheth99b}.  In the
first bottom subpanel we show the  relative difference -- in log space
-- of the  two fits for the halo mass  function tail.  For comparison,
the other two subpanels show the relative residuals of the data points
with no error bars to the two corresponding fits.

\begin{figure}
  \includegraphics[width=\hsize]{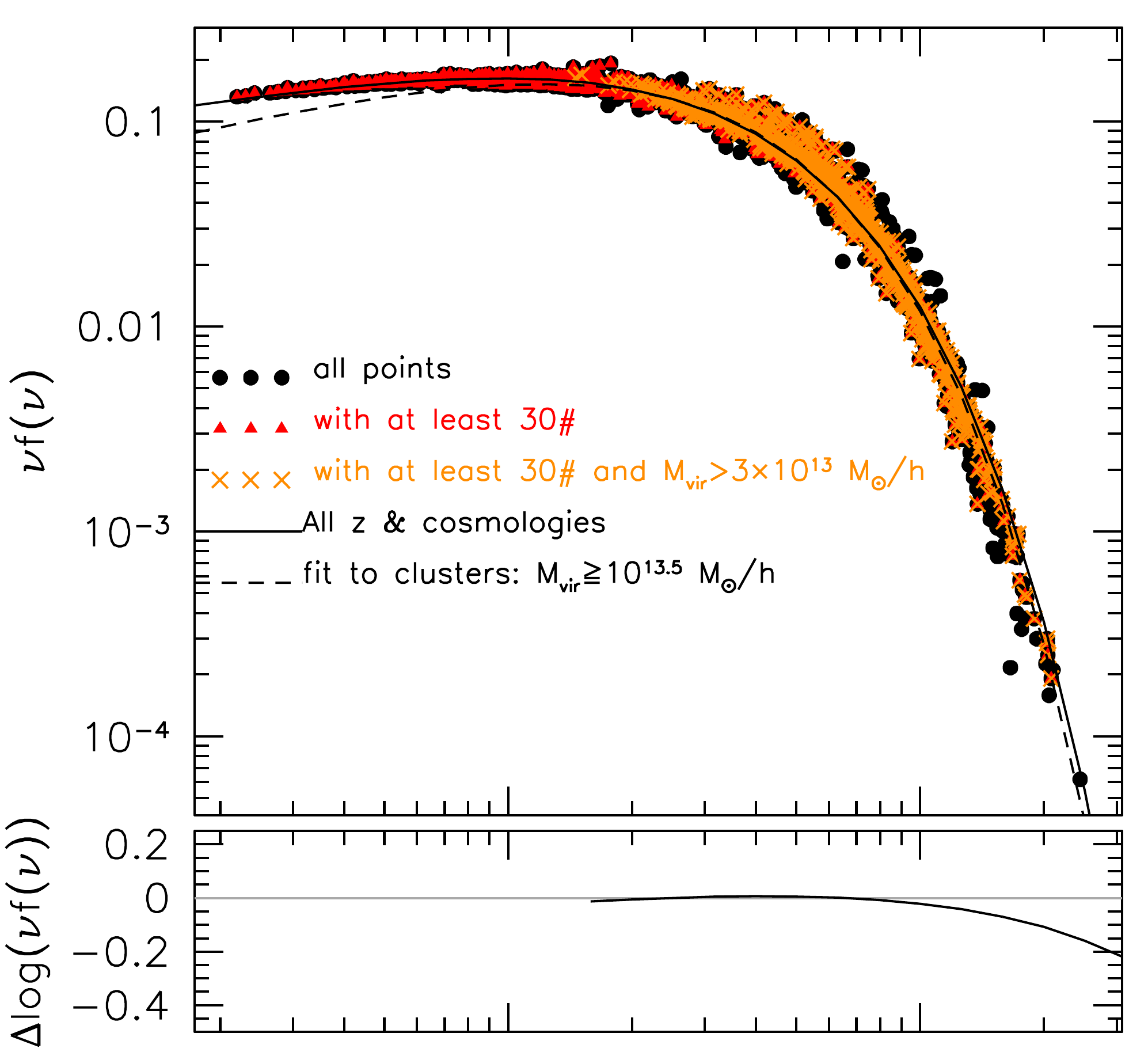}
  \includegraphics[width=\hsize]{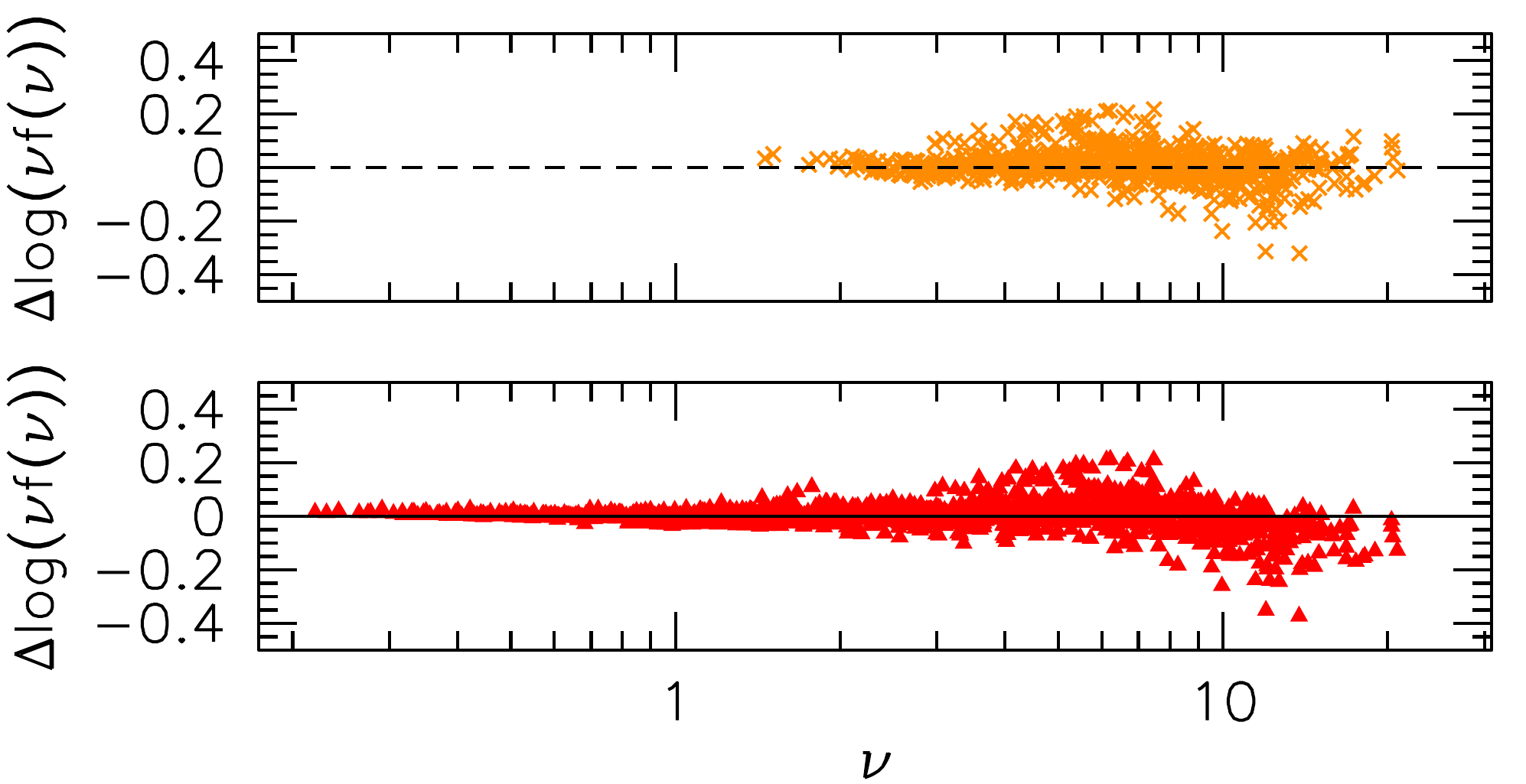}  
  \caption{Halo mass  function for all  considered cosmologies
      and  redshifts  extracted  from   our  simulation  suite  (black
      circles). Red triangles show bins  with at least 30 counts while
      the orange crosses  also require that the halos  be more massive
      than $3 \times 10^{13}M_{\odot}/h$.  The solid and dashed curves
      represent the best  fit to the red triangles  and orange crosses
      -- accounting for the associated  Poisson error bars.  The first
      subpanel shows the relative difference -- in log space -- of the
      two fits, while the other two present the relative difference of
      the data  points with  respect to  their corresponding  best fit
      curve.\label{figClMf}}
\end{figure}

A relative difference in the number  of clusters per square degree may
appear using  the fitting  function computed  using only  the clusters
(hereafter Cluster Mass  Function) or all haloes over  a broader range
of  masses  (Halo Mass  Function).   To  quantify this  difference  in
Fig.~\ref{figCosmoClMF} we present the  relative difference in cluster
counts -- $M_{vir} > 3 \times 10^{13}M_{\odot}/h$ -- per square degree
in  the  $\Omega_m$-$\sigma_8$  plane  between the  CMF  and  the  HMF
fits. To compute the count $N$ we have integrated the HMF and CMF fits
over $M_{vir} > 3 \times 10^{13}M_{\odot}/h$ and comoving volume.  The
figure shows  that in  our reference  Planck13 cosmology  the relative
difference in the  cluster counts between the two fits  is $\sim 4\%$.
It increases  to $\sim 10\%$ for  small values of both  $\Omega_m$ and
$\sigma_8$.  At high $\Omega_m$  and $\sigma_8$ the difference between
the HMF and CMF-derived counts is smaller.
 
\begin{figure}
  \includegraphics[width=\hsize]{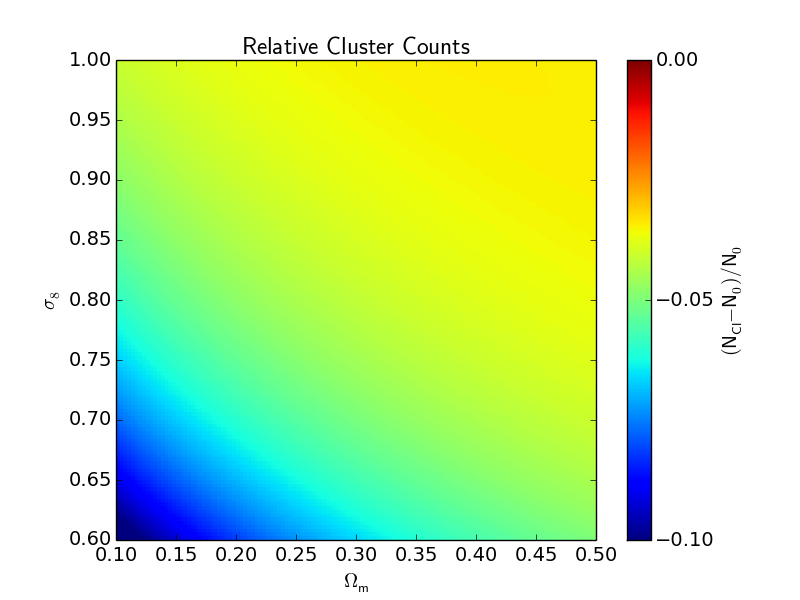}
\caption{Relative difference in the cluster counts -- systems with $M_{vir} > 3 \times 10^{13} M_{\odot}/h$ per square  degree -- in the $\Omega_m$-$\sigma_8$ plane, resulting from the best-fit parameters obtained for clusters (CMF) and all haloes (HMF) as presented in Fig.~\ref{univ_all}.
  \label{figCosmoClMF}}
\end{figure}

In  Fig.~\ref{figCosmoClMFChi} we  present  the relative  cosmological
constraints for the cluster  counts in the $\Omega_m$-$\sigma_8$ plane
between the  HMF and the CMF.   The relative difference is  well below
$10\%$  in  the   whole  range  and  the  diagonal   shape  shows  the
degeneration region in the parameter space.

This  suggests  that,  notwithstanding  the  small  difference
  between the  two fits, caution may  be necessary when using  fits to
  the halo mass function for precision cosmology.  For cluster counts,
  the CMF  reported in  the last row  of Table~\ref{tab_massf}  may be
  more  appropriate  than the  HMF,  and  may  yield better  than  4\%
  accuracy.

\begin{figure}
  \includegraphics[width=\hsize]{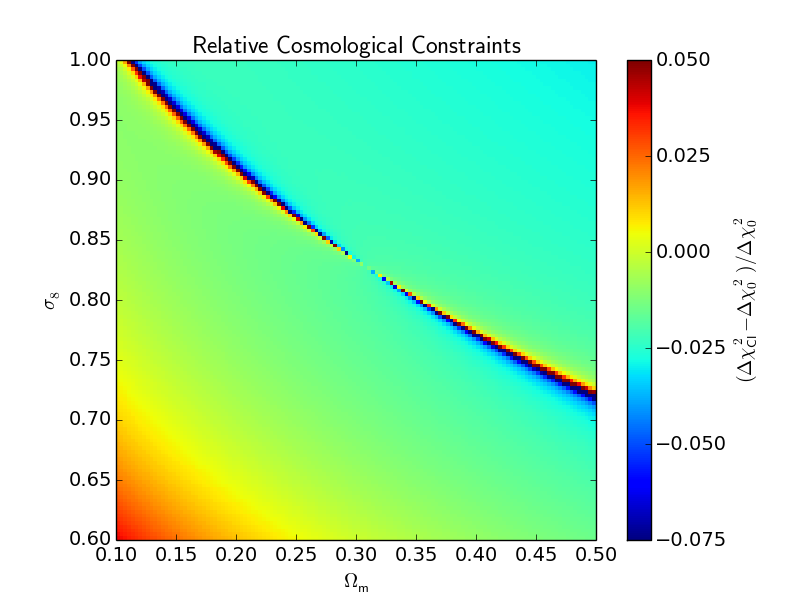}
\caption{Relative cosmological constraints between the CMF and
    the HMF in the $\Omega_m$-$\sigma_8$ plane for systems with $M_{vir} > 3 \times 10^{13} M_{\odot}/h$.
  \label{figCosmoClMFChi}}
\end{figure}

\bibliographystyle{mn2e}               
\bibliography{paper}

\begin{thebibliography}{}

\bibitem[\protect\citeauthoryear{Angulo, Springel, White, Jenkins, Baugh \&
  Frenk}{Angulo et~al.}{2012}]{angulo12}
Angulo R.~E.,  Springel V.,  White S. D.~M.,  Jenkins a.,  Baugh C.~M.,
  Frenk C.~S.,  2012, \mnras, 426, 2046

\bibitem[\protect\citeauthoryear{{Bocquet}, {Saro}, {Dolag} \&
  {Mohr}}{{Bocquet} et~al.}{2015}]{bocquet15}
{Bocquet} S.,  {Saro} A.,  {Dolag} K.,    {Mohr} J.~J.,  2015, ArXiv e-prints

\bibitem[\protect\citeauthoryear{{Boldrin}, {Giocoli}, {Meneghetti},
  {Moscardini}, {Tormen} \& {Biviano}}{{Boldrin} et~al.}{2015}]{boldrin15}
{Boldrin} M.,  {Giocoli} C.,  {Meneghetti} M.,  {Moscardini} L.,  {Tormen} G.,
    {Biviano} A.,  2015, ArXiv e-prints

\bibitem[\protect\citeauthoryear{{Bonamigo}, {Despali}, {Limousin}, {Angulo},
  {Giocoli} \& {Soucail}}{{Bonamigo} et~al.}{2015}]{bonamigo15}
{Bonamigo} M.,  {Despali} G.,  {Limousin} M.,  {Angulo} R.,  {Giocoli} C.,
  {Soucail} G.,  2015, \mnras, 449, 3171

\bibitem[\protect\citeauthoryear{{Bond}, {Cole}, {Efstathiou} \&
  {Kaiser}}{{Bond} et~al.}{1991}]{bond91}
{Bond} J.~R.,  {Cole} S.,  {Efstathiou} G.,    {Kaiser} N.,  1991, \apj, 379,
  440

\bibitem[\protect\citeauthoryear{{Borgani} \& {Kravtsov}}{{Borgani} \&
  {Kravtsov}}{2011}]{borgani11}
{Borgani} S.,  {Kravtsov} A.,  2011, Advanced Science Letters, 4, 204

\bibitem[\protect\citeauthoryear{{Bryan} \& {Norman}}{{Bryan} \&
  {Norman}}{1998}]{bryan98}
{Bryan} G.~L.,  {Norman} M.~L.,  1998, \apj, 495, 80

\bibitem[\protect\citeauthoryear{{Carroll}, {Press} \& {Turner}}{{Carroll}
  et~al.}{1992}]{carroll92}
{Carroll} S.~M.,  {Press} W.~H.,    {Turner} E.~L.,  1992, ARA\&A, 30, 499

\bibitem[\protect\citeauthoryear{{Castorina}, {Sefusatti}, {Sheth},
  {Villaescusa-Navarro} \& {Viel}}{{Castorina} et~al.}{2014}]{castorina14}
{Castorina} E.,  {Sefusatti} E.,  {Sheth} R.~K.,  {Villaescusa-Navarro} F.,
  {Viel} M.,  2014, JCAP, 2, 49

\bibitem[\protect\citeauthoryear{{Corasaniti} \& {Achitouv}}{{Corasaniti} \&
  {Achitouv}}{2011}]{corasaniti11}
{Corasaniti} P.~S.,  {Achitouv} I.,  2011, \prd, 84, 023009

\bibitem[\protect\citeauthoryear{{Courtin}, {Rasera}, {Alimi}, {Corasaniti},
  {Boucher} \& {F{\"u}zfa}}{{Courtin} et~al.}{2011}]{courtin11}
{Courtin} J.,  {Rasera} Y.,  {Alimi} J.-M.,  {Corasaniti} P.-S.,  {Boucher} V.,
     {F{\"u}zfa} A.,  2011, \mnras, 410, 1911

\bibitem[\protect\citeauthoryear{{Crocce}, {Fosalba}, {Castander} \&
  {Gazta{\~n}aga}}{{Crocce} et~al.}{2010}]{crocce10}
{Crocce} M.,  {Fosalba} P.,  {Castander} F.~J.,    {Gazta{\~n}aga} E.,  2010,
  \mnras, 403, 1353

\bibitem[\protect\citeauthoryear{{Crocce}, {Pueblas} \& {Scoccimarro}}{{Crocce}
  et~al.}{2006}]{crocce06}
{Crocce} M.,  {Pueblas} S.,    {Scoccimarro} R.,  2006, \mnras, 373, 369

\bibitem[\protect\citeauthoryear{{Cui}, {Baldi} \& {Borgani}}{{Cui}
  et~al.}{2012}]{cui12}
{Cui} W.,  {Baldi} M.,    {Borgani} S.,  2012, \mnras, 424, 993

\bibitem[\protect\citeauthoryear{{Cui}, {Borgani} \& {Murante}}{{Cui}
  et~al.}{2014}]{cui14}
{Cui} W.,  {Borgani} S.,    {Murante} G.,  2014, \mnras, 441, 1769

\bibitem[\protect\citeauthoryear{{Davis}, {Efstathiou}, {Frenk} \&
  {White}}{{Davis} et~al.}{1985}]{davis85}
{Davis} M.,  {Efstathiou} G.,  {Frenk} C.~S.,    {White} S.~D.~M.,  1985, \apj,
  292, 371

\bibitem[\protect\citeauthoryear{{Del Popolo} \& {Gambera}}{{Del Popolo} \&
  {Gambera}}{1998}]{delpopolo98}
{Del Popolo} A.,  {Gambera} M.,  1998, \aap, 337, 96

\bibitem[\protect\citeauthoryear{{Del Popolo} \& {Gambera}}{{Del Popolo} \&
  {Gambera}}{1999}]{delpopolo99}
{Del Popolo} A.,  {Gambera} M.,  1999, \aap, 344, 17

\bibitem[\protect\citeauthoryear{{Despali}, {Giocoli} \& {Tormen}}{{Despali}
  et~al.}{2014}]{despali14}
{Despali} G.,  {Giocoli} C.,    {Tormen} G.,  2014, \mnras, 443, 3208

\bibitem[\protect\citeauthoryear{{Despali}, {Tormen} \& {Sheth}}{{Despali}
  et~al.}{2013}]{despali13}
{Despali} G.,  {Tormen} G.,    {Sheth} R.~K.,  2013, \mnras, 431, 1143

\bibitem[\protect\citeauthoryear{{Dutton} \& {Macci{\`o}}}{{Dutton} \&
  {Macci{\`o}}}{2014}]{dutton14}
{Dutton} A.~A.,  {Macci{\`o}} A.~V.,  2014, \mnras, 441, 3359

\bibitem[\protect\citeauthoryear{{Einasto}}{{Einasto}}{1965}]{einasto65}
{Einasto} J.,  1965, Trudy Astrofizicheskogo Instituta Alma-Ata, 5, 87

\bibitem[\protect\citeauthoryear{{Eke}, {Cole} \& {Frenk}}{{Eke}
  et~al.}{1996}]{eke96}
{Eke} V.~R.,  {Cole} S.,    {Frenk} C.~S.,  1996, \mnras, 282, 263

\bibitem[\protect\citeauthoryear{{Ettori}, {Morandi}, {Tozzi}, {Balestra},
  {Borgani}, {Rosati}, {Lovisari} \& {Terenziani}}{{Ettori}
  et~al.}{2009}]{ettori09}
{Ettori} S.,  {Morandi} A.,  {Tozzi} P.,  {Balestra} I.,  {Borgani} S.,
  {Rosati} P.,  {Lovisari} L.,    {Terenziani} F.,  2009, \aap, 501, 61

\bibitem[\protect\citeauthoryear{{Evrard}, {Bialek}, {Busha}, {White}, {Habib},
  {Heitmann}, {Warren}, {Rasia}, {Tormen}, {Moscardini}, {Power}, {Jenkins},
  {Gao}, {Frenk}, {Springel}, {White} \& {Diemand}}{{Evrard}
  et~al.}{2008}]{evrard08}
{Evrard} A.~E.,  {Bialek} J.,  {Busha} M.,  {White} M.,  {Habib} S.,
  {Heitmann} K.,  {Warren} M.,  {Rasia} E.,  {Tormen} G.,  {Moscardini} L.,
  {Power} C.,  {Jenkins} A.~R.,  {Gao} L.,  {Frenk} C.~S.,  {Springel} V.,
  {White} S.~D.~M.,    {Diemand} J.,  2008, \apj, 672, 122

\bibitem[\protect\citeauthoryear{{Giocoli}, {Marulli}, {Baldi}, {Moscardini} \&
  {Metcalf}}{{Giocoli} et~al.}{2013}]{giocoli13}
{Giocoli} C.,  {Marulli} F.,  {Baldi} M.,  {Moscardini} L.,    {Metcalf} R.~B.,
   2013, \mnras, 434, 2982

\bibitem[\protect\citeauthoryear{{Giocoli}, {Meneghetti}, {Ettori} \&
  {Moscardini}}{{Giocoli} et~al.}{2012c}]{giocoli12c}
{Giocoli} C.,  {Meneghetti} M.,  {Ettori} S.,    {Moscardini} L.,  2012c,
  \mnras, 426, 1558

\bibitem[\protect\citeauthoryear{{Giocoli}, {Pieri} \& {Tormen}}{{Giocoli}
  et~al.}{2008}]{giocoli08a}
{Giocoli} C.,  {Pieri} L.,    {Tormen} G.,  2008, \mnras, 387, 689

\bibitem[\protect\citeauthoryear{{Jenkins}, {Frenk}, {White}, {Colberg},
  {Cole}, {Evrard}, {Couchman} \& {Yoshida}}{{Jenkins}
  et~al.}{2001}]{jenkins01}
{Jenkins} A.,  {Frenk} C.~S.,  {White} S.~D.~M.,  {Colberg} J.~M.,  {Cole} S.,
  {Evrard} A.~E.,  {Couchman} H.~M.~P.,    {Yoshida} N.,  2001, \mnras, 321,
  372

\bibitem[\protect\citeauthoryear{{Kitayama} \& {Suto}}{{Kitayama} \&
  {Suto}}{1996}]{kitayama96}
{Kitayama} T.,  {Suto} Y.,  1996, \mnras, 280, 638

\bibitem[\protect\citeauthoryear{{Knebe}, {Knollmann}, {Muldrew}, {Pearce},
  {Aragon-Calvo}, {Ascasibar}, {Behroozi}, {Ceverino} \& {et~al.}}{{Knebe}
  et~al.}{2011}]{knebe11}
{Knebe} A.,  {Knollmann} S.~R.,  {Muldrew} S.~I.,  {Pearce} F.~R.,
  {Aragon-Calvo} M.~A.,  {Ascasibar} Y.,  {Behroozi} P.~S.,  {Ceverino} D.,
  {et~al.} 2011, \mnras, 415, 2293

\bibitem[\protect\citeauthoryear{{Knebe}, {Pearce}, {Lux}, {Ascasibar},
  {Behroozi}, {Casado}, {Moran}, {Diemand} \& {et~al.}}{{Knebe}
  et~al.}{2013}]{knebe13}
{Knebe} A.,  {Pearce} F.~R.,  {Lux} H.,  {Ascasibar} Y.,  {Behroozi} P.,
  {Casado} J.,  {Moran} C.~C.,  {Diemand} J.,    {et~al.} 2013, \mnras, 435,
  1618

\bibitem[\protect\citeauthoryear{{Komatsu}, {Smith}, {Dunkley}, {Bennett},
  {Gold}, {Hinshaw}, {Jarosik}, {Larson}, {Nolta} \& {et~al.}}{{Komatsu}
  et~al.}{2011}]{komatsu11}
{Komatsu} E.,  {Smith} K.~M.,  {Dunkley} J.,  {Bennett} C.~L.,  {Gold} B.,
  {Hinshaw} G.,  {Jarosik} N.,  {Larson} D.,  {Nolta} M.~R.,    {et~al.} 2011,
  \apjs, 192, 18

\bibitem[\protect\citeauthoryear{{Lacey} \& {Cole}}{{Lacey} \&
  {Cole}}{1993}]{lacey93}
{Lacey} C.,  {Cole} S.,  1993, \mnras, 262, 627

\bibitem[\protect\citeauthoryear{{Lacey} \& {Cole}}{{Lacey} \&
  {Cole}}{1994}]{lacey94}
{Lacey} C.,  {Cole} S.,  1994, \mnras, 271, 676

\bibitem[\protect\citeauthoryear{{Laureijs}, {Amiaux}, {Arduini},
  {Augu{\`e}res}, {Brinchmann}, {Cole}, {Cropper}, {Dabin}, {Duvet} \& et
  al.}{{Laureijs} et~al.}{2011}]{euclidredbook}
{Laureijs} R.,  {Amiaux} J.,  {Arduini} S.,  {Augu{\`e}res} J.~.,  {Brinchmann}
  J.,  {Cole} R.,  {Cropper} M.,  {Dabin} C.,  {Duvet} L.,    et al. 2011,
  ArXiv e-prints

\bibitem[\protect\citeauthoryear{Lewis, Challinor \& Lasenby}{Lewis
  et~al.}{2000}]{camb}
Lewis A.,  Challinor A.,    Lasenby A.,  2000, Astrophys. J., 538, 473

\bibitem[\protect\citeauthoryear{{Longair}}{{Longair}}{1998}]{longair98}
{Longair} M.~S.,  ed. 1998, {Galaxy formation}

\bibitem[\protect\citeauthoryear{{Ludlow}, {Navarro}, {Boylan-Kolchin}, {Bett},
  {Angulo}, {Li}, {White}, {Frenk} \& {Springel}}{{Ludlow}
  et~al.}{2013}]{ludlow13}
{Ludlow} A.~D.,  {Navarro} J.~F.,  {Boylan-Kolchin} M.,  {Bett} P.~E.,
  {Angulo} R.~E.,  {Li} M.,  {White} S.~D.~M.,  {Frenk} C.,    {Springel} V.,
  2013, \mnras, 432, 1103

\bibitem[\protect\citeauthoryear{{Macci{\`o}}, {Dutton} \& {van den
  Bosch}}{{Macci{\`o}} et~al.}{2008}]{maccio08}
{Macci{\`o}} A.~V.,  {Dutton} A.~A.,    {van den Bosch} F.~C.,  2008, \mnras,
  391, 1940

\bibitem[\protect\citeauthoryear{{Macci{\`o}}, {Dutton}, {van den Bosch},
  {Moore}, {Potter} \& {Stadel}}{{Macci{\`o}} et~al.}{2007}]{maccio07}
{Macci{\`o}} A.~V.,  {Dutton} A.~A.,  {van den Bosch} F.~C.,  {Moore} B.,
  {Potter} D.,    {Stadel} J.,  2007, \mnras, 378, 55

\bibitem[\protect\citeauthoryear{{Manera}, {Sheth} \& {Scoccimarro}}{{Manera}
  et~al.}{2010}]{manera10}
{Manera} M.,  {Sheth} R.~K.,    {Scoccimarro} R.,  2010, \mnras, 402, 589

\bibitem[\protect\citeauthoryear{{Mo}, {van den Bosch} \& {White}}{{Mo}
  et~al.}{2010}]{mo10}
{Mo} H.,  {van den Bosch} F.~C.,    {White} S.,  2010, {Galaxy Formation and
  Evolution}

\bibitem[\protect\citeauthoryear{{More}, {Kravtsov}, {Dalal} \&
  {Gottl{\"o}ber}}{{More} et~al.}{2011}]{more11}
{More} S.,  {Kravtsov} A.~V.,  {Dalal} N.,    {Gottl{\"o}ber} S.,  2011, \apjs,
  195, 4

\bibitem[\protect\citeauthoryear{{Moreno}, {Giocoli} \& {Sheth}}{{Moreno}
  et~al.}{2008}]{moreno08}
{Moreno} J.,  {Giocoli} C.,    {Sheth} R.~K.,  2008, \mnras, 391, 1729

\bibitem[\protect\citeauthoryear{{Murray}, {Power} \& {Robotham}}{{Murray}
  et~al.}{2013}]{murray13}
{Murray} S.~G.,  {Power} C.,    {Robotham} A.~S.~G.,  2013, \mnras, 434, L61

\bibitem[\protect\citeauthoryear{{Musso} \& {Sheth}}{{Musso} \&
  {Sheth}}{2012}]{musso12}
{Musso} M.,  {Sheth} R.~K.,  2012, ArXiv e-prints

\bibitem[\protect\citeauthoryear{{Musso} \& {Sheth}}{{Musso} \&
  {Sheth}}{2014a}]{musso14a}
{Musso} M.,  {Sheth} R.~K.,  2014a, \mnras, 439, 3051

\bibitem[\protect\citeauthoryear{{Musso} \& {Sheth}}{{Musso} \&
  {Sheth}}{2014b}]{musso14b}
{Musso} M.,  {Sheth} R.~K.,  2014b, \mnras, 438, 2683

\bibitem[\protect\citeauthoryear{{Navarro}, {Frenk} \& {White}}{{Navarro}
  et~al.}{1996}]{navarro96}
{Navarro} J.~F.,  {Frenk} C.~S.,    {White} S.~D.~M.,  1996, \apj, 462, 563

\bibitem[\protect\citeauthoryear{{Paranjape}, {Sheth} \&
  {Desjacques}}{{Paranjape} et~al.}{2013}]{paranjape13}
{Paranjape} A.,  {Sheth} R.~K.,    {Desjacques} V.,  2013, \mnras, 431, 1503

\bibitem[\protect\citeauthoryear{{Pillepich}, {Porciani} \&
  {Reiprich}}{{Pillepich} et~al.}{2012}]{pillepich12}
{Pillepich} A.,  {Porciani} C.,    {Reiprich} T.~H.,  2012, \mnras, 422, 44

\bibitem[\protect\citeauthoryear{{Planck Collaboration}, {Ade}, {Aghanim},
  {Armitage-Caplan}, {Arnaud}, {Ashdown}, {Atrio-Barandela}, {Aumont},
  {Baccigalupi}, {Banday} \& et al.}{{Planck Collaboration}
  et~al.}{2013b}]{planckxvi}
{Planck Collaboration} {Ade} P.~A.~R.,  {Aghanim} N.,  {Armitage-Caplan} C.,
  {Arnaud} M.,  {Ashdown} M.,  {Atrio-Barandela} F.,  {Aumont} J.,
  {Baccigalupi} C.,  {Banday} A.~J.,    et al. 2013b, ArXiv e-prints

\bibitem[\protect\citeauthoryear{{Planck Collaboration}, {Ade}, {Aghanim},
  {Armitage-Caplan}, {Arnaud}, {Ashdown}, {Atrio-Barandela}, {Aumont},
  {Baccigalupi}, {Banday} \& et al.}{{Planck Collaboration}
  et~al.}{2013a}]{planckxx}
{Planck Collaboration} {Ade} P.~A.~R.,  {Aghanim} N.,  {Armitage-Caplan} C.,
  {Arnaud} M.,  {Ashdown} M.,  {Atrio-Barandela} F.,  {Aumont} J.,
  {Baccigalupi} C.,  {Banday} A.~J.,    et al. 2013a, ArXiv e-prints

\bibitem[\protect\citeauthoryear{{Press} \& {Schechter}}{{Press} \&
  {Schechter}}{1974}]{press74}
{Press} W.~H.,  {Schechter} P.,  1974, \apj, 187, 425

\bibitem[\protect\citeauthoryear{{Reed}, {Smith}, {Potter}, {Schneider},
  {Stadel} \& {Moore}}{{Reed} et~al.}{2013}]{reed13}
{Reed} D.~S.,  {Smith} R.~E.,  {Potter} D.,  {Schneider} A.,  {Stadel} J.,
  {Moore} B.,  2013, \mnras, 431, 1866

\bibitem[\protect\citeauthoryear{{Retana-Montenegro}, {van Hese}, {Gentile},
  {Baes} \& {Frutos-Alfaro}}{{Retana-Montenegro}
  et~al.}{2012}]{retana-montenegro12}
{Retana-Montenegro} E.,  {van Hese} E.,  {Gentile} G.,  {Baes} M.,
  {Frutos-Alfaro} F.,  2012, \aap, 540, A70

\bibitem[\protect\citeauthoryear{{Rozo}, {Wechsler}, {Rykoff}, {Annis},
  {Becker}, {Evrard}, {Frieman}, {Hansen} \& {et~al.}}{{Rozo}
  et~al.}{2010}]{rozo10}
{Rozo} E.,  {Wechsler} R.~H.,  {Rykoff} E.~S.,  {Annis} J.~T.,  {Becker} M.~R.,
   {Evrard} A.~E.,  {Frieman} J.~A.,  {Hansen}   {et~al.} 2010, \apj, 708, 645

\bibitem[\protect\citeauthoryear{{Sartoris}, {Biviano}, {Fedeli}, {Bartlett},
  {Borgani}, {Costanzi}, {Giocoli}, {Moscardini}, {Weller}, {Ascaso},
  {Bardelli}, {Maurogordato} \& {Viana}}{{Sartoris} et~al.}{2015}]{sartoris15}
{Sartoris} B.,  {Biviano} A.,  {Fedeli} C.,  {Bartlett} J.~G.,  {Borgani} S.,
  {Costanzi} M.,  {Giocoli} C.,  {Moscardini} L.,  {Weller} J.,  {Ascaso} B.,
  {Bardelli} S.,  {Maurogordato} S.,    {Viana} P.~T.~P.,  2015,
  arXiv:1505.02165

\bibitem[\protect\citeauthoryear{{Sheth}, {Mo} \& {Tormen}}{{Sheth}
  et~al.}{2001}]{sheth01b}
{Sheth} R.~K.,  {Mo} H.~J.,    {Tormen} G.,  2001, \mnras, 323, 1

\bibitem[\protect\citeauthoryear{{Sheth} \& {Tormen}}{{Sheth} \&
  {Tormen}}{1999}]{sheth99b}
{Sheth} R.~K.,  {Tormen} G.,  1999, \mnras, 308, 119

\bibitem[\protect\citeauthoryear{{Sheth} \& {Tormen}}{{Sheth} \&
  {Tormen}}{2002}]{sheth02}
{Sheth} R.~K.,  {Tormen} G.,  2002, \mnras, 329, 61

\bibitem[\protect\citeauthoryear{{Springel}}{{Springel}}{2005}]{springel05a}
{Springel} V.,  2005, \mnras, 364, 1105

\bibitem[\protect\citeauthoryear{{Springel}, {White}, {Jenkins}, {Frenk},
  {Yoshida}, {Gao}, {Navarro}, {Thacker}, {Croton}, {Helly}, {Peacock}, {Cole},
  {Thomas}, {Couchman}, {Evrard}, {Colberg} \& {Pearce}}{{Springel}
  et~al.}{2005}]{springel05b}
{Springel} V.,  {White} S.~D.~M.,  {Jenkins} A.,  {Frenk} C.~S.,  {Yoshida} N.,
   {Gao} L.,  {Navarro} J.,  {Thacker} R.,  {Croton} D.,  {Helly} J.,
  {Peacock} J.~A.,  {Cole} S.,  {Thomas} P.,  {Couchman} H.,  {Evrard} A.,
  {Colberg} J.,    {Pearce} F.,  2005, Nature, 435, 629

\bibitem[\protect\citeauthoryear{{Springel}, {White}, {Tormen} \&
  {Kauffmann}}{{Springel} et~al.}{2001b}]{springel01b}
{Springel} V.,  {White} S.~D.~M.,  {Tormen} G.,    {Kauffmann} G.,  2001b,
  \mnras, 328, 726

\bibitem[\protect\citeauthoryear{{Tinker}, {Kravtsov}, {Klypin}, {Abazajian},
  {Warren}, {Yepes}, {Gottl{\"o}ber} \& {Holz}}{{Tinker}
  et~al.}{2008}]{tinker08}
{Tinker} J.,  {Kravtsov} A.~V.,  {Klypin} A.,  {Abazajian} K.,  {Warren} M.,
  {Yepes} G.,  {Gottl{\"o}ber} S.,    {Holz} D.~E.,  2008, \apj, 688, 709

\bibitem[\protect\citeauthoryear{{Tormen}}{{Tormen}}{1998}]{tormen98a}
{Tormen} G.,  1998, \mnras, 297, 648

\bibitem[\protect\citeauthoryear{{Tormen}, {Moscardini} \& {Yoshida}}{{Tormen}
  et~al.}{2004}]{tormen04}
{Tormen} G.,  {Moscardini} L.,    {Yoshida} N.,  2004, \mnras, 350, 1397

\bibitem[\protect\citeauthoryear{{Velliscig}, {Cacciato}, {Schaye}, {Bower},
  {Crain}, {van Daalen}, {Dalla Vecchia}, {Frenk}, {Furlong}, {McCarthy},
  {Schaller} \& {Theuns}}{{Velliscig} et~al.}{2015}]{velliscig15}
{Velliscig} M.,  {Cacciato} M.,  {Schaye} J.,  {Bower} R.~G.,  {Crain} R.~A.,
  {van Daalen} M.~P.,  {Dalla Vecchia} C.,  {Frenk} C.~S.,  {Furlong} M.,
  {McCarthy} I.~G.,  {Schaller} M.,    {Theuns} T.,  2015, ArXiv e-prints

\bibitem[\protect\citeauthoryear{{Vikhlinin}, {Kravtsov}, {Burenin}, {Ebeling},
  {Forman}, {Hornstrup}, {Jones}, {Murray} \& {et~al.}}{{Vikhlinin}
  et~al.}{2009}]{vikhlinin09}
{Vikhlinin} A.,  {Kravtsov} A.~V.,  {Burenin} R.~A.,  {Ebeling} H.,  {Forman}
  W.~R.,  {Hornstrup} A.,  {Jones} C.,  {Murray}   {et~al.} 2009, \apj, 692,
  1060

\bibitem[\protect\citeauthoryear{{Warren}, {Abazajian}, {Holz} \&
  {Teodoro}}{{Warren} et~al.}{2006}]{warren06}
{Warren} M.~S.,  {Abazajian} K.,  {Holz} D.~E.,    {Teodoro} L.,  2006, \apj,
  646, 881

\bibitem[\protect\citeauthoryear{{Watson}, {Iliev}, {D'Aloisio}, {Knebe},
  {Shapiro} \& {Yepes}}{{Watson} et~al.}{2013}]{watson13}
{Watson} W.~A.,  {Iliev} I.~T.,  {D'Aloisio} A.,  {Knebe} A.,  {Shapiro} P.~R.,
     {Yepes} G.,  2013, \mnras, 433, 1230

\bibitem[\protect\citeauthoryear{{Wu}, {Zentner} \& {Wechsler}}{{Wu}
  et~al.}{2010}]{wu10}
{Wu} H.-Y.,  {Zentner} A.~R.,    {Wechsler} R.~H.,  2010, \apj, 713, 856

\bibitem[\protect\citeauthoryear{{Zhao}, {Jing}, {Mo} \& {Bn{\"o}rner}}{{Zhao}
  et~al.}{2009}]{zhao09}
{Zhao} D.~H.,  {Jing} Y.~P.,  {Mo} H.~J.,    {Bn{\"o}rner} G.,  2009, \apj,
  707, 354

\end{thebibliography}
\label{lastpage}
\end{document}